\documentclass[10pt]{iopart}

\usepackage{iopams}  

\expandafter\let\csname equation*\endcsname\relax
\expandafter\let\csname endequation*\endcsname\relax
 \usepackage{amsmath}

\usepackage{graphicx}  
\usepackage{bbm}        
\usepackage{amssymb}
\usepackage{bm}
\usepackage{dsfont}
\usepackage{caption}
\usepackage{xcolor}
\usepackage{comment}
\usepackage{amssymb}
\usepackage{amsthm}

\newcommand{\grafe}[1]{\left\{ #1 \right\}}
\newcommand{\tonde}[1]{\left( #1 \right)}
\newcommand{\quadre}[1]{\left[ #1 \right]}

\makeatletter
\renewcommand\@appendixstar{\@@par
 \ifnumbysec 
 \@addtoreset{table}{section}
 \@addtoreset{figure}{section}\fi
 \setcounter{section}{0}
 \setcounter{subsection}{0}
 \setcounter{subsubsection}{0}
 \setcounter{equation}{0}
 \setcounter{figure}{0}
 \setcounter{table}{0}
 \def\thesection{\Alph{section}} 
 \def\theequation{\ifnumbysec
      \Alph{section}.\arabic{equation}\else
      \Alph{section}\arabic{equation}\fi}
 \def\thetable{\ifnumbysec
      \Alph{section}\arabic{table}\else
      A\arabic{table}\fi}
 \def\thefigure{\ifnumbysec
      \Alph{section}\arabic{figure}\else
      A\arabic{figure}\fi}}
\makeatother

\makeatletter
\let\old@mathcal=\mathcal
\usepackage{dutchcal}
\let\old@dutchcal=\mathcal
\renewcommand{\mathcal}[1]{%
 \if A#1
     \old@mathcal{A}
  \fi
  \if a#1
     \old@dutchcal{a}
  \fi
   \if D#1
     \old@mathcal{D}
  \fi
  \if d#1
     \old@dutchcal{d}
  \fi
   \if I#1
     \old@mathcal{I}
  \fi
  \if i#1
     \old@dutchcal{i}  
  \fi
    \if J#1
     \old@mathcal{J}
  \fi
  \if j#1
     \old@dutchcal{j}  
  \fi
     \if G#1
     \old@mathcal{G}
  \fi
  \if g#1
     \old@dutchcal{g}  
  \fi
 \if N#1
     \old@mathcal{N}
  \fi
  \if n#1
     \old@dutchcal{n}
  \fi
   \if P#1
     \old@mathcal{P}
  \fi
  \if p#1
     \old@dutchcal{p}
  \fi
    \if V#1
     \old@mathcal{V}
  \fi
  \if v#1
     \old@dutchcal{v}
  \fi
   \if R#1
     \old@mathcal{R}
  \fi
  \if r#1
     \old@dutchcal{r}
  \fi
  \if Y#1
     \old@mathcal{Y}
  \fi
  \if y#1
     \old@dutchcal{y}
  \fi
  \if X#1
     \old@mathcal{X}
  \fi
  \if x#1
     \old@dutchcal{x}
  \fi
}

\makeatother

\begin{document}

\title[]{Quenched complexity of equilibria for asymmetric Generalized Lotka-Volterra equations}

\author{Valentina Ros}
\address{Universit\'e Paris--Saclay, CNRS, LPTMS, Orsay, France}
\ead{valentina.ros@universite-paris-saclay.fr}

\author{Felix Roy}
\address{Laboratoire de Physique de l’Ecole Normale Sup\'{e}rieure, ENS, Universit\'{e} PSL,
CNRS, Sorbonne Universit\'{e}, Universit\'{e} de Paris, F-75005 Paris, France}

\author{Giulio Biroli}
\address{Laboratoire de Physique de l’Ecole Normale Sup\'{e}rieure, ENS, Universit\'{e} PSL,
CNRS, Sorbonne Universit\'{e}, Universit\'{e} de Paris, F-75005 Paris, France}

\author{Guy Bunin}
\address{Department of Physics, Technion-Israel Institute of Technology, Haifa 32000, Israel}


\vspace{10pt}

\begin{abstract}
We consider the Generalized Lotka-Volterra system of equations with all-to-all, random asymmetric interactions describing high-dimensional, very diverse and well-mixed ecosystems. We analyze the multiple equilibria phase of the model and compute its quenched complexity, i.e., the expected value of the logarithm of the number of equilibria of the dynamical equations. We discuss the resulting distribution of equilibria as a function of their diversity, stability and average abundance. We obtain the quenched complexity by means of the replicated Kac-Rice formalism, and compare the results with the same quantity obtained within the annealed approximation, as well as with the results of the cavity calculation and, in the limit of symmetric interactions, of standard methods to compute the complexity developed in the context of glasses. 
\end{abstract}

\tableofcontents

%
%
%
%

\section{Introduction}

When modelling complex systems in biology, ecology or economics one is typically forced to give up the framework of conservative systems, and more generally of physical systems that are isolated or in contact with a heat bath; indeed, the interactions in these systems are often directional and asymmetric, and thus the dynamical equations describing the evolution of species, agents or individual constituents do not take a gradient form.
In these types of problems, the asymmetry is a distinctive feature which adds to the other standard ingredients in complex systems modelling, such as  the high-dimensionality and randomness.

Some of the interesting phenomenology of conservative, high-dimensional systems with random interactions persists in presence of asymmetry; for instance, the competition between random and deterministic contributions tends to generate transitions between complex and simple phases. In conservative systems, these transitions can be understood in terms of the structure of the high-dimensional energy landscapes associated to the system: they separate regimes in which the landscape is very non-convex and rugged (as the energy landscape of toy models of glasses \cite{mezard1987spin, PUZbook}) from regimes in which the landscape has a much smoother or even convex shape. They are referred to as \emph{topology trivialization transitions} \cite{fyodorov2004complexity,fyodorov2007replica,FyoNad12,FLD2013}. In non-conservative systems this landscape interpretation can not be invoked. Nevertheless, these transitions still have a meaning in terms of  the dynamical equations associated to the system: complex phases are characterized by the presence of plenty of equilibria (fixed points or, more generally, dynamical attractors) with different stability properties, while simple phases are characterized by few, in some cases a single one, of these equilibria \cite{may1977thresholds,amari1972characteristics, sompolinsky1988chaos,ispolatov2015chaos}. When the dimensionality is high, in the complex phase the number of equilibria  can diverge exponentially with the dimensionality. The interest lies in characterizing this complexity quantitatively, i.e. in estimating the number of equilibria and in classifying them in terms of their dynamical stability or other relevant properties \cite{hong2021exclusion, lischke2017finding, logofet2005stronger}.

 Despite the analogy with glasses, for which these questions have been tackled extensively in the last decades, getting a quantitative understanding of the complex, multiple-equilibria  phases in non-conservative systems is in general a harder task. One of the reasons for this is that several of the tools developed in the context of glasses are precluded, as they rely on variations of equilibrium-like calculations \cite{monasson1995structural, franz1995recipes} and are therefore specifically conceived for systems associated to energy and free energy landscapes. Alternative approaches not relying explicitly on the landscape formulation have also been developed in the theoretical literature on glasses \cite{cavagna1998stationary}. They essentially make use of the so called Kac-Rice formalism, that has been revisited and developed extensively in the recent years (see \cite{ReviewLandscape} for a review); in particular, it has been pointed out that the counting problem for random systems can essentially be formulated as a problem of random matrix theory. This observation is at the root of most of the more recent developments in the field, and it also opened the door to a mathematically rigorous formulation of the problem. Complex phases and trivialization transitions in non-gradient systems have been studied within the Kac-Rice formalism in~\cite{WT13,FK16,F2016,Ipsen,BAFK21,FFI21,LCTF22}. 
 
These recent works, however, address the problem within the so called \emph{annealed} approximation, which allows one to  determine the asymptotics (in the system's dimension) of the \emph{average} number of equilibria of the dynamical equations. In high-dimension, however, the number of equilibria in the complex phase is in general a \emph{broadly} distributed random variable, whose average is  dominated by realizations of the random interactions occurring with extremely small probability. In this setting, the interest lies in characterizing the \emph{typical} value of the number of equilibria, rather than its average. This problem, which is well known in the theory of glasses as it applies there to the random partition function of the system, requires to go beyond the annealed setting and to perform \emph{quenched} calculations. The replica trick, which entails that typical values can be extracted from the calculation of the higher moments, is the well-known tool to address this problem. In this work, we embed the replica trick into the Kac-Rice formalism to perform a quenched calculation of the number of equilibria. We follow the approach introduced in Ref.~\cite{MioSegnale}, applying it here to a non-conservative dynamical system. 
 
 We focus on the Generalized Lotka-Volterra system of equations with all-to-all, random asymmetric interactions. These equations have been used quite extensively recently in their high-dimensional setting to model  diverse, well-mixed ecosystems such as bacterial ecosystems \cite{Hu2022}. We consider asymmetric pairwise random interaction couplings with Gaussian statistics, and compute the typical number of equilibria of the system of equations as a function of the parameters of the model, as well as of the diversity (fraction on coexisting species) of the equilibria. We show explicitly how the annealed calculation gives a non-tight upper bound to the \emph{typical} number of equilibria in the complex phase, and discuss also the connection with other approaches discussed in the literature, based on the cavity formalism \cite{mezard1987sk, GuyCavity, barbier2017cavity}.

 The work is structured as follows: in \Sref{sec:Model} we introduce the model, we summarize previous results and we discuss the difference between quenched and annealed calculations. In \Sref{sec:StructureKR} we present the structure of the quenched calculation of the typical number of equilibria, and derive the self-consistent equations defining the problem. In \Sref{sec:SCE} we focus on the case of totally uncorrelated interactions and describe in detail the solution procedure of the self-consistent equations. In \Sref{sec:Results} we present our results for the uncorrelated case, and in Section \ref{sec:GeneralGamma} we discuss generalizations to correlated but asymmetric interactions. Finally, in \Sref{sec:Discussion} we present our conclusions. Further technical details on the calculations are given in the Appendices. This work presents a detailed account of the  structure of the quenched calculation: for a more concise summary of the results and their interpretation, we refer the readers to Ref.~\cite{ros2022generalized}.

\section{The model and its multiple equilibria phase}\label{sec:Model}
We consider the generalized Lotka-Volterra equations describing the evolution of species in well-mixed ecosystems, interacting pairwise with random couplings. We are interested in ecosystems with a large number of species: we let $i=1, \cdots, S \gg 1$ label the different species belonging to a species pool. $N_i(t) \geq 0$ denotes the abundance of species $i$ at a given time $t$, and $\vec{N}(t)= (N_1(t), \cdots, N_S(t))$ a configuration of the ecosystem. We neglect the  discreteness of the $N_i$, setting  $N_i \in [0, +\infty)$. The generalized Lotka-Volterra equations describing the evolution of a community read:
\begin{equation}\label{eq:SystemEqs}
\frac{d{N}_i(t)}{dt}=N_i(t) \tonde{\kappa_i -  N_i(t) - \sum_{j =1}^S   \alpha_{ij} N_j(t)} \quad  \forall \;  i, \quad \quad N_i(t)
 \geq 0\end{equation}
where $\kappa_i$ is the carrying capacity of the species $i$, and $\alpha_{ij}$ are components of a random matrix encoding the interactions between the different species. We choose each $\alpha_{ij}$ to be a Gaussian random variable with a statistics characterized by three parameters $\mu, \sigma, \gamma$: 
\begin{equation} 
\langle \alpha_{i j} \rangle=\frac{\mu}{S}, \quad \quad  \langle \alpha_{ij} \alpha_{kl} \rangle_c=\langle \alpha_{ij} \alpha_{kl} \rangle-\langle \alpha_{ij} \rangle \langle \alpha_{kl} \rangle=\frac{\sigma^2}{S}[ \delta_{i k} \delta_{jl} + \gamma \delta_{il} \delta_{jk}],
\end{equation}
and denote with $ \mathbb{P} (\grafe{\alpha_{ij}}_{ij}) $ the joint distribution of these variables, and with $\langle \cdot \rangle$ the average with respect to it \footnote{In this formulation, the couplings $\alpha_{ii}$ are non-zero and have the same statistics as the $\alpha_{ij}$ with $i \neq j$. We remark that one may absorb these terms in the carrying capacities $\kappa_i$ and work with $\alpha_{ii}=0$.  These different choices do not affect any result on the complexity derived in this work. }.
The parameter  $\gamma \in [-1,1]$ encodes the (a-)symmetry properties of the interaction matrix, while $\mu, \sigma$ measure the average strength of the interactions, and their variability. We set:
\begin{equation}\label{eq:CenteredMatrix}
\alpha_{ij}= \frac{\mu}{S} + \frac{\sigma}{\sqrt{S}} a_{ij}, \quad \quad \langle a_{i j} \rangle=0, \quad \quad  \langle a_{ij} a_{kl} \rangle=\delta_{i k} \delta_{jl} + \gamma \delta_{il} \delta_{jk}.
\end{equation} 
We focus on the case in which carrying capacities are equal for all species, $\kappa_i = \kappa$, even though the analysis can be easily generalized to the case of inhomogeneous carrying capacities. The case of symmetric interactions $\gamma=1$ has been discussed quite extensively in previous literature, both in the dense \cite{GuyCavity,LVMarginality,AltieriRoyTemperature} and sparse \cite{marcus2021local} case. Here we are rather interested in asymmetric interactions $\gamma<1$; we focus in particular on the \emph{uncorrelated} case $\gamma=0$, when the entries $\alpha_{ij}$ and $\alpha_{ji}$ are independent random variables. 

For general $\gamma$, this system of equations has been studied analytically in the large-$S$ limit by means of the so called {cavity method} \cite{GuyCavity}. The cavity analysis and the numerical simulation of the dynamics~\cite{RoyDMFT,rieger1989solvable} have revealed the existence of three distinct regimes: (i) a \emph{unique equilibrium} regime in which any arbitrary initialization $N_i(0)$ of the population vector converges to a fixed equilibrium value $N^*_i$,  (ii) a \emph{multiple equilibria} regime in which the dynamics is attracted by a succession of different configurations, the ecosystem remaining in their vicinity for some time before being pushed away along unstable directions in the species space, and (iii) an \emph{unbounded} regime, where the abundance of some species diverges as a function of time. The system is driven from one regime to the others by tuning the variability $\sigma$, at fixed $\mu, \gamma$, see Ref.~\cite{GuyCavity} for a phase diagram. For $\mu>0$ and $\gamma \in (-1,1]$, the unique and multiple equilibria regimes are separated by a sharp transition line  at $\sigma_c=\sqrt{2}(1+ \gamma)^{-1}$~\cite{rieger1989solvable}.  The cavity approach captures exactly the equilibrium of the system in the regime in which one single equilibrium exists, but it is only approximate within the multiple equilibria regime. The latter has been studied in more detail in the symmetric case $\gamma=1$ \cite{diederich1989replicators,biscari1995replica,GuyCavity,LVMarginality,AltieriRoyTemperature, marcus2022local}, where the model \eref{eq:SystemEqs} can be treated with standard methods of disordered systems applicable to conservative systems, i.e. systems admitting a potential function.
The totally antisymmetric case $\gamma=-1$ in absence of the niche-like term (the term $-N_i(t)$ in the right-hand side of Eq.~\eqref{eq:SystemEqs})
has been looked at in~\cite{pearce2019stabilization}.
Here we aim at characterizing the multiple equilibria regime away from $\gamma=\pm 1$, by computing explicitly the number of equilibria of the dynamical equations \eref{eq:SystemEqs} as a function of their properties defined below. 

\subsection{Equilibria and their properties: diversity, abundance, stability}
Equilibria are special configurations $\vec{N}^{*}$  satisfying:
\begin{equation}\label{eq:SystemEqs2}
\frac{d N^*_i}{dt}=N^*_i  F_i(\vec{N}^*)=0 \quad  \quad  N_i^* \geq 0  \quad \quad \forall \;  i,
\end{equation}
where we have introduced the vector $\vec{F}$ with components:
\begin{equation}\label{eq:GradDef}
F_i(\vec{N})=\kappa -  N_i - \sum_{j =1}^S   \alpha_{ij} N_j,
\end{equation}
which we refer to as ``vector of forces" in analogy with constraint satisfaction problems~\cite{franz2016simplest}, and whose ecological interpretation is the growth-rate in the configuration $\vec{N}$. 
When many equilibria configurations are present, they can be classified according to their \emph{diversity}, \emph{average abundance} and \emph{stability}. 
Each equilibrium configuration $\vec{N}^*$ is characterized by a certain number of species that are absent,  $N_i^*=0$: the diversity  $\phi$ measures the fraction of species that  coexist in the configuration. To define it, we let $I(\vec{N}^*)=(i_1, \cdots, i_{s})$ with $0 \leq s(\vec{N}^*) \leq S$ be an index set collecting the indices of the species that coexist in the given equilibrium $\vec{N}^*$, i.e. $N_i^*>0$ for $i \in I$ and $N_i^*=0$ for $i \notin I$. We then set:    
\begin{equation}\label{eq:Diversity}
\phi(\vec{N}^*) = \lim_{S \to \infty} \frac{|I(\vec{N}^*)|}{S}= \lim_{S \to \infty} \frac{s(\vec{ N}^*)}{S}  \in \quadre{0,1}. 
\end{equation}
Each of the species that coexist in a given equilibrium contribute to its average abundance, which we define through another intensive parameter $m$ defined by:
\begin{equation}\label{eq:Mass}
 m(\vec{ N}^*) = \lim_{S \to \infty} \frac{1}{S} \sum_{i=1}^S N^*_i.
\end{equation}
We classify equilibria according to two different notions of dynamical stability; first, we consider the stability with respect to the species that are absent ($N_i^*=0$), and define the equilibrium \emph{uninvadable} if it is stable with respect to the re-introduction of these species from the species pool. This corresponds to the requirement that its growth-rate is negative if the species are introduced in small numbers, i.e., $F_i(\vec{ N}^*)<0$ for any $i \notin I$. In addition, we consider the stability with respect to perturbations  $N^*_i \to N^*_i + \delta N^*_i$ of the populations of the species that coexist: the equilibrium is stable if the system initialized in $N^*_i + \delta N^*_i$  is driven back to the nearby equilibrium configuration $N^*_i$ by the linearized version of the dynamics \eref{eq:SystemEqs}. This linear stability is controlled~\cite{stone2018feasibility} by the interaction matrix restricted to the subspace of coexisting species, 
\begin{equation}\label{eq:Hessian}
H_{ij}(\vec{ N}^*)=  \tonde{\frac{\delta F_i (\vec{ N}^*)}{dN_j} }_{i, j \in I(\vec{ N}^*)}.
\end{equation}
The equilibrium is linearly stable if the eigenvalues of this matrix have all negative real part. Notice that as it follows from the fact that the interactions couplings $\alpha_{ij}$ are random and asymmetric, the matrices \eref{eq:Hessian} are themselves asymmetric random matrices.
As we shall show below, all equilibria ${\bf N}^*$ with a given diversity $\phi$ are associated with matrices \eqref{eq:Hessian} displaying the same distribution of eigenvalues in the limit $S \to \infty$ (we discuss the role of subleading $1/S$ contributions to the density of eigenvalues in Section \ref{sec:Ginibre}). In particular, the dominant part of the eigenvalue distribution is supported on the negative real sector provided that the diversity does not exceed a critical value given by:
\begin{equation}\label{eq:MayBound}
\phi < \phi_{\rm May}= \frac{1}{\sigma^2 (1+ \gamma)^2}.
\end{equation}
When $\phi= \phi_{\rm May}$, the support of the spectrum touches zero and the corresponding equilibrium is marginally stable; for larger values of $\phi$, the equilibrium is unstable. Not surprisingly, the stability criterion \eref{eq:MayBound} is related to that identified by R.~May when studying the linear stability of random ecosystems assumed to be in the vicinity of an equilibrium configuration \cite{may1972will}. We henceforth refer to it as the \emph{May stability bound}.

\subsection{Counting equilibria: quenched and annealed complexity}\label{sec:CountingComplexity}
 Let $\mathcal{N}_{S}(\phi )$ denote the total number of uninvadable equilibria with fixed diversity $\phi$.
In the unique equilibrium phase, $\mathcal{N}_{S}(\phi )$ is a self-averaging random variable which is expected to be equal to one at the value of diversity corresponding to the equilibrium, and equal to zero otherwise \cite{ros2022generalized}. In the multiple equilibria phase, instead, $\mathcal{N}_{S}(\phi )$ is expected to scale exponentially with $S$, in analogy with the number of stable configurations of complex systems such as spin glasses~\cite{mezard1987spin}. The logarithm of $\mathcal{N}_{S}(\phi )$, or complexity, is self-averaging in the large $S$ limit. We therefore define the \emph{quenched complexity} $\Sigma^{(Q)}_\sigma(\phi )$ of equilibria from:
\begin{equation}\label{eq:QuenchedComp}
\Sigma^{(Q)}_\sigma(\phi )= \lim_{S \to \infty} \frac{1}{S} \left \langle \log \quadre{ \mathcal{N}_{S}(\phi)} \right \rangle,
\end{equation}
where the average is performed with respect to the random interactions $\alpha_{ij}$ at fixed values of the parameters $\mu, \sigma, \gamma$ and $\kappa$ (to simplify the notation, henceforth we neglect the dependence of all quantities on these parameters). 
The quenched complexity controls the exponential scaling of the typical number of equilibria when $S$ is large; in the multiple equilibria phase, it is different from zero in a whole range of diversities $\phi$. By convexity, it is bounded from above by the so called \emph{annealed complexity}: 
\begin{equation}\label{eq:AnnComp}
\Sigma^{(A)}_\sigma(\phi )= \lim_{S \to \infty} \frac{1}{S} \log \left \langle { \mathcal{N}_{S}(\phi )} \right \rangle,
\end{equation}
which controls the exponential scaling of the average number of equilibria. The latter quantity is the one usually considered in the Kac-Rice literature \cite{FyodorovNonlinearAnalogue}. It is easier to compute than the quenched complexity, since it only requires to compute the asymptotic scaling of the average value of the random variable $\mathcal{N}_{S}(\phi )$. In contrast, the calculation of the average of the logarithm in \eref{eq:QuenchedComp} requires to determine the behaviour of arbitrarily high moments of $\mathcal{N}_{S}(\phi )$, from which the logarithmic average can be obtained using the replica trick:
\begin{equation}\label{eq:RepTrick}
 \left \langle \log { \mathcal{N}_{S}(\phi )} \right \rangle = \lim_{n \to 0} \frac{\log \left \langle { \mathcal{N}^n_{S}(\phi )} \right \rangle}{n}.
\end{equation}
In complex disordered systems, the quenched complexity is the relevant quantity to look at in the limit of large $S$, as it characterizes the behaviour of the system for typical realizations of the random couplings. The annealed complexity describes instead the average behavior, and may be dominated by rare instances of the randomness. As we show below, within the multiple equilibria phase quenched and annealed complexity differ for most values of diversities, while they coincide in the unique equilibrium phase.

\section{Getting the quenched complexity: the replicated Kac-Rice calculation}\label{sec:StructureKR}
The replicated Kac-Rice method provides us with a formula for the moments of the random variable $\mathcal{N}_{S}(\phi)$ appearing in~\eref{eq:RepTrick}. More generally, Kac-Rice formulas are used to determine the statistics (in most cases, the average) of the number of solutions of a random system of equations. Let $I$ denote an index set, collecting a fraction of the indices $i=1, \cdots, S$. For the system \eqref{eq:SystemEqs2} at fixed realization of the random couplings $\alpha_{ij}$, the number of uninvadable solutions having diversity $\phi$ can be formally written as: 
\begin{equation}\label{eq:KRdisdep}
\begin{split}
 \mathcal{N}_S(\phi)= \hspace{-.2 cm} \sum_{\begin{small} I:  |I|= S \phi  \end{small}}   \hspace{-.1 cm} \int d{\vec N}\, d {\vec f}  \prod_{i \in I} \theta (N_i) \, \delta( f_i) \prod_{i \notin I} \delta( N_i) \theta( -f_i) \left| {\rm det} \tonde{\frac{\delta F_i}{dN_j}}_{\begin{small}i, j \in I \end{small}}\right|  \delta \hspace{-.1 cm}\tonde{\vec{F}(\vec{N})- \vec{ f}}.
\end{split}
\end{equation}
In this expression, the sum is over all possible choices of configurations $\vec{N}$ such that the species $i \in I$ are present; the delta functions enforce that whenever a species $i \notin I$ is absent, the corresponding component of the force vector takes a negative value, implying uninvadibility; finally, the absolute value of the determinant accounts for the fact that the equilibrium equations \eqref{eq:SystemEqs2} are non-linear in the variables $N_i$, and thus may admit a multiplicity of solutions. To introduce the formula for the $n$-th moment of this quantity, we need to introduce $n$ different configurations $\vec{ N}^{a}$ for $a=1, \cdots n$, which we refer to as \emph{replicas.} Let ${\bf N}= (\vec{ N}^{1}, \cdots, {\vec N}^{n})$ denote the concatenation of configurations of all replicas. For each replica, let  $I_a = I({\vec N}^{a})$ be the index set collecting the indices of coexisting species, such that $|I_a|=S \phi$ for all $a$. We introduce the vectors of forces ${\vec F}^{a}= {\vec F}({\vec N}^{a})$ and ${\bf F}({\bf N})=({\vec F}^{1}, \cdots, {\vec F}^{n})$. Let ${\bf f}$ be the value taken by this random vector, and $\mathcal{P}^{(n)}_{{\bf N}} \tonde{ {\bf f}}$ the joint distribution of the $S$-dimensional vectors ${\vec F}^{a}$ evaluated at ${\vec f}^{a}$,
\begin{equation}
\mathcal{P}^{(n)}_{{\bf N}} \tonde{ {\bf f}}=\int \prod_{i,j=1}^S \,d \alpha_{ij} \mathbb{P} (\grafe{\alpha_{ij}}_{ij}) \, \delta \tonde{{\bf F}({\bf N})- {\bf f}}.
\end{equation}
We also introduce the following conditional expectation value:
\begin{equation}\label{eq:ExpectationJointDet}
 \mathcal{D}^{(n)}_{{\bf N}} \tonde{ {\bf f}}= \left \langle  \tonde{\prod_{a=1}^n  \left| {\rm det} \tonde{\frac{\delta F^{a}_i}{dN_j^{a}}}_{i, j \in I_a}\right| }       \; \; \Big| \; {\bf  F}({\bf N})= {\bf f}\right\rangle.
 \end{equation}
The latter is the expectation of the product of the absolute values of $n$ determinants of the $S \phi \times S \phi$ matrices of derivatives of the components of ${\bf F}$, conditioned to ${\bf F}$ taking value ${\bf f}$.
The Kac-Rice formula for the $n$-th moment of the number  $\mathcal{N}_{S}(\phi)$ of uninvadable equilibria then reads:
\begin{equation}\label{eq:KRmoments}
\begin{split}
\left \langle \mathcal{N}^n_S(\phi) \right \rangle= &\sum_{\begin{array}{c}  I_1\\  |I_1|= S \phi  \end{array}} \cdots  \sum_{\begin{array}{c}  I_n\\  |I_n|= S \phi  \end{array}}  \prod_{a=1}^n  \int d{\vec N}^{a}\, d {\vec f}^{a}  \prod_{i \in I_a} \theta (N_i^{a}) \, \delta( f_i^{a})\\
&\times \prod_{i \notin I_a} \delta( N_i^{a}) \theta( -f_i^{a}) \mathcal{D}^{(n)}_{{\bf N}} \tonde{ {\bf f}} \mathcal{P}^{(n)}_{{\bf N}} \tonde{ {\bf f}}.
\end{split}
\end{equation} 
In  \eqref{eq:KRmoments}  and henceforth, $\theta(x)$ denotes the Heaviside step function, taking value one if $x>0$ and zero otherwise. By inspecting \eqref{eq:KRmoments}, one realizes that the Kac-Rice formula is obtained averaging powers of \eqref{eq:KRdisdep} over the random variables $\alpha_{ij}$: in particular, the expectation value  \eqref{eq:ExpectationJointDet} arises after enforcing the constraint ${\bf F}({\bf N})={\bf f}$ as a conditioning. In the following, we determine the behaviour of the moments \eref{eq:KRmoments} for generic values of $n$ to leading exponential order in $S$, and extract the complexity from it. The main steps of the calculation are summarized below, while we refer to the Appendices for details.\\

\subsection{ Large-$S$ expansion and order parameters}
The quantities $\mathcal{D}^{(n)}_{\bf N} \tonde{\bf f}$ and $\mathcal{P}^{(n)}_{\bf N} \tonde {\bf f}$ in \eref{eq:KRmoments}  depend on the vectors ${\bf N}^{a}$ and ${\bf f}^{a}$:  as we show explicitly in \Sref{sec:CalculationMoments}, however, the dependence on these vectors enters only through their scalar products. For $a,b=1, \cdots, n$ we can therefore introduce a set of \emph{order parameters} defined as follows:
\begin{equation}\label{eq:OrderParameters}
\begin{split}
  S q_{ab}= \vec{ N}^a \cdot \vec{ N}^b,\quad 
   S \xi_{ab}= \vec{ f}^a \cdot \vec{ f}^b,\quad
   S z_{ab}= \vec{ N}^a \cdot \vec{ f}^b,\quad
   S m_a= \vec{ N}^a \cdot \vec{ 1},\quad
   S p_a= \vec{ f}^a \cdot \vec{ 1},
  \end{split}
\end{equation}
where $\vec{ 1}= (1, \cdots, 1)^T$ is an $S$-dimensional vector with all entries equal to one.  
Let ${\bf x}$ denote the collection of all of these order parameters. We can re-write $\mathcal{P}^{(n)}_{\bf N} \tonde {\bf f} \to  \mathcal{P}_{n} ({\bf x}, \phi)$,  $\mathcal{D}^{(n)}_{\bf N} \tonde {\bf f} \to  \mathcal{D}_{n}( {\bf x}, \phi)$ and thus
\begin{equation}
\left \langle \mathcal{N}^n_S(\phi) \right \rangle = \int d {\bf x} \,{v}_n({\bf x}, \phi) \, \mathcal{D}_n( {\bf x}, \phi)\, \mathcal{P}_n( {\bf x}, \phi)
\end{equation}
where $ {v}_n({\bf x}, \phi)$ is a compact notation for the “volume" term: 
\begin{equation}
\begin{split}
{v}_n({\bf x}, \phi) &=S^{2n(n+1)}\sum_{\begin{subarray}{c}  I_1\\  |I_1|= S \phi  \end{subarray}} \cdots  \sum_{\begin{subarray}{c}  I_n\\  |I_n|= S \phi  \end{subarray}}  \prod_{a=1}^n  \int d {\vec N}^{a}\, d {\vec f}^{a}  \prod_{i \in I_a} \theta (N_i^{a}) \, \delta( f_i^{a}) \prod_{i \notin I_a} \delta( N_i^{a}) \theta( -f_i^{a})\\
&\times \prod_{a, b=1}^n \delta \tonde{\vec{ N}^a \cdot \vec{ N}^b-S q_{ab}} 
\prod_{a,b=1}^n \delta \tonde{\vec{ f}^a \cdot \vec{ f}^b- S \xi_{ab}} \prod_{a \neq b=1}^{n} \delta \tonde{ \vec{ N}^a \cdot \vec{ f}^b-S z_{ab}} \\
&\times\prod_{a=1}^n \delta \tonde{\vec{ N}^a \cdot \vec{ 1}- S m_a} \delta \tonde{ \vec{ f}^a \cdot \vec{ 1}-S p_a},
\end{split}
\end{equation}
where we used the fact that the equilibrium constraint imposes $z_{aa}=\vec{ N}^a \cdot \vec{ f}^a=0$ for any $a$\footnote{Indeed, the constraint that $\vec{ N}^a$ is an equilibrium configuration satisfying \eqref{eq:SystemEqs2} implies that for each $i=1, \cdots, S$, either $N_i^a$ vanishes or $F^a_i(\vec N^a)$, which we imposed to take value $f_i^a$, vanishes. From this it follows that $\vec{ N}^a \cdot \vec{ f}^a=0$ }. To compute the volume term, it is convenient to introduce integral representations of the constraints via \emph{conjugate parameters}, e.g.:
\begin{equation}
 \delta \tonde{\vec{ N}^a \cdot \vec{ N}^b-S q_{ab}} = \int \frac{d \hat{q}_{ab}}{2 \pi} e^{i \hat q_{ab} \tonde{\vec{ N}^a \cdot \vec{ N}^b-S q_{ab}}},
\end{equation}
and similarly for all other order parameters.  For each replica, we introduce a $S$-dimensional vector $\vec{\tau}^a$ with components $0,1$ such that $\tau^a_i= 1$ if $i \in I_a$ (meaning, if the species labelled by $i$ is present in the configuration ${\vec N}^a$), while $\tau^a_i= 0$ if $i \notin I_a$. The constraint on the diversity $\phi$ implies $\vec{\tau}^a \cdot \vec{\tau}^a= S \phi$ for any $a$: we enforce this constraint in a similar way as above, introducing a conjugate parameter $\hat \phi_a$ for each replica. We let $\hat {\bf x}$ denote the collection of all these conjugate parameters. After rotating on the complex plane of the conjugate variables, we can write:
\begin{equation}\label{eq:bSP}
\left \langle \mathcal{N}^n_S(\phi) \right \rangle =\frac{1}{(2 \pi)^{2 n^2 + 3 n}} \int d {\bf x} \, i d  \hat{\bf x}\; e^{S g_n ({\bf x}, \hat{\bf x}, \phi)}\, \mathcal{V}_n({\bf x}, \hat {\bf x}) \, \mathcal{D}_n( {\bf x}, \phi)\, \mathcal{P}_n( {\bf x}, \phi)
\end{equation}
where now
\begin{equation}
g_n ({\bf x}, \hat{\bf x}, \phi)=\sum_{a=1}^n \tonde{\hat m_a m^a + \hat p_a p_a + \hat \phi_a \phi}+ \sum_{a,b=1}^n \tonde{\hat q_{ab}  q_{ab} +\hat \xi_{ab}  \xi_{ab}}+ \sum_{a \neq b} \hat z_{ab} z_{ab}
\end{equation}
and 
\begin{equation}\label{eq:startAppVol}
\begin{split}
&\mathcal{V}_n({\bf x},\hat {\bf x}) =S^{4n(n+1)+n}\sum_{{\tau^1_i =0,1}} \cdots  \sum_{\tau^n_i=0,1} \;    \int  \prod_{a=1}^n \, d {\vec N}^{a}\, d {\vec f}^{a} \, e^{- \hat m_a \,{\vec{ N}^a \cdot \vec{ 1}}  - \hat p_a \,{\vec{ f}^a \cdot \vec{ 1}} -  \hat \phi_a \, \vec \tau^a \cdot \vec \tau^a}\\
&\times   \prod_{a \neq b=1}^n e^{- \hat z_{ab} \, \vec{ N}^a \cdot \vec{ f}^b} 
\prod_{a, b=1}^n   \,e^{-\hat q_{ab} \,\vec{ N}^a \cdot \vec{ N}^b- \hat \xi_{ab} \,\vec{ f}^a \cdot \vec{ f}^b} \prod_{a=1}^n \tonde{\prod_{i:  \tau_i^a=1} \theta (N_i^{a}) \, \delta( f_i^{a}) \prod_{i: \tau_i^a=0} \delta( N_i^{a}) \theta( -f_i^{a})}.
\end{split}
\end{equation}

 The formula \eref{eq:bSP} holds for arbitrary {integer} value of $n$, and for arbitrary $S$. In order to extract the quenched complexity \eref{eq:QuenchedComp}, as well as the annealed one \eref{eq:AnnComp}, we exploit the saddle point approximation, which requires first to determine the behaviour of the integrand in \eref{eq:bSP} to leading order in large $S$. More precisely, we aim at determining a function $ \mathcal{A}_n({\bf x} , \hat{\bf x}, \phi)$ such that:
\begin{equation}\label{eq:bSP-rs}
\left \langle \mathcal{N}^n_S(\phi) \right \rangle = \int d {\bf x} \,  id \hat{\bf x}\; e^{S \, \mathcal{A}_n({\bf x} , \hat{\bf x}, \phi) + o(S)}.
\end{equation}
Given this function, the annealed complexity can be obtained setting $n=1$ and optimizing $\mathcal{A}_1({\bf x} , \hat{\bf x}, \phi)$ over the order and conjugate parameters: defining the annealed saddle point values ${\bf x}_1, \hat{\bf x}_1$ from:
 \begin{equation}\label{eq:VariationalAnnealed}
\frac{\delta \mathcal{A}_1({\bf x} , \hat{\bf x}, \phi) }{\delta {\bf x}} \Big|_{{\bf x}_1, \hat{\bf x}_1}=0, \quad \quad \frac{\delta \mathcal{A}_1({\bf x} , \hat{\bf x}, \phi)}{\delta \hat {\bf x}} \Big|_{{\bf x}_1, \hat{\bf x}_1}=0,
\end{equation}
we readily obtain from  \eref{eq:AnnComp} that:
\begin{equation}\label{eq:AnnComp2}
\Sigma^{(A)}_\sigma(\phi)= \mathcal{A}_1({\bf x}_1 , \hat{\bf x}_1, \phi).
\end{equation}
The calculation of the quenched complexity via the replica trick \eref{eq:RepTrick} requires an additional step, that is the analytic continuation of the function $\mathcal{A}_n({\bf x} , \hat{\bf x}, \phi)$  to arbitrary real values of $n$, in order to take the $n \to 0$ limit in \eref{eq:RepTrick}. In particular, within the saddle point scheme this can be achieved by expanding around $n=0$, 
\begin{equation}\label{eq:LinearExpansion}
\mathcal{A}_n({\bf x} , \hat{\bf x}, \phi)= n\,  \bar{\mathcal{A}}({\bf x} , \hat{\bf x}, \phi) + o(n).
\end{equation}
Given this expansion, the quenched variational parameters ${\bf x}_\star, \hat{\bf x}_\star$ are obtained from
\begin{equation}\label{eq:VariationalQuenched}
\frac{\delta \bar{\mathcal{A}}({\bf x} , \hat{\bf x}, \phi) }{\delta {\bf x}} \Big|_{{\bf x}_\star, \hat{\bf x}_\star}=0, \quad \quad \frac{\delta \bar{\mathcal{A}}({\bf x} , \hat{\bf x}, \phi)}{\delta \hat {\bf x}} \Big|_{{\bf x}_\star, \hat{\bf x}_\star}=0,
\end{equation}
and the quenched complexity from:
\begin{equation}
\Sigma^{(Q)}_\sigma(\phi)= \bar{\mathcal{A}}({\bf x}_\star , \hat{\bf x}_\star, \phi).
\end{equation}
We remark that the vectors of order and conjugate parameters ${\bf x}, \hat{\bf x}$ have different sizes depending on wether one is considering the annealed or the quenched setting, since the number of order and conjugate parameters when $n=1$ is smaller. As a consequence, the solutions of the saddle point equations are different in the two cases. To proceed with the calculation and determine $\mathcal{A}_n({\bf x} , \hat{\bf x}, \phi)$ explicitly, we have to make some assumption on the structure of these order parameters at the saddle point: this will be discussed in the following subsection.\\

\subsection{The replica symmetric assumption and the analytic continuation}\label{sec:RS}
In \eref{eq:OrderParameters} we introduced $2 n (n+1) $ order parameters, with a similar number of conjugate parameters, to be determined via saddle point equations. In order to proceed and determine the behaviour of the three terms appearing in \eref{eq:bSP} for generic values of $n$, we need to make some assumption on these parameters. In the following, we assume that the order parameters and the conjugate ones are symmetric  with respect to permutations of the replicas, setting:
\begin{equation}\label{eq:RSa}
\begin{split}
\centering
 q_{ab}= \delta_{ab} q_1+ (1-\delta_{ab}) q_0,\quad  &\hat q_{ab}= \delta_{ab} \hat q_1+ (1-\delta_{ab}) \hat q_0,  \quad \xi_{ab}=\delta_{ab} \xi_1+ (1-\delta_{ab}) \xi_0\\
 \hat   \xi_{ab}=\delta_{ab} \hat \xi_1+ (1-\delta_{ab})\hat  \xi_0,  \quad  z_{ab}=& (1-\delta_{ab}) z, \quad   \hat  z_{ab}= (1-\delta_{ab}) \hat z, \quad  m^a= m,\\
 \hat m^a= \hat m, \quad & p^a= p, \quad  \hat p^a= \hat p, \quad \hat \phi_a= \hat \phi.
  \end{split}
\end{equation}

 This choice corresponds to assuming a Replica Symmetric (RS) structure of the Kac-Rice saddle-point. 
On general grounds, this assumption may turn out not to be exact. It is known from the theory of conservative disordered systems that more complicated variational ansätze might be necessary to correctly capture the thermodynamic behaviour of the system, described by its partition function (in conservative systems, one can define a potential energy and discuss the equilibrium properties of the system, encoded in the partition function). This is the case of the symmetric $\gamma=1$ Lotka-Volterra equations, whose equilibrium properties at low temperatures have to be described by a more complicated structure of the order parameters (a so called full-RSB structure) \cite{LVMarginality, AltieriRoyTemperature}. In the language of standard equilibrium calculations for conservative systems, the approximation \eref{eq:RSa} to the Kac-Rice saddle point is equivalent to a 1-step breaking of symmetry in the thermodynamic calculation \footnote{The calculation of the system's partition function at inverse temperature $\beta$ within a  1-step replica symmetry breaking scheme involves the introduction of three different overlap parameters: a self-overlap  $q_d(\beta)$ measuring the typical overlap of a configuration with itself, and two additional overlaps, $q_1(\beta)$ and $q_0(\beta)$, measuring the typical similarity between replicas belonging to the same or to different \emph{states}, respectively.  In the limit of zero temperature ($\beta \to \infty$) the equilibrium calculation captures the properties of the deepest minima of the energy functional associated to the system; in this limit, states collapse into isolated minima and one finds $q_d \to q_1$. The saddle point values of   $q_1(\infty)$ and $q_0(\infty)$ are also solutions of the Kac-Rice saddle point equations obtained within the RS scheme, when conditioning the counting of the equilibria (stationary points of the energy) to the deepest minima of the energy landscape. For an explicit check of this correspondence in a conservative model, see Ref.~\cite{MioSegnale}.}. This is expected to provide a rather good approximation even in the symmetric case \cite{LVMarginality}: for instance, it identifies with good accuracy the diversity of the equilibria dominating the thermodynamics at zero temperature, see Sec.~\ref{sec:ComparisonReplicas}. In the asymmetric case, no equilibrium calculation is available to compare with, and \eref{eq:RSa} are assumptions the exactness of which has to be confirmed via a stability analysis of the variational manifold in which the saddle point is taken. We leave this analysis to future work, and work within the RS formalism in the rest of this work. \\
 
 Within the RS framework, the number of order parameters ${\bf x}$ is reduced to $7$ for any value of $n$, with $8$ conjugate parameters $\hat {\bf x}$. 
Under this assumption, we can derive (see \Sref{sec:CalculationMoments} for the calculation) the explicit expression of $\mathcal{A}_n({\bf x}, \hat{\bf x}, \phi)$, 
\begin{equation}\label{eq:ActionRS}
\begin{split}
&\mathcal{A}_n({\bf x}, \hat{\bf x}, \phi)=\mathcal{p}_n({\bf x})+ n\, \mathcal{d}(\phi)+ n \tonde{\hat{q}_1 q_1+\hat{\xi}_1 \xi_1+ \hat m m + \hat p p + \hat \phi \phi}\\
&+ \frac{n(n-1)}{2} \tonde{\hat{q}_0 q_0+\hat{\xi}_0 \xi_0+2i\hat{z} z}+  \mathcal{J}_n( \hat{\bf x}).
\end{split}
\end{equation}
In this formula, the term $ \mathcal{p}_n({\bf x})$ is the leading order contribution at the exponential scale of $\mathcal{P}_n({\bf x}, \phi)$ in \eref{eq:bSP}, and equals to:
\begin{equation}\label{eq:GradAsy}
\begin{split}
& \mathcal{p}_n({\bf x})=-\frac{1}{2 \sigma^2 (1+\gamma)}\frac{n \, U_n({\bf x})}{ (q_1-q_0)^2 \, [q_1+(n-1) q_0]^2} \\
&-\frac{n}{2} \log (2 \pi \sigma^2) - \frac{n-1}{2} \log(q_1-q_0)- \frac{1}{2}\log [q_1+ (n-1)q_0] 
   \end{split}
\end{equation}
with:
\begin{equation}\label{eq:Fulln}
\begin{split}
U_n({\bf x})&=(\kappa -\mu  m)^2 (q_1-q_0)^2 \grafe{(1+\gamma) [q_1+(n-1) q_0] -\gamma n m^2 }\\
&-2  (\kappa-\mu  m) (q_1-q_0)^2\grafe{m [q_1+(n-1)(q_0-\gamma  z)]+(1+\gamma) p [q_1+(n-1) q_0]}\\
&+(1+\gamma) \xi_1 (q_1-q_0) [q_1+(n-2) q_0] [q_1+(n-1)
   q_0]-\gamma  (n-1) z^2 [q_1^2+(n-1) q_0^2]\\
    &+(q_1-q_0)^2 [q_1+(n-1) q_0]^2- (n-1)(1+\gamma)  \xi_0 q_0 (q_1-q_0)[q_1+ (n-1) q_0]\\
    &-2 (n-1) q_0 z (q_1-q_0)[q_1+(n-1) q_0].
      \end{split}
\end{equation}
 We stress that the order parameters satisfy $q_0 \leq q_1$, as it follows from a  Cauchy–Schwarz inequality and from the positivity of the components of $\vec{N}$: therefore, the above expressions are well-defined.
The term $\mathcal{d}(\phi)$ is instead the contribution coming from the expectation value of the determinants $\mathcal{D}_n({\bf x}, \phi)$, and it reads:
\begin{equation}\label{eq:DetSmall}
\begin{split}
&\mathcal{d}(\phi)=\frac{\phi}{\pi} \int_{-1}^1 dx \int_0^{\sqrt{1-x^2}} dy \log \grafe{\quadre{ \sigma \sqrt{\phi} (1+\gamma)x+1}^2 + \sigma^2 \phi (1-\gamma)^2 y^2}\\
&=\begin{cases}
\frac{1}{4 \gamma \sigma^2} \tonde{1- \sqrt{1- 4 \gamma \sigma^2 \phi}}+ \phi \log \tonde{1+ \sqrt{1- 4 \gamma \sigma^2 \phi}}- \phi \tonde{\frac{1}{2}+\log 2} &\quad \phi \leq  \phi_{\rm May}\\
\frac{1}{2 \sigma^2} \frac{1}{1+\gamma} -\frac{\phi}{2}+\frac{\phi}{2} \log (\sigma^2 \phi) &\quad \phi> \phi_{\rm May}
     \end{cases}
   \end{split}
\end{equation}
where $\phi_{\rm May}$ is given in \eref{eq:MayBound}, and where the parameters $\sigma$ and $\gamma$ encode for the variability and for the asymmetry of the interactions, respectively. Finally, the remaining term is the contribution coming from the volume $\mathcal{V}_n(\hat {\bf x}, {\bf x})$, and can be compactly written as:
\begin{equation}\label{eq:Volumen}
 \mathcal{J}_n( \hat{\bf x})= \log \quadre{ \sum_{k=0}^n \binom{n}{k} e^{-i\hat{\phi} \,k  } \int d{\bf y} \prod_{a=1}^n \theta(y_a) \text{exp} \tonde{-\frac{1}{2} \; {\bf y}^T \cdot  \mathbb{A}_k[\hat{\bf x}] \cdot {\bf y} - {\bm \mu}_k[\hat{\bf x}] \cdot {\bf y} }},
\end{equation}
where we introduce the following $n \times n$ matrices and $n \times 1$ vectors :
 \begin{equation}
 \mathbb{A}_k[\hat{\bf x}]= \begin{pmatrix}
 \stackrel{k \times k}{ \hat{\mathbb{Q}}} &   - \stackrel{k \times (n-k)}{\hat{\mathbb{Z}}}\\
 - \stackrel{(n-k) \times k}{\hat{\mathbb{Z}}} & \stackrel{(n-k) \times (n-k)}{\hat{\mathbb{X}}}
              \end{pmatrix}, \quad \quad {\bm \mu}_k[\hat{\bf x}]=\begin{pmatrix} \stackrel{k \times 1}{\hat{\bf m}} \\
             \stackrel{ (n-k) \times 1 }{- \hat{\bf p} }
              \end{pmatrix},
 \end{equation}
 with
 \begin{equation}
 \hat{\mathbb{Q}}_{ab} =2 \delta_{ab}\,   \hat{q}_{1} + (1-\delta_{ab}) \hat{q}_{0}, \quad   \hat{\mathbb{Z}}_{ab}=\hat{z}, \quad  \hat{\mathbb{X}}_{ab} =2\delta_{ab}\,   \hat{\xi}_{1} + (1-\delta_{ab})\hat{\xi}_{0}.
  \end{equation}
and with 
 \begin{equation}
\hat{ m}_{a}= \hat m, \quad \hat{p}_{a}= \hat p.
  \end{equation}
The derivation of these terms is given in full detail in Appendix \ref{sec:CalculationMoments}. From these expressions, we can obtain the explicit form of the functionals to be optimized in the quenched and annealed calculation, respectively. 

\subsubsection{The quenched case. }
Expanding  \eref{eq:ActionRS} to linear order in $n$ we obtain:
{\thinmuskip=1mu
\thickmuskip=1mu
\begin{equation}\label{eq:ActionRSn0}
\begin{split}
\bar{\mathcal{A}}({\bf x}, \hat{\bf x}, \phi)=\bar{\mathcal{p}}({\bf x})+ \mathcal{d}(\phi)+ \hat{q}_1 q_1+\hat{\xi}_1 \xi_1+ \hat m m + \hat p p + \hat \phi \phi- \frac{1}{2} \tonde{\hat{q}_0 q_0+\hat{\xi}_0 \xi_0}-\hat{z} z + \bar{\mathcal{J}} ( \hat{\bf x}),
\end{split}
\end{equation}}
where:
\begin{equation}\label{eq:Forc}
\begin{split}
\bar{ \mathcal{p}}({\bf x})&=\frac{\tonde{\kappa-\mu  m}}{\sigma^2 (1+\gamma)} \frac{m (q_1-q_0+z \gamma)}{(q_1-q_0)^2}+\frac{\tonde{\kappa-\mu  m}}{\sigma^2} \frac{ p }{(q_1-q_0)}- \frac{\gamma}{2 \sigma^2 (1+\gamma)} \frac{ z^2 (q_1+ q_0) }{ (q_1-q_0)^3}\\
&- \frac{\xi_1 }{ 2 \sigma^2(q_1-q_0)}
- \frac{q_0 (\xi_0- \xi_1)}{2 \sigma^2 (q_1-q_0)^2}
 -\frac{1}{2 \sigma^2 (1+\gamma)} \quadre{1+ \frac{2 q_0 z}{(q_1-q_0)^2}}-\frac{1}{2 \sigma^2} \frac{\left(\kappa-\mu 
   m\right)^2}{q_1-q_0} \\
   &-\frac{\log[2 \pi \sigma^2 (q_1-q_0)]}{2}  -\frac{q_0}{2[q_1-q_0]},
   \end{split}
\end{equation}
and where $\bar{\mathcal{J}} ( \hat{\bf x})$ admits the following integral representation:
\begin{equation}\label{eq:VolumeQuenched}
\begin{split}
&\bar{ \mathcal{J}} (\hat{\bf x})=\int \frac{du_1 du_2}{2 \pi \; \sqrt{\hat q_0 \hat \xi_0- \hat z^2}}  \text{exp}\quadre{\frac{\hat{\xi}_0 u_1^2 + \hat{q}_0 u_2^2 - 2 \hat{z} u_1 u_2}{2 (\hat{q}_0 \hat{\xi}_0- \hat{z}^2)} } \times\\
 &\times \log \quadre{e^{- \hat{\phi}} \sqrt{\frac{\pi}{2 [2 \hat q_1- \hat q_0]}} \Pi\tonde{\frac{\hat m -u_1}{\sqrt{2 (2 \hat q_1 -\hat q_0)}}} + \sqrt{\frac{\pi}{2 [2 \hat \xi_1- \hat \xi_0]}} \Pi \tonde{\frac{-[\hat p -u_2]}{\sqrt{2 (2 \hat \xi_1 -\hat \xi_0)}}} },
 \end{split}
\end{equation}
with
\begin{equation}\label{eq:DefPi}
\Pi(x)=e^{x^2} \text{Erfc}(x), \quad \quad \text{Erfc}(x)=\frac{2}{\sqrt{\pi}} \int_x^{\infty} e^{-t^2} dt.
\end{equation}
The integral representation \eqref{eq:VolumeQuenched} is derived under the assumptions:
\begin{equation}\label{eq:Assumptions}
{\begin{split}
2 \hat q_1 - \hat q_0 >0, \quad 2 \hat \xi_1- \hat \xi_0 >0, \quad \hat q_0 <0  \quad \hat \xi_0<0,  \quad \hat q_0 \hat \xi_0- \hat z^2>0.
\end{split}}
\end{equation}
In this case, the function depends on  $7$ order parameters $q_1,q_0, \xi_1, \xi_0, z, m, p$ and $8$ conjugate parameters $\hat q_1,\hat q_0,\hat  \xi_1, \hat \xi_0, \hat z,\hat m,\hat  p, \hat  \phi$ to be determined via the saddle-point calculation. 

\subsubsection{The annealed case. }
By choosing $n=1$, we obtain instead:
\begin{equation}\label{eq:ActionRSn1}
\begin{split}
\mathcal{A}_1({\bf x}, \hat{\bf x}, \phi)=\mathcal{p}_1({\bf x})+  \mathcal{d} (\phi)+ \tonde{\hat{q}_1 q_1+\hat{\xi}_1 \xi_1+ \hat m m + \hat p p + \hat \phi \phi}+  \mathcal{J}_1( \hat{\bf x}),
\end{split}
\end{equation}
with 
\begin{equation}\label{eq.p1ann}
\begin{split}
 &\mathcal{p}_1({\bf x})=-\frac{1}{2 \sigma^2 q_1^2 }\quadre{ (\kappa -\mu  m)^2  \tonde{q_1 -\frac{\gamma\, m^2 }{1+ \gamma} }-2  (\kappa-\mu  m) q_1\tonde{p+\frac{m}{1+\gamma} }+ \xi_1  q_1} \\
 &-\frac{1}{2} \log (2 \pi \sigma^2 \, q_1) -\frac{1}{2 \sigma^2 (1+\gamma)}
   \end{split}
\end{equation}
and 
\begin{equation}\label{eq:VolumeAnnealed}
\begin{split}
 \mathcal{J}_1(\hat{\bf x})&=    \log \quadre{\frac{1}{2} \sqrt{\frac{\pi}{\hat{\xi}_1}} e^{\frac{\hat{p}^2}{4 \hat{\xi}_1}} \text{Erfc} \tonde{-\frac{\hat{p}}{2 \sqrt{\hat{\xi}_1}}}+\frac{e^{- \hat{\phi}}}{2} \sqrt{\frac{\pi}{\hat{q}_1}} e^{\frac{\hat{m}^2}{4 \hat{q}_1}} \text{Erfc} \tonde{\frac{\hat{m}}{2 \sqrt{\hat{q}_1}}}}.
  \end{split}
\end{equation}
As expected, this functional does not depend on $q_0, \xi_0, z$ and on the associated conjugate parameters, that have a meaning only whenever more than one replica is present. One is left therefore with $4$ order parameters $q_1, \xi_1, m, p$ and $5$ conjugate parameters $\hat q_1,\hat  \xi_1, \hat m,\hat  p, \hat  \phi$ to determine via the saddle-point calculation.\\

\subsection{ The linear stability matrices, their spectrum and the May bound}\label{sec:Ginibre}
Comparing the expressions \eref{eq:ActionRS} and \eref{eq:ActionRSn1}, one notices that the contribution of the expectation value of the product of determinants \eref{eq:ExpectationJointDet} at the exponential scale in $S$ equals to $n \mathcal{d} (\phi)$, where $\mathcal{d} (\phi)$ is the contribution one gets for $n=1$. This implies that at the exponential scale in $S$, the contribution of $n$ replicas is simply $n$-times the contribution of one single replica. There are essentially two reasons for this: (i)  to leading (exponential) order in $S$, the conditional expectation value of the product of determinants factorizes into the product of the conditional expectation values, and (ii) the conditioning to ${\bf F}({\bf N})={\bf f}$ is irrelevant to leading (exponential) order in $S$. We argue for these facts in Appendix \ref{app:Determinants}, and here we just briefly discuss the statistics of one of these linear stability matrices prior to conditioning, given that this is what matters for the calculation of the complexity.

 The $N \phi \times N \phi$ linear stability matrices \eqref{eq:Hessian} can be decomposed, using \eqref{eq:CenteredMatrix}, as:
\begin{equation}\label{eq:HessianDecomposedB}
H_{ij} \equiv \tonde{\frac{\partial F_i(\vec{N})}{\partial N_j}}=
-\tonde{\frac{\sigma \sqrt{\phi}}{\sqrt{S \, \phi}} a _{ij}+\frac{\mu }{S} + \delta_{ij} } \quad \quad i,j \in I.
\end{equation}
The first term in \eqref{eq:HessianDecomposedB} is a random matrix of the real elliptic type, with variance $v^2= \sigma^2 \phi$ and with asymmmetry parameter $\gamma$  \cite{mehta2004random}. The elliptic ensemble takes its name from the fact that the asymptotic density of eigenvalues of such matrices is given by a uniform distribution on the complex plane, having an ellipse as support \cite{girko1986elliptic, nguyen2015elliptic}.  More precisely, the empirical spectral measures  $\mu_M(\lambda)$  of $M \times M$ real elliptic matrices with variance $v$ and asymmmetry parameter $\gamma$ converges almost surely (when $M \to \infty$) to a deterministic measure $d\overline \mu (\lambda)= \overline \rho(\lambda) d\lambda$ with density \cite{CrisantiSommersStein}:
\begin{equation}\label{eq:AsyDensB}
\overline \rho(\lambda) =\frac{1}{\pi v^2 (1-\gamma^2)}\,  \mathbbm{1}_{{\lambda \in S_{v, \gamma} }}, \quad  \quad {S}_{v, \gamma}= \left\{ \frac{(\Re \lambda)^2}{v^2 (1+ \gamma)^2} +\frac{(\Im \lambda)^2}{v^2 (1- \gamma)^2}   \leq 1 \right\}.
 \end{equation}
For $\gamma=0$, the ensemble is known as real Ginibre ensemble \cite{ginibre1965statistical}. Its limiting density, known as ``circular law", was first derived in \cite{edelman1997probability} for matrices with real entries. As it happens with the semicircular law for real symmetric matrices, the convergence to the elliptic law is universal \cite{nguyen2015elliptic}, and moreover the limiting form \eqref{eq:AsyDensB} is not affected by finite rank perturbations to the matrix. Finite rank perturbations may give rise to outliers that do not belong to the support ${S}_{v, \gamma}$; however, the spectral weight associated to these isolated eigenvalues is suppressed at large $M$ with respect to the contribution of the bulk density $\overline \rho(\lambda) $ (see \cite{o2014low} for the explicit calculation of these outliers for elliptic matrices with real entries subject to finite-rank additive perturbations).  This implies that the second constant term in \eqref{eq:HessianDecomposed} does not modify \eqref{eq:AsyDensB} to $O(1)$ in $S$, as it corresponds to a rank-one additive perturbation of strength $\mu \phi$ along the direction of the $S \phi$ - dimensional vector $(1, \cdots, 1)^T$. 
The asymptotic density \eqref{eq:AsyDensB} is the only quantity needed to compute the conditional expectation value of the determinant to leading order in the dimensionality $S$, and taking into account the shift given by the identity matrix in \eref{eq:HessianDecomposedB} one gets the integral expression \eref{eq:DetSmall}, see Appendix \ref{app:Determinants} for the precise derivation. The role of the May diversity $\phi_{\rm May}$ as a stability threshold is then clear: the asymptotic density of the matrix \eref{eq:HessianDecomposedB} is supported on a shifted ellipse, whose upper edge on the real axis is given by $v (1+\gamma)-1= \sigma \sqrt{\phi} (1+\gamma)-1$: therefore, at $\phi=\phi_{\rm May}=[\sigma (1+\gamma)]^{-2}$ the edge of the support touches zero, corresponding to marginal stability of the associated equilibria. \\

\subsection{The variational problem: general route}
Given the explicit form of the functionals $\mathcal{A}_1$ and $\bar{\mathcal{A}}$, the general route to determine the complexity is as follows. 
In the quenched case, taking the variation of $\bar{\mathcal{A}}({\bf x}, \hat{\bf x}, \phi)$ with respect to the $15$ order and conjugate parameters we obtain two sets of equations of the form ${\bf x}= F_1[\hat {\bf x}]$ and $\hat {\bf x}= F_2[ {\bf x}]$, respectively. These equations couple the $7$ order parameters ${\bf x}$ with the $8$ conjugate parameters $\hat {\bf x}$: inverting one of these sets, one can express the order parameters as a function of the conjugate parameters, ${\bf x}= f_3[\hat {\bf x}]$. The latter can then be fixed by solving the set of coupled self-consistent equations $\hat {\bf x}= F_2[ f_3[\hat {\bf x}]]$: once the self-consistent values of the conjugate parameters $\hat {\bf x}$ are found, the order parameters can be determined and the quenched complexity can be obtained. The annealed calculation is formally analogous. This scheme can be implemented for generic values of $\gamma$:  we report the generic saddle point equations obtained from the variational procedure in \Sref{sec:VariationalEquations}, and focus on the uncorrelated case $\gamma=0$ in the following.

\section{The uncorrelated $\gamma=0$ case: solving the self-consistent equations}\label{sec:SCE}
To illustrate the strategy to solve the saddle-point problem, we focus on the uncorrelated case $\gamma=0$, when several simplifications occur which allow us to reduce the number of equations to be solved. We begin the discussion from the quenched case.

\subsection{The quenched self-consistent equations}\label{sec:QuenchedEqs}
The equations for the order and conjugate parameters obtained taking the variation of $\bar{\mathcal{A}}$ are given in \Sref{sec:VariationalEquations}. For $\gamma=0$, one sees that the equation for $\hat z$ and that for $\hat \xi_0$ are identical, implying $\hat z =\hat \xi_0$. 
The remaining conjugate parameters satisfy self-consistent equations that are more concisely written in terms of this new set of variables:
{\medmuskip=1mu
\thinmuskip=1mu
\thickmuskip=1mu
\begin{equation}\label{eq:NewParametersQ}
\begin{split}
x_1=\frac{\hat m}{\sqrt{2 \hat q_1- \hat q_0}}, \quad 
x_2=\frac{\hat p}{\sqrt{2 \hat \xi_1- \hat \xi_0}}, \quad 
y={\sqrt{2 \hat \xi_1- \hat \xi_0}}, \quad 
r =\sqrt{\frac{2 \hat q_1- \hat q_0}{2 \hat \xi_1- \hat \xi_0}}, \quad \beta_1=\frac{\hat q_0}{y^2}, \quad \beta_2=\frac{\hat \xi_0}{y^2}.
\end{split}
\end{equation}}
We recall that the expressions in the previous section are derived under the assumptions \eqref{eq:Assumptions}, which imply $y>0$. We therefore assume $y>0$, and comment in Sec.~\ref{sec:unbounded} on the meaning of $y \to 0$ (as we shall see, this is related to the emergence of the unbounded regime). As we derive in Appendix~\ref{sec:EqsUncorrelated}, the relations $\hat {\bf x}= F_2[ {\bf x}]$  can be rewritten as:
\begin{equation}\label{eq:SetA}
\begin{split}
&(a) \quad x_2=-\kappa \, y + \mu \; m y, \\
&(b) \quad r x_1= (1+\mu){x_2}+{\mu}\tonde{{y m+ y p}},\\
&(c) \quad 1=\sigma^2 y^2 (q_1-q_0),\\
&(d) \quad \beta_2=1- \sigma^2 y^2 q_1= - \sigma^2 y^2 q_0,\\
&(e) \quad r^2=\sigma^2-\sigma^2(\xi_1-\xi_0- 2z) y^2,\\
&(f) \quad \beta_1=r^2 \beta_2+\frac{\mu + 2}{\mu} \sigma^2 x_2^2 -\frac{2}{\mu}\sigma^2 x_1 x_2 r +\sigma^4 q_1 y^2-\sigma^2  \quadre{ r^2 q_1 y^2+  \xi_1 y^2}.
  \end{split}
\end{equation}
The equation defining implicitly the remaining conjugate parameter $\hat \phi$ can be written as:
\begin{equation}\label{eq:HatPhi}
\phi=\int du_1 du_2 \mathcal{G}_{\hat {\bf x}}(u_1, u_2) \frac{ e^{\frac{(u_1-x_1)^2}{2 }} \text{Erfc} \tonde{\frac{x_1 -u_1}{\sqrt{2 }}}}{ \mathcal{R}_{\hat {\bf x}}(u_1, u_2)},
\end{equation}
where we introduced the functions:
 \begin{equation}
 \begin{split}
 &\mathcal{G}_{\hat {\bf x}}(u_1, u_2)=\frac{r}{2 \pi}\frac{ e^{-\frac{1}{2}\frac{\tonde{ u_1^2  r^2 -2  r u_1 u_2 + u_2^2 \frac{\beta_1}{\beta_2}}}{\beta_2- \beta_1} } }{ \sqrt{\beta_1 \beta_2- \beta_2^2}},\\
& \mathcal{R}_{\hat {\bf x}}(u_1, u_2)=e^{\frac{(u_1-x_1)^2}{2}} \text{Erfc} \tonde{\frac{x_1 -u_1}{\sqrt{2 }}} +e^{\hat{\phi}} r e^{\frac{(u_2-x_2)^2}{2 }} \text{Erfc} \tonde{-\frac{[x_2 -u_2]}{\sqrt{2}}}.
\end{split}
\end{equation}

 In turn, the order parameters appearing in \eqref{eq:SetA} are themselves functions of the conjugate parameters $x_1, x_2, \beta_1, \beta_2, r, \hat \phi$, given by the following convolutions: 
\begin{equation}\label{eq:A}
\begin{split}
&m y=\int du_1 du_2\, \mathcal{G}_{\hat {\bf x}}(u_1, u_2) \tonde{  \frac{1}{r} \frac{\sqrt{\frac{2}{\pi}} - (x_1 -u_1) e^{\frac{(x_1-u_1)^2}{2}} \text{Erfc} \tonde{\frac{x_1 -u_1}{\sqrt{2 }}}}{ \mathcal{R}_{\hat {\bf x}}(u_1, u_2)}}\\
&p y =\int du_1 du_2\, \mathcal{G}_{\hat {\bf x}}(u_1, u_2)  \tonde{- r e^{\hat \phi} \frac{\sqrt{\frac{2}{\pi}} + (x_2 -u_2) e^{\frac{(x_2-u_2)^2}{2}} \text{Erfc} \tonde{-\frac{x_2 -u_2}{\sqrt{2 }}}}{\mathcal{R}_{\hat {\bf x}}(u_1, u_2)}}
 \end{split}
\end{equation}
and 
\begin{equation}\label{eq:B}
\begin{split}
&q_1 y^2=\int du_1 du_2\, \mathcal{G}_{\hat {\bf x}}(u_1, u_2) \tonde{ \frac{-\sqrt{\frac{2}{\pi}}(x_1 -u_1) +[1+ (x_1 -u_1)^2] e^{\frac{(x_1-u_1)^2}{2}} \text{Erfc} \tonde{\frac{x_1 -u_1}{\sqrt{2 }}}}{r^2 \; \mathcal{R}_{\hat {\bf x}}(u_1, u_2)}}\\
&\xi_1 y^2=\int du_1 du_2\, \mathcal{G}_{\hat {\bf x}}(u_1, u_2) \tonde{ e^{\hat \phi} r \frac{\sqrt{\frac{2}{\pi}}(x_2 - u_2) +[1+(x_2 -u_2)^2]e^{\frac{(u_2- x_2)^2}{2}} \text{Erfc} \tonde{-\frac{x_2 -u_2}{\sqrt{2 }}}}{ \mathcal{R}_{\hat {\bf x}}(u_1, u_2)}}\\
 \end{split}
\end{equation}
and finally
\begin{equation}\label{eq:C}
\begin{split}
q_0 y^2&=\int  du_1 du_2\, \mathcal{G}_{\hat {\bf x}}(u_1, u_2) \; \tonde{\frac{1}{r}  \frac{\sqrt{\frac{2}{\pi}} - (x_1 -u_1) e^{\frac{(x_1-u_1)^2}{2}} \text{Erfc} \tonde{\frac{x_1 -u_1}{\sqrt{2 }}}}{ \mathcal{R}_{\hat {\bf x}}(u_1, u_2)}}^2\\
\xi_0 y^2&=\int  du_1 du_2\, \mathcal{G}_{\hat {\bf x}}(u_1, u_2) \; \tonde{- r e^{\hat \phi} \frac{\sqrt{\frac{2}{\pi}} + (x_2 -u_2) e^{\frac{(x_2-u_2)^2}{2}} \text{Erfc} \tonde{-\frac{x_2 -u_2}{\sqrt{2 }}}}{ \mathcal{R}_{\hat {\bf x}}(u_1, u_2)}}^2\\
z y^2&=  \int  du_1 du_2\, \mathcal{G}_{\hat {\bf x}}(u_1, u_2) \; \tonde{ \frac{1}{r} \frac{\sqrt{\frac{2}{\pi}} - (x_1 -u_1) e^{\frac{(x_1-u_1)^2}{2}} \text{Erfc} \tonde{\frac{x_1 -u_1}{\sqrt{2 }}}}{\mathcal{R}_{\hat {\bf x}}(u_1, u_2)}} \times\\
&\times \tonde{- r e^{\hat \phi} \frac{\sqrt{\frac{2}{\pi}} + (x_2 -u_2) e^{\frac{(x_2-u_2)^2}{2}} \text{Erfc} \tonde{-\frac{x_2 -u_2}{\sqrt{2 }}}}{ \mathcal{R}_{\hat {\bf x}}(u_1, u_2)}}.
 \end{split}
\end{equation}

 Plugging these expressions into \eqref{eq:SetA} and \eqref{eq:HatPhi}, one gets therefore a set of coupled self-consistent equations for the rescaled conjugate parameters \eqref{eq:NewParametersQ} together with $\hat \phi$. The motivation for introducing the rescaled parameters \eqref{eq:NewParametersQ} is that the equations expressed in terms of these variables can be partially decoupled. 
Inspecting the equations one notices indeed that the convolutions on the right-hand side of equations \eqref{eq:A}, \eqref{eq:B} and \eqref{eq:C} do not depend on $y$ explicitly; on the other hand, in \eqref{eq:SetA} the order parameters appear multiplied by the suitable power of $y$ which appears also to the left-hand side of the equations \eqref{eq:A}, \eqref{eq:B} and \eqref{eq:C}. Therefore, the value of $y$ can be fixed at the end of the calculation using Eq. (\ref{eq:SetA} a), (where the product $m y$ is given in \eqref{eq:A} and does not depend on $y$ itself), once the values of the other conjugate variables are determined. Moreover, the fixed parameter $\phi$ appears only in the equation \eqref{eq:HatPhi}; therefore, one can tune it by tuning its conjugate parameter $\hat \phi$. As a consequence, one can solve the coupled equations for  $x_1, x_2, \beta_1, \beta_2, r$ at fixed value of $\hat \phi$, and then use \eqref{eq:HatPhi} to determine the diversity $\phi$ corresponding to the chosen $\hat \phi$. Repeating this procedure for different values of $\hat \phi$, one can parametrically resolve the complexity as a function of $\phi$. This procedure then leaves us with $5$ coupled self-consistent equations for the parameters $x_1, x_2, \beta_1, \beta_2, r$. \\
Exploiting an additional relation between the convolutions in \eqref{eq:A}, \eqref{eq:B} and \eqref{eq:C}, we now show that the number of relevant equations can be further reduced by one.
The simplification comes from noticing that the integral expression are related by the following identity:
\begin{equation}\label{eq:EqSum}
r^2 [q_1 y^2]+  [\xi_1 y^2]= -r x_1 \,[m y]- x_2 \,[p y] + 1 - \beta_1 \tonde{[q_1 y^2]-[ q_0 y^2]} - \beta_2 \tonde{[\xi_1 y^2]- [\xi_0 y^2] - 2 [z y^2]},
\end{equation}
where the expressions within the brackets have to be replaced by the integral representations \eqref{eq:A}, \eqref{eq:B} and \eqref{eq:C}. This identity follows from integration by parts, see Appendix~\ref{sec:IdentityByParts}. It holds for arbitrary values of $\gamma$. Plugging it into Eq.~(\ref{eq:SetA} f) and using that $\sigma^4 q_1 y^2= \sigma^2 (1-\beta_2)$, $\sigma^2 \beta_1 \tonde{q_1 y^2- q_0 y^2}= \beta_1$ together with Eq.~(\ref{eq:SetA} e), we are left with:
\begin{equation}
x_2^2 + \frac{2}{\mu} x_2^2 -\frac{2}{\mu} r x_1 x_2 +  r x_1 \, m y+ x_2 \,p y =0.
\end{equation}
Using now Eq.~(\ref{eq:SetA} b) to eliminate $ p y$ we find that Eq.~(\ref{eq:SetA} f) reduces to:
\begin{equation}
\tonde{\mu \,m y -x_2}(r \, x_1-x_2)=\kappa \, y\, (r \, x_1-x_2)=0,
\end{equation}
which for $y \neq 0$ is simply solved by $x_2= x_1 r$. Plugging $x_2=r x_1$ into the remaining equations, we are left with 
 4 self-consistent equations to solve simultaneously for $x_1, r, \beta_1, \beta_2$ at fixed value of $\hat \phi$:
 \begin{equation}\label{eq:SetSelfCons}
\begin{split}
&(a) \quad r x_1=-\tonde{y m+ y p},\\
&(b) \quad \beta_2=1- \sigma^2 y^2 q_1= - \sigma^2 y^2 q_0,\\
&(c) \quad 1=\sigma^2 y^2 (q_1-q_0),\\
&(d) \quad r^2=\sigma^2-\sigma^2(\xi_1-\xi_0- 2z) y^2.
  \end{split}
\end{equation}
Explicitly, these equations read: 
{\medmuskip=1mu
\thinmuskip=1mu
\thickmuskip=1mu
\begin{equation}
\begin{split}
 r x_1+ \int d{\bf u} \,\mathcal{G}_{\hat {\bf x}}(u_1, u_2) \;  {  \frac{ \frac{1}{r} \quadre{\sqrt{\frac{2}{\pi}} - (x_1 -u_1) \Pi \tonde{\frac{x_1 -u_1}{\sqrt{2 }}}}- r e^{\hat \phi} \quadre{ \sqrt{\frac{2}{\pi}} +(r x_1 -u_2) \Pi \tonde{-\frac{r x_1 -u_2}{\sqrt{2 }}}}}{ \mathcal{R}_{\hat {\bf x}}(u_1, u_2)}  }=0,
\end{split}
\end{equation}
}

and
\begin{equation}
\begin{split}
\beta_2 +\sigma^2 \int d{\bf u} \,\mathcal{G}_{\hat {\bf x}}(u_1, u_2) \;  \quadre{\frac{1}{r}  \frac{\sqrt{\frac{2}{\pi}} - (x_1 -u_1) \Pi \tonde{\frac{x_1 -u_1}{\sqrt{2 }}}}{ \mathcal{R}_{\hat {\bf x}}(u_1, u_2)}}^2=0,
\end{split}
\end{equation}
and 
\begin{equation}
\begin{split}
1-\beta_2- \sigma^2  \int d{\bf u} \,\mathcal{G}_{\hat {\bf x}}(u_1, u_2) \;  {  \frac{1}{r^2} \frac{[1+ (x_1 -u_1)^2] \; \Pi \tonde{\frac{x_1 -u_1}{\sqrt{2 }}}-\sqrt{\frac{2}{\pi}}(x_1 -u_1) }{\mathcal{R}_{\hat {\bf x}}(u_1, u_2)}}=0,
\end{split}
\end{equation}
and finally
\begin{equation}
\begin{split}
&r^2-\sigma^2+  \sigma^2\int d{\bf u} \,\mathcal{G}_{\hat {\bf x}}(u_1, u_2) \; e^{\hat \phi} r \frac{\sqrt{\frac{2}{\pi}}(r x_1 - u_2) +[1+(r x_1 -u_2)^2] \Pi  \tonde{-\frac{r x_1 -u_2}{\sqrt{2 }}}}{ \mathcal{R}_{\hat {\bf x}}(u_1, u_2)}\\
&+ \sigma^2 \int d{\bf u} \,\mathcal{G}_{\hat {\bf x}}(u_1, u_2) 
\;\tonde{ r e^{\hat \phi} \frac{\sqrt{\frac{2}{\pi}} + (r x_1 -u_2) \Pi\tonde{-\frac{r x_1 -u_2}{\sqrt{2 }}}}{ \mathcal{R}_{\hat {\bf x}}(u_1, u_2)}} \times \\
&\tonde{ \frac{\frac{2}{r} \quadre{\sqrt{\frac{2}{\pi}} - (x_1 -u_1) \Pi\tonde{\frac{x_1 -u_1}{\sqrt{2 }}}}- r e^{\hat \phi} \quadre{\sqrt{\frac{2}{\pi}} + (r x_1 -u_2) \Pi \tonde{-\frac{r x_1 -u_2}{\sqrt{2 }}}}}{\mathcal{R}_{\hat {\bf x}}(u_1, u_2)}}=0,
\end{split}
\end{equation}
where we recall that the function $\Pi$ is defined in \eqref{eq:DefPi}.
Once these equations are solved (for instance, by iteration), one can determine the parameters $\phi, y$ through:
\begin{equation}\label{eq:QuenchedPhi}
\begin{split}
\phi&=\int d{\bf u} \,\mathcal{G}_{\hat {\bf x}}(u_1, u_2)  \frac{ \Pi \tonde{\frac{x_1 -u_1}{\sqrt{2 }}}}{ \mathcal{R}_{\hat {\bf x}}(u_1, u_2)},
\end{split}
\end{equation}
and 
\begin{equation}
\begin{split}\label{eq:QuenchedY}
\kappa \,y&=-r\, x_1 + \mu \int d{\bf u} \,\mathcal{G}_{\hat {\bf x}}(u_1, u_2)  \tonde{  \frac{1}{r} \frac{\sqrt{\frac{2}{\pi}} - (x_1 -u_1) \Pi \tonde{\frac{x_1 -u_1}{\sqrt{2 }}}}{ \mathcal{R}_{\hat {\bf x}}(u_1, u_2)}}.
\end{split}
\end{equation}
Once also $y$ is fixed, one can use Eqs. \eqref{eq:A}, \eqref{eq:B} and \eqref{eq:C} to solve for the order parameters. \\

 It is quite straightforward to check that the quenched complexity can be expressed as a function of the conjugate parameters \eqref{eq:NewParametersQ} as:
{\medmuskip=1mu
\thinmuskip=1mu
\thickmuskip=1mu
\begin{equation}\label{eq:QuenchedCompConj}
\Sigma^{(Q)}_\sigma =\frac{1}{2} \quadre{1-r^2 (q_0 y^2) -(\xi_1 y^2)- \frac{1}{ \sigma^2} }  +  \int d{\bf u}  \,\mathcal{G}_{\hat {\bf x}}(u_1, u_2)  \log \quadre{  \frac{\mathcal{R}_{\hat {\bf x}}(u_1, u_2)}{2r}}- \hat \phi (1-\phi)+  \mathcal{d}(\phi)
\end{equation}}
where the quantities $q_0 y^2,\xi_1 y^2 $ and $\phi$ are again given by the integral representations and where for $\gamma=0$ the contribution of the determinant reads:
\begin{equation}\label{eq:Det}
\mathcal{d}(\phi)=  \begin{cases}
  0 &\text{  if  }\quad \sigma \sqrt{\phi} <1\\
 \frac{1}{2 \sigma^2}-\frac{\phi}{2}+\frac{\phi}{2}\log (\sigma^2 \phi) &\text{  if  }\quad \sigma \sqrt{\phi} >1
   \end{cases}
 \end{equation}
 Notice that \eqref{eq:QuenchedCompConj} does not depend explicitly on $y$.
 
 \subsection{The annealed self-consistent equations}\label{sec:AnnealedEqs}
 For the annealed case, we can proceede similarly as for the quenched, see again Sec. \ref{sec:EqsUncorrelated} for details.  We introduce a new set of variables that plays the same role as above, but which we denote with curly symbols to signify that they take different values at the saddle-point with respect to the corresponding quenched quantities:
 \begin{equation}\label{eq:NewParametersA}
\begin{split}
 \mathcal{x}_1=\frac{\hat m}{\sqrt{2 \hat q_1}}, \quad 
  \mathcal{x}_2=\frac{\hat p}{\sqrt{2 \hat \xi_1}}, \quad 
 \mathcal{y}={\sqrt{2 \hat \xi_1}}, \quad 
\mathcal{r} =\sqrt{\frac{\hat q_1}{ \hat \xi_1}}.    
\end{split}
\end{equation}
The variational equations obtained taking the derivatives of \eqref{eq:ActionRSn1} read in this case:
\begin{equation}\label{eq:SetAann}
\begin{split}
&(a) \quad   \mathcal{x}_2=-\kappa \,   \mathcal{y}+ \mu \; m  \mathcal{y}, \\
&(b) \quad  \mathcal{r}   \mathcal{x}_1= (1+\mu){ \mathcal{x}_2}+{\mu}\tonde{m \mathcal{y} +  p \mathcal{y}},\\
&(c) \quad 1=\sigma^2  q_1\, \mathcal{y}^2,\\
&(d) \quad  \mathcal{r}^2=\sigma^2-\sigma^2 \,\xi_1\,  \mathcal{y}^2+ \sigma^2  \mathcal{x}_2^2 - \frac{2}{\mu} \sigma^2   \mathcal{x}_2 (\mathcal{r}  \mathcal{x}_1- \mathcal{x}_2).
  \end{split}
\end{equation}
Taking the derivatives with respect to the conjugate parameters, we obtain instead:
\begin{equation}\label{eq:Aann}
\begin{split}
&m \mathcal{y}=   \frac{1}{\mathcal{r}} \frac{\sqrt{\frac{2}{\pi}} -  \mathcal{x}_1 \, e^{\frac{ \mathcal{x}_1^2}{2}} \text{Erfc} \tonde{\frac{ \mathcal{x}_1}{\sqrt{2 }}}}{e^{\frac{ \mathcal{x}_1^2}{2}} \text{Erfc} \tonde{\frac{ \mathcal{x}_1}{\sqrt{2 }}} +e^{\hat{\phi}} \mathcal{r}e^{\frac{ \mathcal{x}_2^2}{2 }} \text{Erfc} \tonde{-\frac{ \mathcal{x}_2}{\sqrt{2}}} }\\
&p \mathcal{y} =- \mathcal{r} e^{\hat \phi} \frac{\sqrt{\frac{2}{\pi}} +  \mathcal{x}_2 \,e^{\frac{ \mathcal{x}_2^2}{2}} \text{Erfc} \tonde{-\frac{ \mathcal{x}_2}{\sqrt{2 }}}}{e^{\frac{ \mathcal{x}_1^2}{2}} \text{Erfc} \tonde{\frac{ \mathcal{x}_1}{\sqrt{2 }}} +e^{\hat{\phi}} \mathcal{r} e^{\frac{ \mathcal{x}_2^2}{2 }} \text{Erfc} \tonde{-\frac{ \mathcal{x}_2}{\sqrt{2}}}}
 \end{split}
\end{equation}
and similarly:
\begin{equation}\label{eq:Bann}
\begin{split}
&q_1 \mathcal{y}^2=   \frac{1}{\mathcal{r}^2} \frac{-\sqrt{\frac{2}{\pi}} \, \mathcal{x}_1 +(1+  \mathcal{x}_1^2) e^{\frac{ \mathcal{x}_1^2}{2}} \text{Erfc} \tonde{\frac{ \mathcal{x}_1}{\sqrt{2 }}}}{e^{\frac{ \mathcal{x}_1^2}{2}} \text{Erfc} \tonde{\frac{ \mathcal{x}_1}{\sqrt{2 }}} +e^{\hat{\phi}} \mathcal{r}e^{\frac{ \mathcal{x}_2^2}{2 }} \text{Erfc} \tonde{-\frac{ \mathcal{x}_2}{\sqrt{2}}}}\\
&\xi_1 \mathcal{y}^2= e^{\hat \phi} \mathcal{r} \frac{\sqrt{\frac{2}{\pi}}\,  \mathcal{x}_2  +(1+ \mathcal{x}_2^2)e^{\frac{ \mathcal{x}_2^2}{2}} \text{Erfc} \tonde{-\frac{ \mathcal{x}_2}{\sqrt{2 }}}}{ e^{\frac{ \mathcal{x}_1^2}{2}} \text{Erfc} \tonde{\frac{ \mathcal{x}_1}{\sqrt{2 }}} +e^{\hat{\phi}} \mathcal{r} e^{\frac{ \mathcal{x}_2^2}{2 }} \text{Erfc} \tonde{-\frac{ \mathcal{x}_2}{\sqrt{2}}}}\\
 \end{split}
\end{equation}

 Similarly to the quenched case, by inspecting Eqs. \eqref{eq:Aann} and \eqref{eq:Bann} we see that the following relation holds:
\begin{equation}\label{eq:EqSumann}
\mathcal{r}^2 [q_1  \mathcal{y}^2]+  [\xi_1 y^2]= -\mathcal{r} \mathcal{x}_1 [m  \mathcal{y}]-  \mathcal{x}_2 [p  \mathcal{y}] + 1,
\end{equation}
where again the brackets indicate that the products inside have to be replaced by the corresponding expressions in the right-hand side of Eqs. \eqref{eq:Aann} and \eqref{eq:Bann}. This relation, when substituted into Eq. (\ref{eq:SetAann} d) leads to $ \mathcal{x}_2= \mathcal{r} \mathcal{x}_1$, similarly to the quenched case. Therefore, in the annealed case we are left with two coupled self-consistent equations for $ \mathcal{x}_1$ and $\mathcal{r}$ at fixed $\hat \phi$:
\begin{equation}\label{eq:SetAannSimp}
\begin{split}
&(a) \quad  \mathcal{r}   \mathcal{x}_1= -\tonde{m \mathcal{y} +  p \mathcal{y}},\\
&(b) \quad 1=\sigma^2  q_1\, \mathcal{y}^2
  \end{split}
\end{equation}
or, explicitly:
\begin{equation}\label{eq:CavAnInutile}
\mathcal{r} \mathcal{x}_1+\frac{ \frac{1}{\mathcal{r}}\quadre{\sqrt{\frac{2}{\pi}} -  \mathcal{x}_1 \, e^{\frac{ \mathcal{x}_1^2}{2}} \text{Erfc} \tonde{\frac{ \mathcal{x}_1}{\sqrt{2 }}}} - \mathcal{r} e^{\hat \phi} \quadre{\sqrt{\frac{2}{\pi}} + \mathcal{r} \mathcal{x}_1 \,e^{\frac{ \mathcal{r}^2 \mathcal{x}_1^2}{2}} \text{Erfc} \tonde{-\frac{ \mathcal{r} \mathcal{x}_1}{\sqrt{2 }}}}}{e^{\frac{ \mathcal{x}_1^2}{2}} \text{Erfc} \tonde{\frac{ \mathcal{x}_1}{\sqrt{2 }}} +e^{\hat{\phi}} \mathcal{r} e^{\frac{ r^2 \mathcal{x}_1^2}{2 }} \text{Erfc} \tonde{-\frac{ \mathcal{r} \mathcal{x}_1}{\sqrt{2}}} }=0
\end{equation}
and 
\begin{equation}\label{eq:CavAn}
\mathcal{r}^2-{\sigma^2} \frac{\sqrt{\frac{2}{\pi}} \, \mathcal{x}_1 -(1+  \mathcal{x}_1^2) e^{\frac{ \mathcal{x}_1^2}{2}} \text{Erfc} \tonde{\frac{ \mathcal{x}_1}{\sqrt{2 }}}}{e^{\frac{ \mathcal{x}_1^2}{2}} \text{Erfc} \tonde{\frac{ \mathcal{x}_1}{\sqrt{2 }}} +e^{\hat{\phi}} \mathcal{r} e^{\frac{ \mathcal{r}^2 \mathcal{x}_1^2}{2 }} \text{Erfc} \tonde{-\frac{\mathcal{r} \mathcal{x}_1}{\sqrt{2}}}}=0.
\end{equation}
Once  $ \mathcal{x}_1$ and $\mathcal{r}$ are determined,  $\phi$ and $\mathcal{y}$ can be fixed via:
\begin{equation}\label{eq:AnnealedPhi}
\phi=\frac{e^{\frac{ \mathcal{x}_1^2}{2}} \text{Erfc} \tonde{\frac{ \mathcal{x}_1}{\sqrt{2 }}}}{e^{\frac{ \mathcal{x}_1^2}{2}} \text{Erfc} \tonde{\frac{ \mathcal{x}_1}{\sqrt{2 }}} +e^{\hat{\phi}} \mathcal{r} e^{\frac{ \mathcal{r}^2 \mathcal{x}_1^2}{2 }} \text{Erfc} \tonde{-\frac{ \mathcal{r} \mathcal{x}_1}{\sqrt{2}}} },
\end{equation}
and
\begin{equation}\label{eq:AnnealedY}
\kappa \, \mathcal{y}= - \mathcal{r} \mathcal{x}_1 +  \frac{\mu }{\mathcal{r}} \; \frac{\sqrt{\frac{2}{\pi}} -  \mathcal{x}_1 \, e^{\frac{ \mathcal{x}_1^2}{2}} \text{Erfc} \tonde{\frac{ \mathcal{x}_1}{\sqrt{2 }}}}{e^{\frac{ \mathcal{x}_1^2}{2}} \text{Erfc} \tonde{\frac{ \mathcal{x}_1}{\sqrt{2 }}} +e^{\hat{\phi}} \mathcal{r}e^{\frac{\mathcal{r}^2 \mathcal{x}_1^2}{2 }} \text{Erfc} \tonde{-\frac{\mathcal{r} \mathcal{x}_1}{\sqrt{2}}} }.
\end{equation}
The annealed complexity can be expressed as a function of the solutions of the saddle-point equations as follows:
\begin{equation}\label{eq:AnnealedCompConj}
\Sigma^{(A)}_\sigma =\frac{1}{2} \tonde{1- \xi_1 \mathcal{y}^2- \frac{1}{ \sigma^2}}+  \log \quadre{ \frac{e^{\frac{ \mathcal{x}_1^2}{2}} \text{Erfc} \tonde{\frac{ \mathcal{x}_1}{\sqrt{2 }}} +e^{\hat{\phi}} \mathcal{r}e^{\frac{ \mathcal{x}_2^2}{2 }} \text{Erfc} \tonde{-\frac{ \mathcal{x}_2}{\sqrt{2}}}}{2  \mathcal{r}}}- \hat \phi (1-\phi)+  \mathcal{d}(\phi),
\end{equation}
where $\mathcal{d}(\phi)$ is still given by \eqref{eq:Det}. Again, \eqref{eq:AnnealedCompConj} does not depend on $\mathcal{y}$.

  \subsection{When quenched and annealed coincide: the “cavity" matching point}\label{sec:CavPoint}
We now discuss a particular point in the space of solutions of the quenched self-consistent equations, which corresponds to $\hat \phi=0$. We define this as the “cavity" point since, as we shall show, when this point is reached the quenched equations map into the annealed one, and become equivalent to the equations obtained within the cavity method recalled in Appendix~\ref{sec:Cavity}. From the solution of the quenched self-consistent equations one sees that when $\hat \phi \to 0^-$ the equations become singular, since it holds: 
 \begin{equation}\label{eq:CavityLimit}
 r \to 1, \quad \quad \beta_1 \to \beta_2.
  \end{equation}
  We therefore introduce the following scaling parameters:
 \begin{equation}
 \Delta= \beta_2 - \beta_1, \quad \quad R= \frac{\beta_1}{\beta_2}.
 \end{equation}
 The point $\hat \phi =0$ has to be approached as a limit since the convolutions in \eqref{eq:A},\eqref{eq:B} and \eqref{eq:C} are derived under the assumptions $ \Delta>0$ and $R>1$; in the limit we are considering,  
  \begin{equation}\label{eq:CavityLimit2}
 \Delta \to 0, \quad \quad R \to 1, \quad \quad \frac{ \Delta^2}{ R -1} \to b , \quad \quad r \to 1,
  \end{equation}
 the integrands remain regular while the Gaussian measure
 \begin{equation}
 \mathcal{G}_{\hat {\bf x}}(u_1, u_2)=\frac{r \sqrt{R-1}}{2 \pi \, \Delta^2}\, e^{-\frac{1}{2}\frac{\tonde{ u_1^2  r^2 -2  r u_1 u_2 + R \,u_2^2}}{\Delta^2} } 
 \end{equation}
becomes singular, since the quadratic form at the exponent displays a divergent eigenvalue. Diagonalizing the quadratic form and taking the above limits, we see that: 
 \begin{equation}\label{eq:DeltaG}
 \mathcal{G}_{\hat {\bf x}}(u_1, u_2) \to \frac{1}{\sqrt{2 \pi \, b}}e^{-\frac{1}{2 b} u_1^2} \delta \tonde{u_1-u_2}.
   \end{equation}
Plugging this into \eqref{eq:A} and using that $x_2=r x_1$ at the saddle point, we get :
\begin{equation}
\begin{split}
m y  &\to \sqrt{1+ b} \frac{\sqrt{\frac{2}{\pi}} - \frac{x_1}{ \sqrt{1+b}} e^{\frac{x_1^2}{2 (1+b)}  } \text{Erfc}\tonde{\frac{x_1}{\sqrt{2 (1+b)}}}}{2\,  e^{\frac{x_1^2}{2 (1+b)}  }}\\
p y  &\to - \sqrt{1+ b} \frac{\sqrt{\frac{2}{\pi}} + \frac{x_1}{ \sqrt{1+b}} e^{\frac{x_1^2}{2 (1+b)}  } \text{Erfc}\tonde{-\frac{x_1}{\sqrt{2 (1+b)}}}}{2\,  e^{\frac{x_1^2}{2 (1+b)}  }}
\end{split}
\end{equation}
and similarly from \eqref{eq:B}  we get:
\begin{equation}
\begin{split}
q_1 y^2  &\to (1+ b) \frac{-\sqrt{\frac{2}{\pi}}\frac{x_1}{ \sqrt{1+b}} + \quadre{1+\frac{x_1^2}{1+b}}  e^{\frac{x_1^2}{2 (1+b)}  } \text{Erfc}\tonde{\frac{x_1}{\sqrt{2 (1+b)}}}}{2\,  e^{\frac{x_1^2}{2 (1+b)}  }}\\
\xi_1 y^2  &\to (1+ b) \frac{\sqrt{\frac{2}{\pi}}\frac{x_1}{ \sqrt{1+b}} + \quadre{1+\frac{x_1^2}{1+b}}  e^{\frac{x_1^2}{2 (1+b)}  } \text{Erfc}\tonde{-\frac{x_1}{\sqrt{2 (1+b)}}}}{2\,  e^{\frac{x_1^2}{2 (1+b)}  }}.\\
\end{split}
\end{equation}

It appears that in this limit, the parameter $b$ can be re-absorbed by rescaling $x_1, y$, and the annealed expressions for the order parameters can be recovered. More precisely, comparing these expressions with \eqref{eq:Aann} and \eqref{eq:Bann} we see that the quenched expressions for the order parameters $m,p, q_1, \xi_1$ reproduce the annealed expressions  (using again that $\mathcal{r} \mathcal{x}_1= \mathcal{x}_2$), provided that we identify:
\begin{equation}\label{eq:CavityRescaling}
y= \mathcal{y}   \sqrt{1+ b}, \quad \quad x_1 = \mathcal{x}_1   \sqrt{1+ b}, \quad \quad r= \mathcal{r}=1.
\end{equation}
Moreover, with these identifications the first two of Eqs. \eqref{eq:SetSelfCons} map into Eqs. \eqref{eq:SetAannSimp}, and similarly \eqref{eq:QuenchedPhi} and \eqref{eq:QuenchedY} map into \eqref{eq:AnnealedPhi} and \eqref{eq:AnnealedY}. Therefore, in this limit the quenched calculation reproduces the annealed one, and the physical order parameters $m,p,q_1, \xi_1$ computed in the two schemes coincide. As we show explicitly in Appendix~\ref{sec:Cavity}, the value of $\phi$ corresponding to this point is exactly the same obtained within the cavity approximation: we denote it with $\phi_{\rm cav}$. Also the values of the order parameters $m, q_1$ coincide with those obtained within the cavity formalism.\\

 The quenched prescription provides us with two additional equations, the last two ones in Eqs. \eqref{eq:SetSelfCons}. 
It is easy to check, using the expressions \eqref{eq:C} and using Eq. (\ref{eq:SetSelfCons}  c),  that Eq. (\ref{eq:SetSelfCons}  d) is automatically satisfied at the cavity point (see Sec. \ref{sec:Cavity}). Finally, we remark that once the annealed self-consistent equations are solved and $\mathcal{x}_1$ is determined,  Eq. (\ref{eq:SetSelfCons}  c) gives a self-consistent equations for the parameter $b$, which reads:
 \begin{equation}\label{eq:Selfb}
 b= \sigma^2 \, \int du\, \frac{e^{-\frac{u^2}{2 b}}}{\sqrt{2 \pi b}}  \quadre{\frac{\sqrt{\frac{2}{\pi}} - \tonde{x_1(b)-u} e^{\frac{(x1(b)-u)^2}{2}} \text{Erf} \tonde{\frac{x_1(b)-u}{\sqrt{2}}} }{ 2 e^{\frac{(x1(b)-u)^2}{2}}}}^2, \quad \quad x_1(b)= \frac{\mathcal{x}_1}{\sqrt{1+b}}.
 \end{equation}
This parameter $b$ is in general not equal to zero, which in turn implies that at this point $q_0 \neq 0$, given that the two sets of equations, in particular Eq. (\ref{eq:SetSelfCons}  c),  imply the following relation at the cavity matching point:
\begin{equation}\label{eq:Qu}
q_0 = \frac{b}{1+ b} q_1.
\end{equation}
Therefore, within the quenched calculation the annealed limit is not attained when $q_0, \xi_0, z \to 0$, as one might naively expect.\\

In the cavity limit the two functions \eqref{eq:QuenchedCompConj} and \eqref{eq:AnnealedCompConj} coincide. Indeed, using \eqref{eq:DeltaG} and $r=1$ one finds that:
\begin{equation}
\int du_1 du_2 \,\mathcal{G}_{\hat {\bf x}}(u_1, u_2)  \log \quadre{  \mathcal{R}_{\hat {\bf x}}(u_1, u_2)} \to \log 2 + \frac{b + x_1^2}{2}.
\end{equation}
and thus from \eqref{eq:CavityRescaling} we get: 
\begin{equation}\label{eq:DiffCompCav}
\Sigma^{(Q)}_\sigma - \Sigma^{(A)}_\sigma  \to - \frac{1+b}{2}\quadre{   (q_0 \mathcal{y}^2) +(\xi_1 \mathcal{y}^2)} + \frac{b + (1+ b) \mathcal{x}_1^2}{2}- \frac{\mathcal{x}_1^2}{2}=0,
\end{equation}
as it follows from \eqref{eq:EqSumann}, \eqref{eq:Qu} and \eqref{eq:SetAannSimp}.

\section{The uncorrelated $\gamma=0$ case: the resulting complexity}\label{sec:Results}
We have discussed in the previous section the structure of the self-consistent equations obtained in the uncorrelated case $\gamma=0$. In this section, we present what the solutions of these equations entail for the quenched and annealed complexity of the model. Some of these results are also discussed in Ref.~\cite{ros2022generalized}.

\subsection{Quenched complexity, annealed complexity and the cavity matching point}
\begin{figure}[ht]
\centering \includegraphics[width=.48\linewidth]{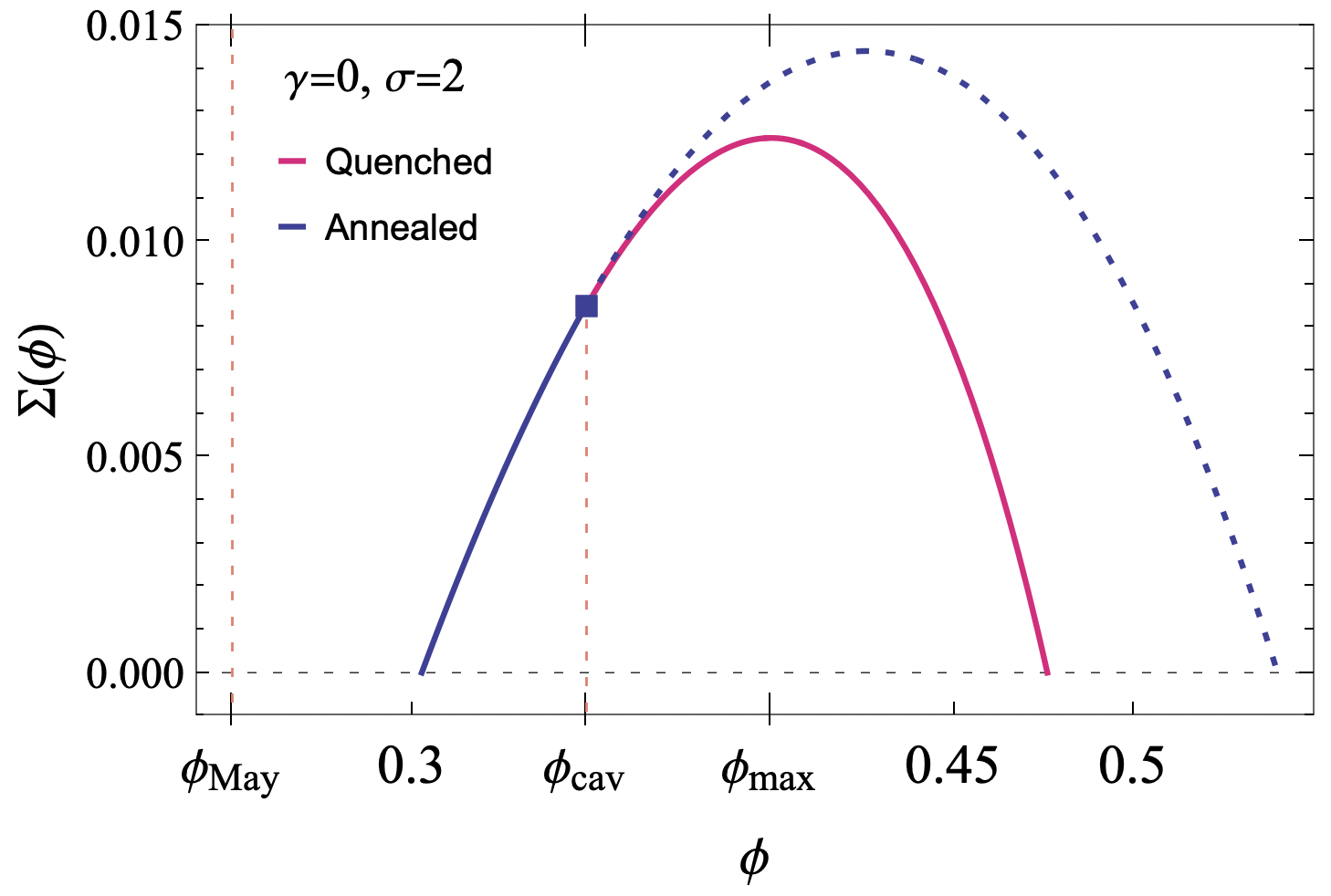}
    \centering \includegraphics[width=.48\linewidth]{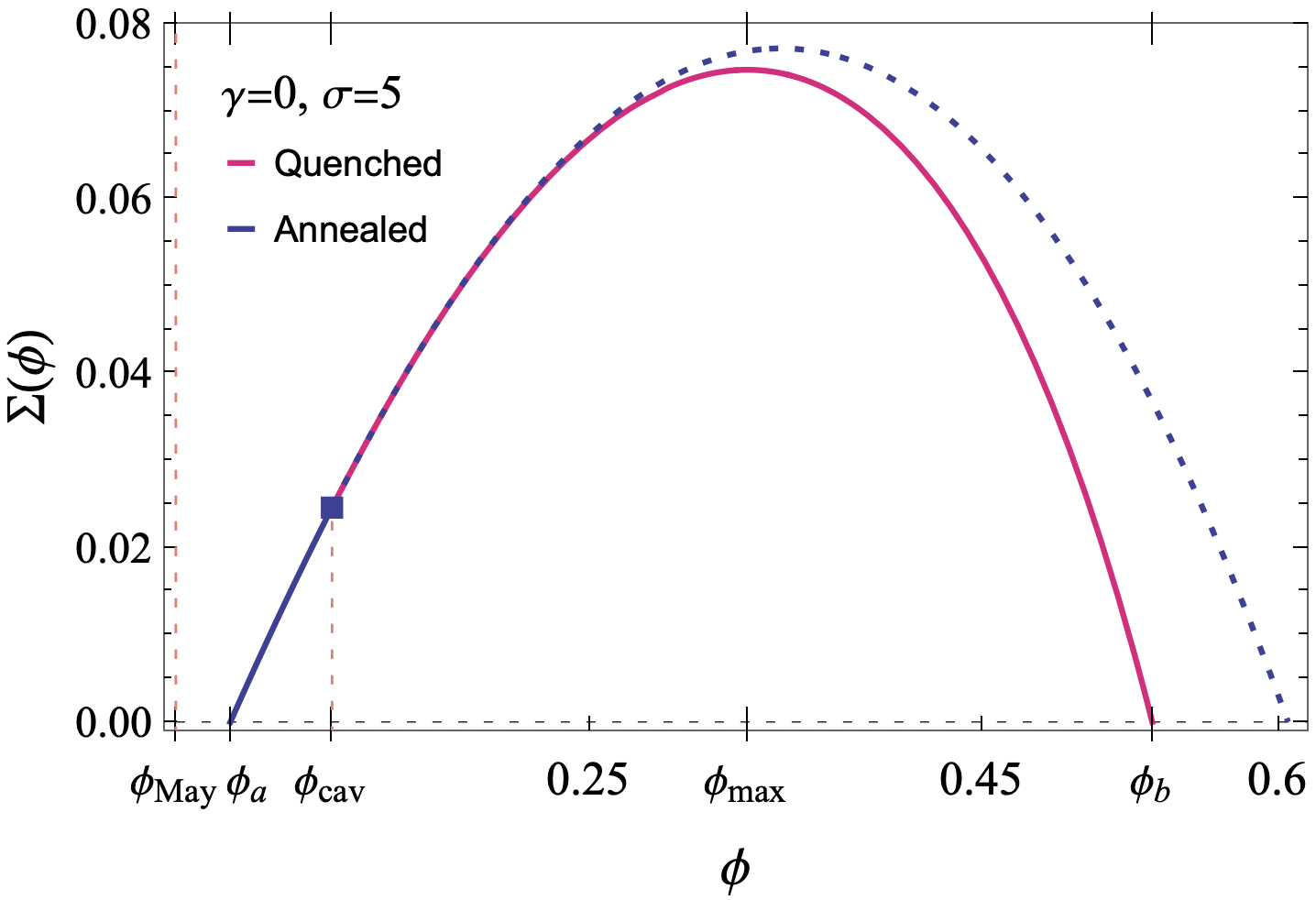}
\caption{ Quenched (magenta) and annealed (blue) complexity of uninvadable equilibria as a function of the diversity $\phi$, for $\gamma=0$ and $\sigma=2,5 > \sigma_c$. For $\phi >\phi_{\rm cav}$ the annealed complexity is strictly larger than the quenched one,  while the two coincide for  $\phi <\phi_{\rm cav}$. Both the annealed and quenched complexities are positive only for values of diversity that are beyond the May bound, meaning that all the equilibria are linearly unstable with respect to perturbations in the populations of species that coexist.}\label{fig:RepresentativePlotComp}
\end{figure}

\begin{figure}[ht]
     \includegraphics[width=.47\linewidth]{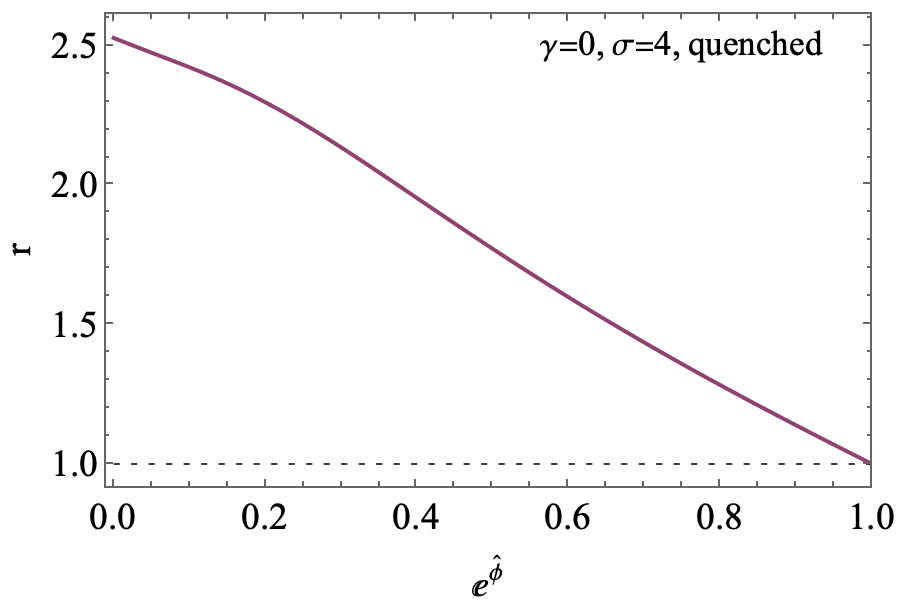}
   \includegraphics[width=.47\linewidth]{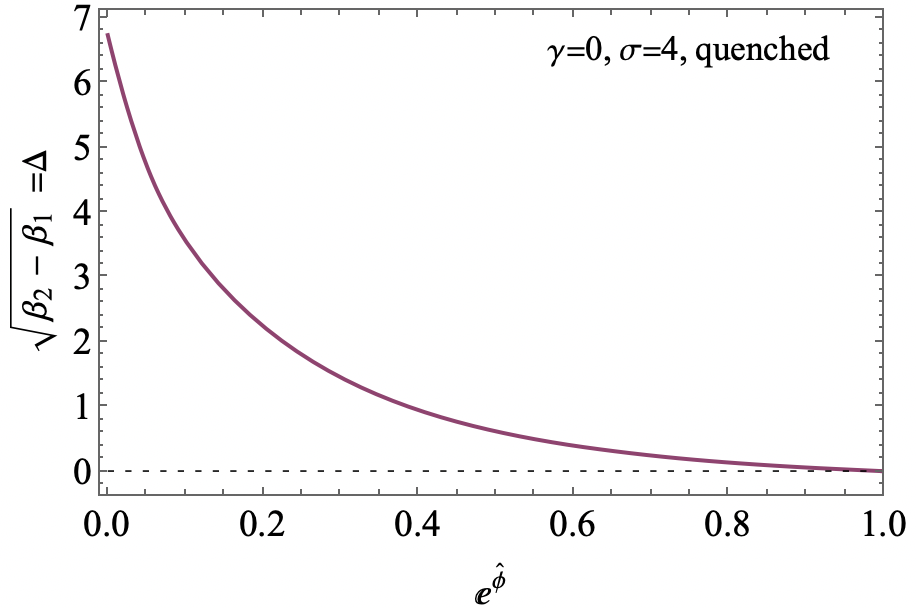}
    \includegraphics[width=.49\linewidth]{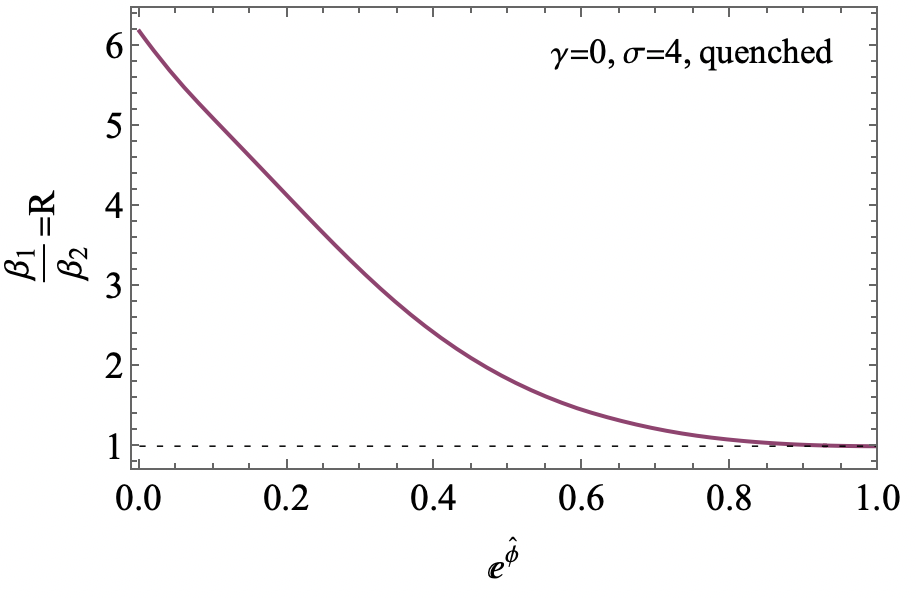}
    \includegraphics[width=.5\linewidth]{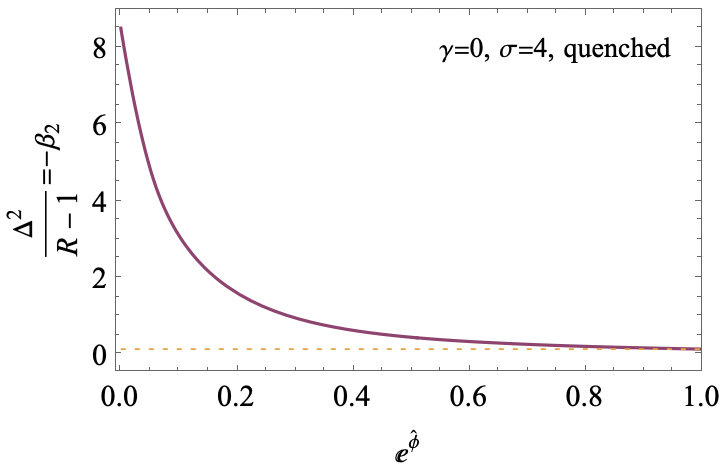}
\caption{Behaviour of the parameters $r, \Delta=\sqrt{\beta_2-\beta_1}, R=\beta_1/ \beta_2$ and $\Delta^2/(R-1)$ obtained solving the quenched self-consistent equations at a fixed value of the conjugate parameter $\hat \phi$. The plot shows for $\hat \phi \to 0^-$, the solutions to the saddle point equations satisfy $r, R \to 1$, and $\Delta \to 0$, which is the limiting behaviour discussed in Sec.~\ref{sec:CavPoint}. Moreover, the ratio $\Delta^2/(R-1)$ approaches a finite value $b=0.146$. For positive values of $\hat \phi$, the annealed self-consistent equations have to be considered. }\label{fig:ParametersLimit}
\end{figure}

A plot of the quenched and annealed complexities $\Sigma_\sigma(\phi)$ of uninvadable equilibria is given in Fig. \ref{fig:RepresentativePlotComp} for $\gamma=0$ and for two representative values of $\sigma > \sigma_c$ in the multiple equilibria phase. The complexity curves are positive for an extensive range of diversities $\phi \in \quadre{\phi_a(\sigma), \phi_b(\sigma)}$, that becomes larger as $\sigma$ increases; therefore, for $\sigma$ large enough the generalized Lotka-Volterra dynamical equations admit an exponentially large number of uninvadable equilibrium configurations with a full distribution of diversities. To each value of $\sigma$ there corresponds a unique value of diversity $\phi_{\rm max}$ which maximizes the complexity curve $\Sigma^{(Q)}_\sigma(\phi)$, giving therefore the diversity of the equilibria that are the exponentially most numerous at the given $\sigma$. All these equilibria are linearly unstable, since they have values of diversities $\phi$ that all exceed the May stability bound, Eq.~ \eqref{eq:MayBound}. Both plots show the special value of the diversity parameter $\phi_{\rm cav}$, such that for $\phi >\phi_{\rm cav}$ the annealed complexity is strictly larger than the quenched one, while for $\phi \leq \phi_{\rm cav}$ the two curves coincide. The diversity $\phi_{\rm cav}$ corresponds exactly to the cavity point discussed in Sec.~\ref{sec:CavPoint}: it is the diversity corresponding to $\hat \phi=0$ in the self-consistent equations. When $\phi \to \phi_{\rm cav}^+$, the solutions of the quenched self-consistent equations satisfy the limiting behaviour \eqref{eq:CavityLimit}, and the quenched equations can be mapped exactly to the annealed one as shown in Sec.~\ref{sec:CavPoint}. We remark that the mapping of Sec.~\ref{sec:CavPoint} holds exactly at the cavity point and not for $\phi<\phi_{\rm cav}$, since it assumes $\mathcal{r} = 1$: the complexity at smaller values of diversity $\phi$ (equivalently, at larger values of $\hat \phi$) must then be obtained solving the annealed self-consistent equations of Sec.~\ref{sec:AnnealedEqs}, since the quenched ones have no meaning in this regime. In Fig.~\ref{fig:ParametersLimit} we show the behaviour of the conjugate parameters obtained solving the quenched equations, to confirm that the limiting behaviour \eqref{eq:CavityLimit2} holds true when $\hat \phi \to 0^-$. \\

\subsection{Role of the average interaction strength $\mu$ and the unbounded phase. }\label{sec:unbounded}
\begin{figure}[ht]
\centering
\includegraphics[width=.42\linewidth]{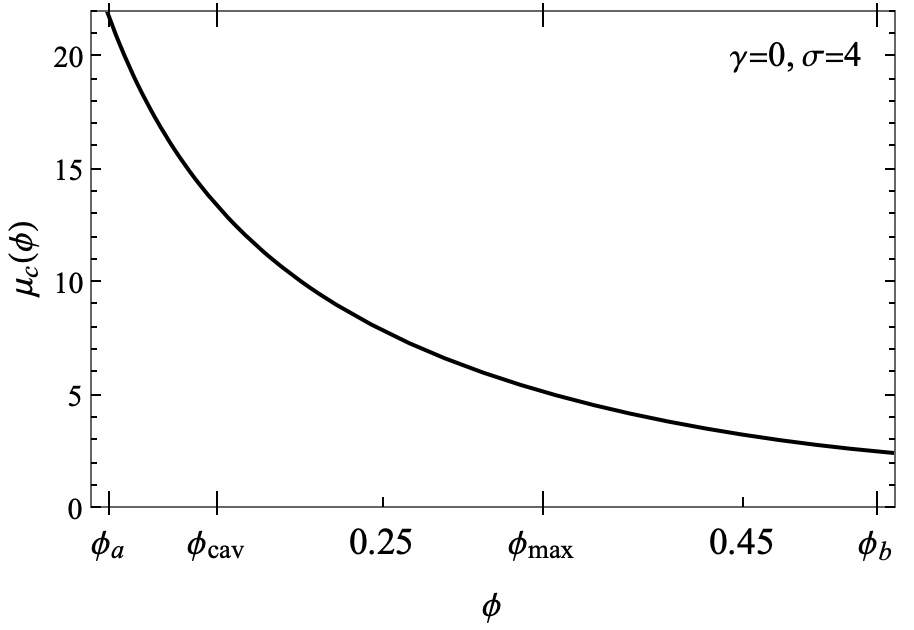}
\includegraphics[width=.45\linewidth]{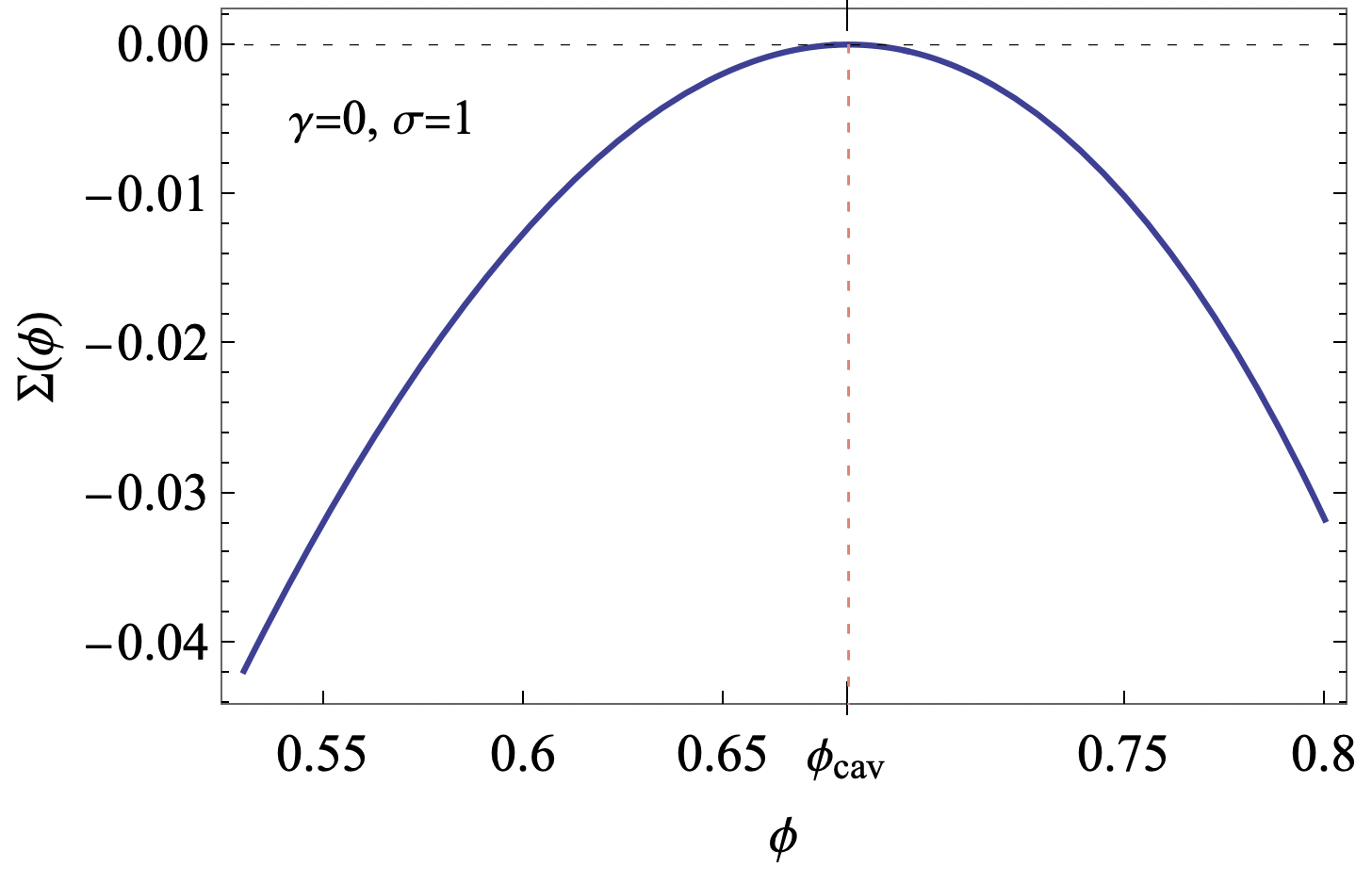}
\caption{\emph{Left.} Critical curve separating the unbounded phase ($\mu < \mu_c$) from the bounded one ($\mu >\mu_c$), as a function of the diversity $\phi$ and for fixed $\sigma$.  \emph{Right.} Complexity of equilibria for $\sigma<\sigma_c= \sqrt{2}$. In this region, annealed and quenched calculations coincide. The complexity is non-zero only at the value of diversity predicted by the cavity formalism. } \label{fig:MuCritico}
\end{figure}

Let us comment on the role of the parameters $\mu, \kappa$. We focus on the quenched case to fix the ideas — the annealed case is analogous. As it follows from the discussion in Sec.~\ref{sec:QuenchedEqs} and Sec.~\ref{sec:AnnealedEqs},  the coupled self-consistent equations for the quenched parameters $r, x_1, \beta_1, \beta_2$ are independent of $\mu, \kappa$; the equations relating the diversity $\phi$ to the conjugate parameter $\hat \phi$ do not depend on $\mu, \kappa$  either.  The parameters $\mu$ and $\kappa$ enter only in the equations for $y$. Given that Eq.~\eqref{eq:QuenchedCompConj} does not depend on $y$, it follows that the complexity curves $\Sigma^{(Q)}_\sigma(\phi)$ at fixed values of $\sigma$ are the same for any value of $\mu, \kappa$.
Changes in $\mu, \kappa$ amount to changing the value of variable $y$, and therefore to rescaling the order parameters: since $\kappa$ just gives a simple linear rescaling of $y$, see Eq. \eqref{eq:QuenchedY}, we set $\kappa=1$. From Eq. \eqref{eq:QuenchedY} it follows that decreasing $\mu$ at fixed $\sigma, \phi$, the variable $y$ decreases and therefore the values of $m, q_1,q_0$ increase. Thus, decreasing $\mu$ one can drive the system toward the unbounded phase, where the order parameters diverge. The unbounded phase is reached whenever $y \to 0$, which is also a limit of stability of our calculation — recall that all the self- consistent equations are obtained under the assumption that $y>0$. For each $\sigma$ one can define a curve of critical values $\mu_c(\phi)$ such that for values of $\mu <\mu_c$, equilibria with diversity $\phi$  are in the unbounded phase. An example of this curve is given in Fig. \ref{fig:MuCritico} (left). Notice that it holds:
\begin{equation}
\mu^* \equiv \max_{\phi: \Sigma(\phi) \geq 0} \mu_c(\phi)=  \mu_c(\phi_{\rm a}) >  \mu_c(\phi_{\rm cav}). 
\end{equation}
Therefore, the prediction of the location of the unbounded regime obtained through the cavity calculation does not account for \emph{all} the equilibria: for $\mu$ slightly larger than $\mu_c(\phi_{\rm cav})$ , there are still equilibria at $\phi<\phi_{\rm cav}$ that are in the unbounded regime. On the other hand, if one defines the phase boundary by requiring that only the most numerous equilibria (those having $\phi=\phi_{\rm max}$) are bounded, one gets a yet different transition line which can be determined explicitly from our calculation. Finally, the divergence of the dynamics might be on yet another different line.
The transition to the unbounded phase can also be characterized in terms of an isolated eigenvalue in the spectrum of the stability matrix: it is expected to occur at those values of parameters for which the isolated eigenvalue crosses zero (in the unique equilibrium phase, this has been argued in Ref.~\cite{baron2022non}.)\\

\subsection{Behaviour of the order parameters } 
\begin{figure}[ht]
 \centering  \includegraphics[width=.45\linewidth]{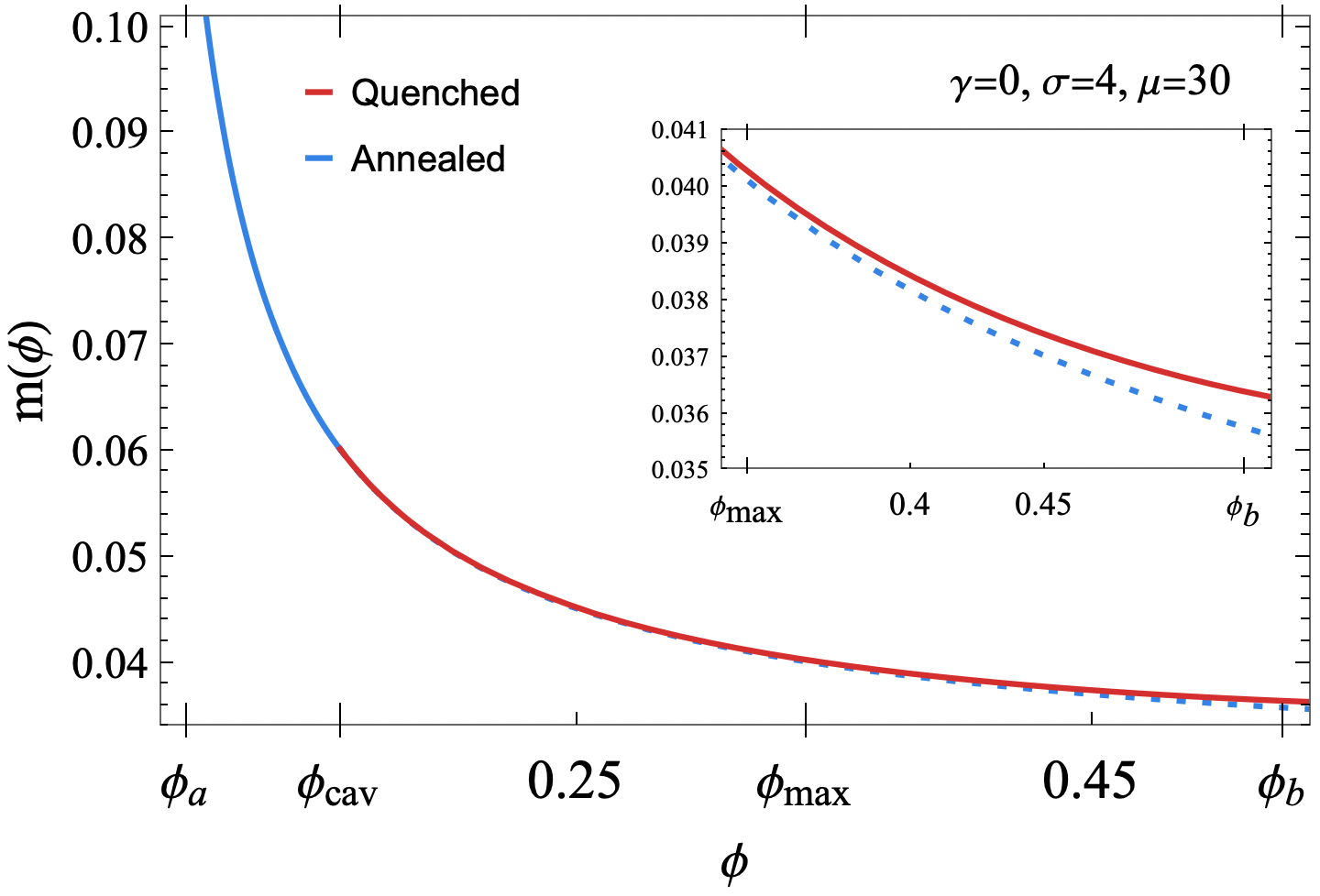}
    \includegraphics[width=.45\linewidth]{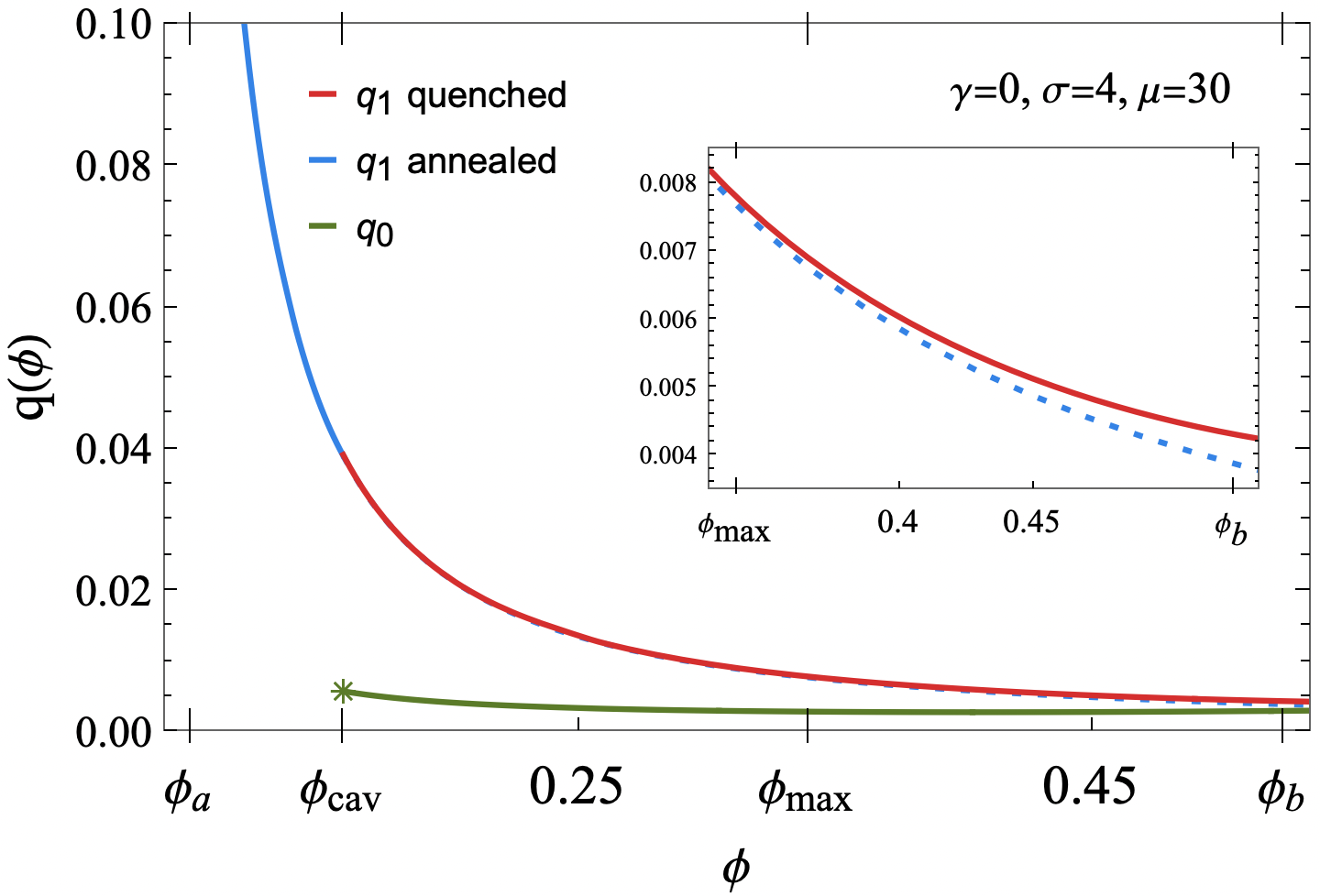}
\caption{Order parameters characterizing the equilibria at fixed diversity $\phi$, for $\sigma=4$ and $\mu=30$.  The insets are zoomed versions of the main plots. More diverse equilibria have a smaller average abundance $m$, and are less correlated to each others ($q_0$ is smaller). For $\phi> \phi_{\rm cav}$, the annealed calculation underestimates the average abundance $m$ and the self-overlap $q_1$ of the equilibria.   }\label{fig:OrderPar}
\end{figure}

We focus on values of $\mu> \mu_c(\phi_a)$ large enough so that none of the equilibria in this range of diversities is unbounded.  
The behaviour of the order parameters is shown in Fig.~\ref{fig:OrderPar} for one such value of $\mu$. One sees that more diverse equilibria have a smaller average abundance $m$, and are less correlated to each others (the typical overlap between them $q_0$ is smaller). The abundance $m$ and the self-overlap $q_1$ obtained within the annealed approximation are a lower bound to the quenched ones, as shown more clearly in the inset of the plots. \\

\subsection{Dependence on $\sigma$ and the topology trivialization  transition} 
\begin{figure}[ht]
 \centering 
 \includegraphics[width=.48\linewidth]{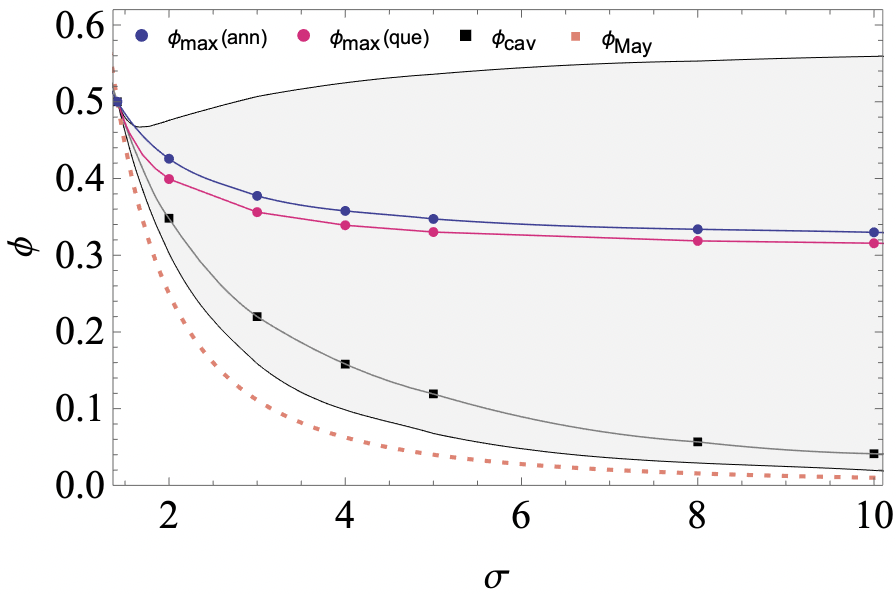}
    \includegraphics[width=.48\linewidth]{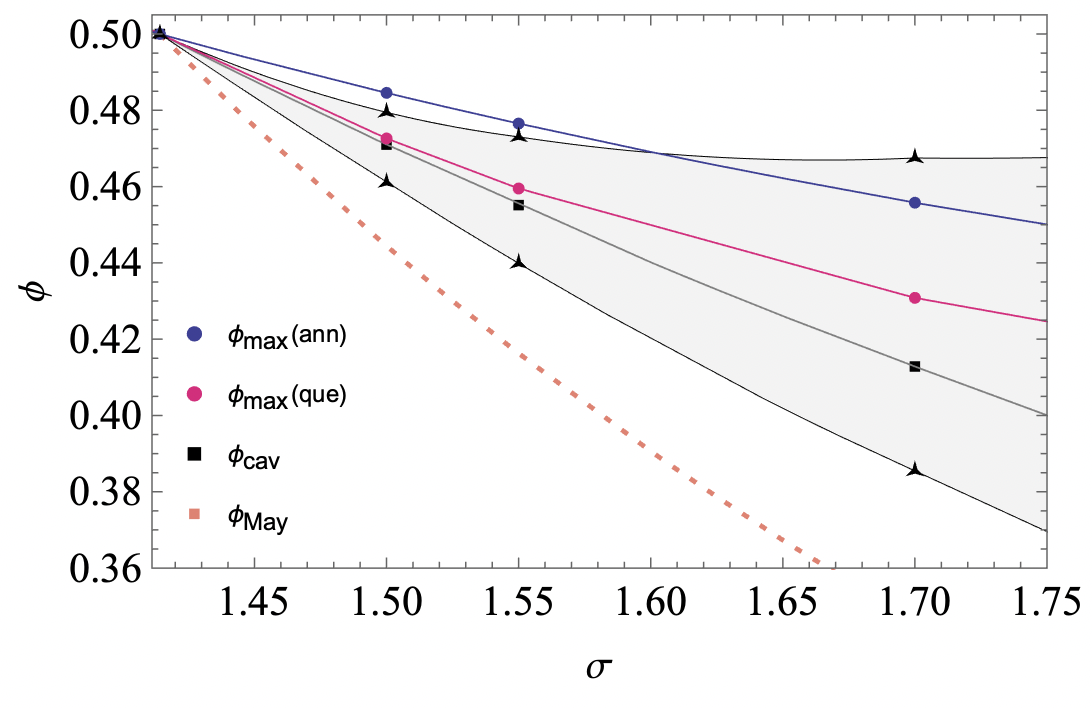}
\caption{\emph{Left. } Relevant diversities as a function of $\sigma$. The two black lines show the edges  $\phi_a(\sigma), \phi_b(\sigma)$ of the support of the quenched complexity, i.e. the boundary of the interval of diversities (grey area) within which the quenched complexity is positive. The dotted lines correspond to the diversity maximizing the annealed (blue) and quenched (pink) complexity, while the black squares give $\phi_{\rm cav}$. Finally, the orange dashed line corresponds to the diversity $\phi_{\rm May}$ above which all equilibria are linearly unstable. \emph{Right. } Zoom of the plot in the vicinity of the critical value $\sigma_c=\sqrt{2}$, where all curves cross. }\label{fig:DivSigmaPreliminary}
\end{figure}

In Fig. \ref{fig:DivSigmaPreliminary} we show $\sigma$-dependence of the relevant diversities; the grey area gives the support of the quenched complexity, which is seen to decrease with decreasing $\sigma$. 

 As the unique-to-multiple equilibria transition is approached, the complexity curves such as those in Fig.~\ref{fig:RepresentativePlotComp} decrease in height, while their support squeezes. At the same time, $\phi_{\rm cav}$ moves towards $\phi_{\rm max}$, see Fig.~\ref{fig:DivSigmaPreliminary} (right). Exactly at $\sigma=\sigma_c=\sqrt{2}$, one finds that the complexity is maximal at  $\phi_{\rm max}=\phi_{\rm cav}=1/2$, and the corresponding complexity vanishes: the unique equilibrium phase is reached. At the transition, the annealed self-consistent equations (to which the quenched ones reduce to) are solved by $\mathcal{r}=1, \mathcal{x}_1=0$, which imply $\mathcal{y}^2=\mu^2/(2 \pi)$ and thus  $m=-p=\mu^{-1}$and $q_1=\xi_1= \pi \mu^{-2}$.
One would naturally expect that the transition corresponds to $b \to 0$ but this can not be concluded from the equation \eqref{eq:Selfb}: indeed, plugging $\mathcal{x}_1=0$ into \eqref{eq:Selfb} one simply finds an identity for any value of $b$. For $\sigma< \sigma_c$, the annealed complexity is non-zero only at $\phi=\phi_{\rm cav}$, see Fig.~\ref{fig:MuCritico}(left), which is indeed the  diversity of the unique equilibrium. For other values of $\phi$, the complexity is negative, signifying that no equilibria of those diversities exist typically (i.e., the probability to find them is exponentially suppressed in $S$, as it follows from the Markov inequality \cite{auffinger2013random}). 

\begin{figure}[ht]
\centering
    \includegraphics[width=.67\linewidth]{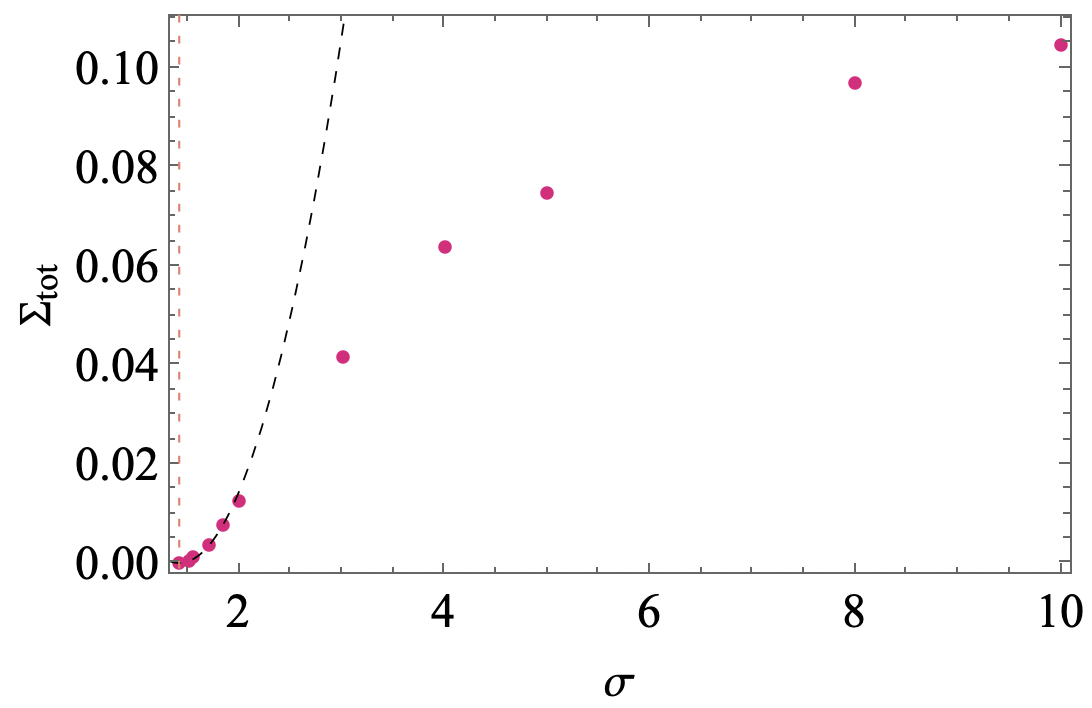} 
\caption{Total quenched complexity $\Sigma_{\sigma}^{\rm tot}= \Sigma^{(Q)}_\sigma(\phi_{\rm max})$ as a function of $\sigma$: the black dotted line is a quadratic fit of the form $\Sigma_{\sigma}^{\rm tot}= a (\sigma- \sigma_c)^2$ with $a \approx 0.037$.  }\label{fig:TotalComp}
\end{figure}

 In Fig.~\ref{fig:TotalComp}, we show the behaviour of the total quenched complexity 
\begin{equation}
\Sigma_{\sigma}^{\rm tot}= \Sigma^{(Q)}_\sigma(\phi_{\rm max}).
\end{equation}
At the trivialization transition $\sigma=\sigma_c= \sqrt{2}$ the total complexity vanishes as $\Sigma_{\sigma}^{\rm tot} \sim (\sigma-\sigma_c)^2$. 
The quadratic vanishing of the complexity at the transition has been observed in other models \cite{LCTF22, arous2021landscape, Toboul} treated within the annealed approximation, and it has been conjectured to be a robust feature. In fact, we find (see Sec. \ref{sec:TopTrivGenGamma}) that the same behavior holds true for general $\gamma$ within the annealed approximation. For $\gamma=0$, this behavior is recovered within the quenched framework, too, as we now show explicitly. Indeed, the total derivative of the quenched complexity  with respect to $\sigma$ is contributed by four terms:
\begin{equation}\label{eq:TotVar}
\frac{d \Sigma_{\sigma}^{\rm tot}}{d \sigma}=\nabla_{ {\bf x}} \bar{\mathcal{A}} (\bf{x}, \hat {\bf x}, \phi) \partial_{\sigma} {\bf x} + \nabla_{ \hat{\bf x}} \bar{\mathcal{A}} (\bf{x}, \hat {\bf x}, \phi) \partial_{\sigma} {\hat {\bf x}} +\partial_{\phi} \bar{\mathcal{A}} (\bf{x}, \hat {\bf x}, \phi) \partial_{\sigma} \phi+ \partial_{\sigma} \bar{\mathcal{A}} (\bf{x}, \hat {\bf x}, \phi) \Big|_{{\bf x}^*, \hat {\bf x}^*, \phi_{\rm max}}.
\end{equation}
The first three terms vanish for any value of $\sigma$, due to the fact that ${\bf x}^*, \hat {\bf x}^*, \phi_{\rm max}$ are precisely chosen to maximize $\bar{\mathcal{A}}$. On the other hand, the derivative with respect to $\sigma$, 
\begin{equation}\label{eq:terms}
\partial_{\sigma} \bar{\mathcal{A}} (\bf{x}, \hat {\bf x}, \phi)= \partial_{\sigma} \bar{\mathcal{p}}({\bf x})\Big|_{{\bf x}^*, \hat {\bf x}^*, \phi_{\rm max}}+ \partial_{\sigma} \mathcal{d}(\phi)\Big|_{{\bf x}^*, \hat {\bf x}^*, \phi_{\rm max}},
\end{equation}
also vanishes when plugging the values of the order parameters at $\sigma=\sigma_c$, since both terms vanish separately. Indeed, taking the derivative of \eqref{eq:Det} one finds
\begin{equation}
\partial_{\sigma}\mathcal{d}(\phi)=  \begin{cases}
  0 &\text{  if  }\quad 0<\sigma \sqrt{\phi} <1\\
- \frac{1}{\sigma^3}+\frac{\phi}{\sigma} &\text{  if  }\quad \sigma \sqrt{\phi} >1,
   \end{cases}
 \end{equation}
 which vanishes at $\phi=\phi_{\rm May}=\sigma^{-2}$, which is the $\phi_{\rm max}$ at $\sigma=\sigma_c$. On the other hand, the partial derivative of \eqref{eq:Forc} for $\kappa=1$ and $\gamma=0$ reads:
 \begin{equation}\label{eq:ParF}
\begin{split}
&\partial_\sigma\bar{ \mathcal{p}}({\bf x})= -\frac{1}{\sigma}\\
&\frac{2}{\sigma^3}\quadre{\frac{1}{2}+ \frac{ q_0 z}{(q_1-q_0)^2}-\frac{(m+p) (1-\mu  m)}{(q_1-q_0)}+ \frac{\xi_1 }{ 2(q_1-q_0)}
+ \frac{q_0 (\xi_0- \xi_1)}{2 (q_1-q_0)^2}
 +\frac{1}{2} \frac{\left(1-\mu 
   m\right)^2}{q_1-q_0}}.
   \end{split}
\end{equation}
At $\sigma=\sigma_c^-$, the annealed calculation implies that the order parameters characterizing one single replica equal to  $m=-p=\mu^{-1}$and $q_1=\xi_1= \pi \mu^{-2}$. Using that these parameters are continuous at $\sigma_c$, plugging them into the quenched equations  \eqref{eq:SetSelfCons} and using that $r=1$ one finds  $\xi_1- \xi_0- 2 z=(q_1-q_0)$. Using these results, one sees that also \eqref{eq:ParF} vanishes at $\sigma=\sigma_c$.

\section{Additional results for $\gamma \neq 0$}\label{sec:GeneralGamma}
We do not present in this work the results for the quenched complexity for $\gamma \neq 0$; however, we discuss in this section some interesting dependence on the asymmetry parameter $\gamma$ that can be deduced from the (much simpler) calculation of the annealed complexity. 

\subsection{On the stability of equilibria for general $\gamma$}
For $\gamma=0$, it follows from the calculation presented above that all the uninvadable equilibria are linearly unstable: for all values of $\sigma$, the complexity is entirely supported in the region $\phi > 	\phi_{\rm May}$.  It is natural to ask whether this remains true for $\gamma>0$.
To get information  on the stability of equilibria  we evaluated the annealed complexity and computed the lower edge of its support, $\phi_a^{A}$, at fixed positive $\gamma, \sigma$.
Given that the annealed complexity is an upper bound to the quenched complexity, the $\phi_a^{A}$ obtained from the annealed calculation is a lower bound to the corresponding diversity obtained within the quenched calculation (for $\gamma=0$ the two quantities coincide). The inequality $\phi^{A}_a > \phi_{\rm May}$ thus implies that {\emph no} linearly stable equilibrium exists for the given  $\sigma, \gamma$. 

\begin{figure}[ht]
 \includegraphics[width=.48\linewidth]{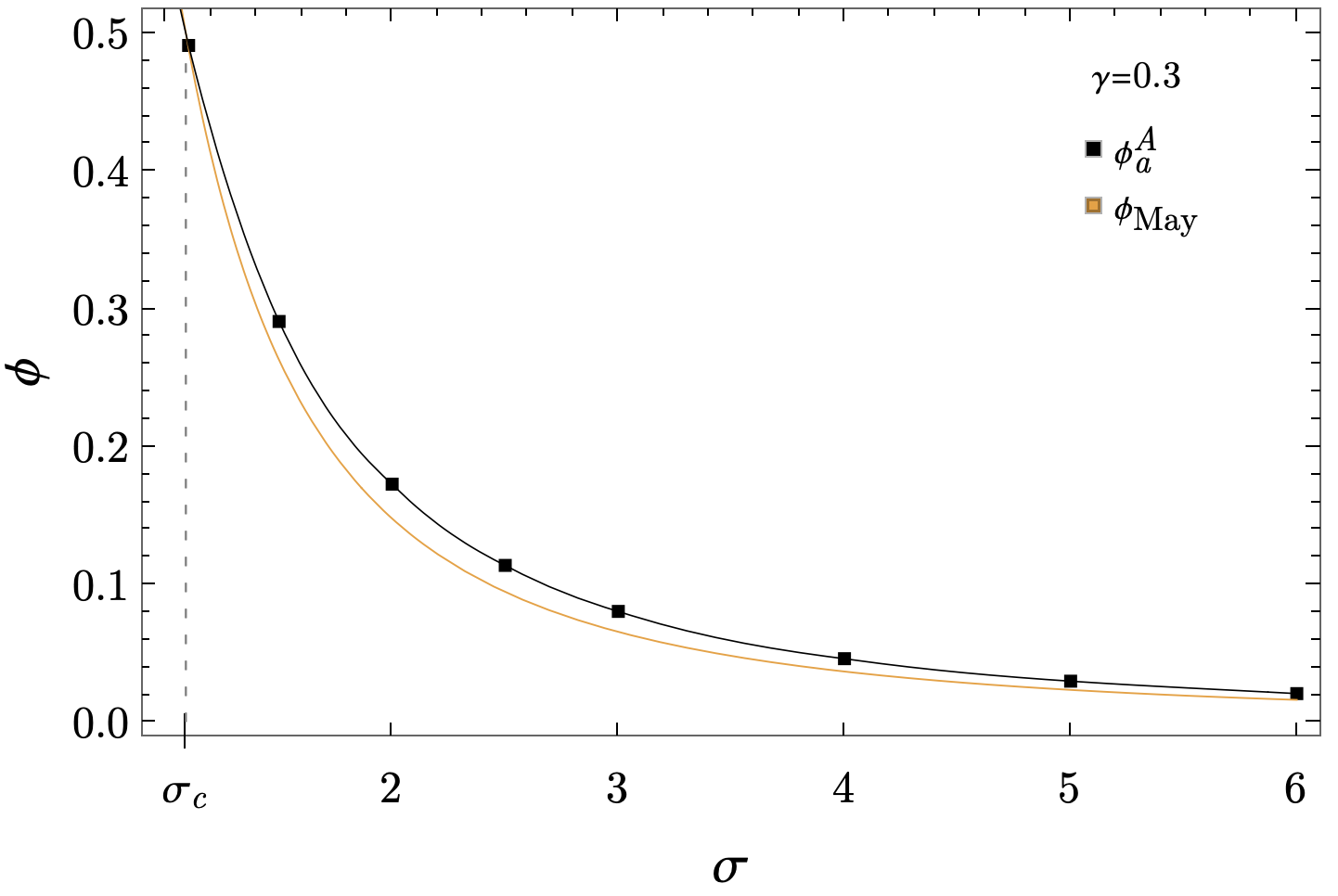} 
 \includegraphics[width=.48\linewidth]{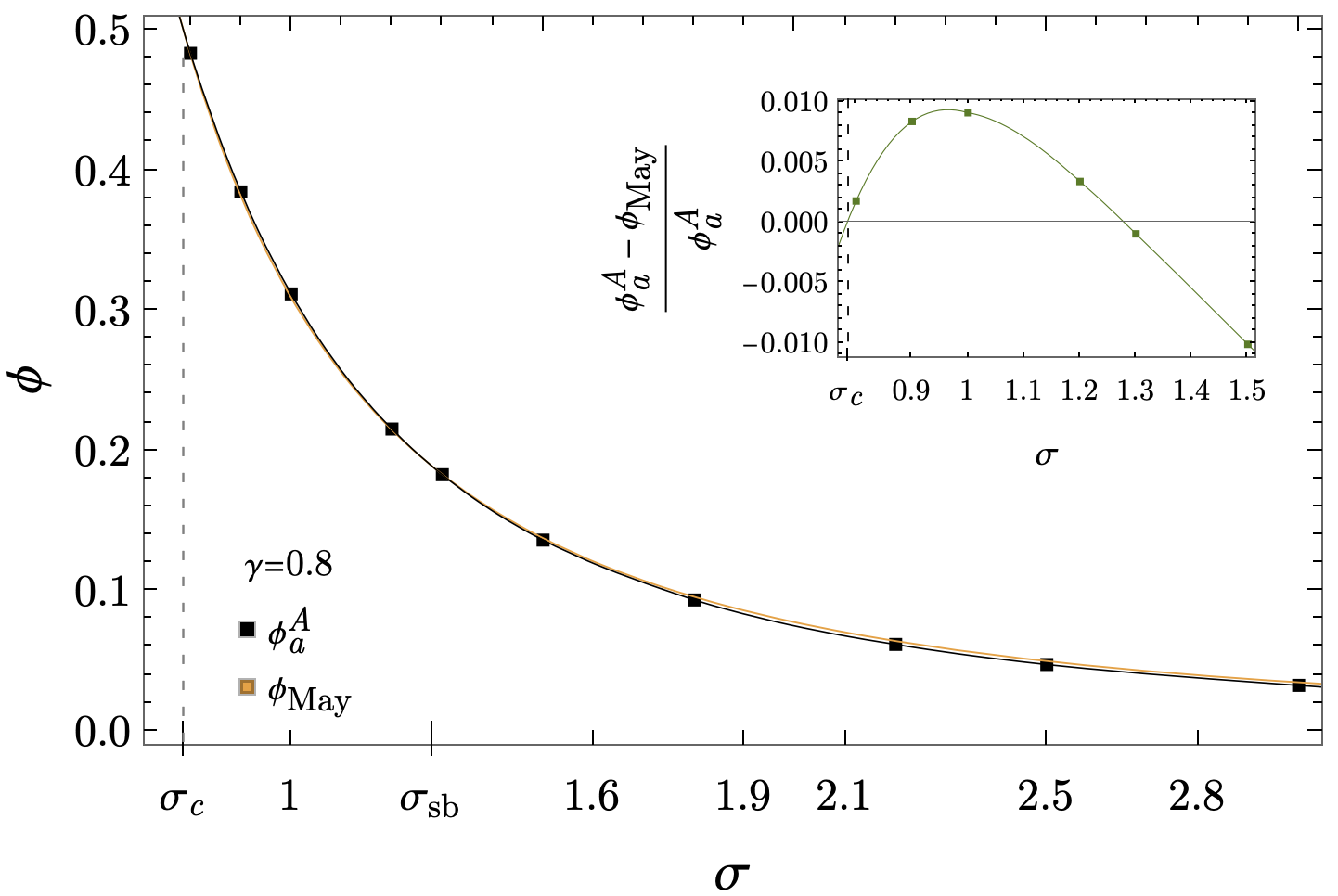} \caption{Comparison between the minimal diversity $\phi_a^A$ at which the annealed complexity is positive and the diversity $\phi_{\rm May}$ above which equilibria are unstable, for asymmetry $\gamma=0.3$ ({\it left}) and $\gamma=0.8$ ({\it right}). For $\gamma=0.8$, the curves cross at $\sigma_{\rm sb}=1.279$.}\label{fig:GammaDependence}
\end{figure}

\begin{figure}[ht]
 \includegraphics[width=.48\linewidth]{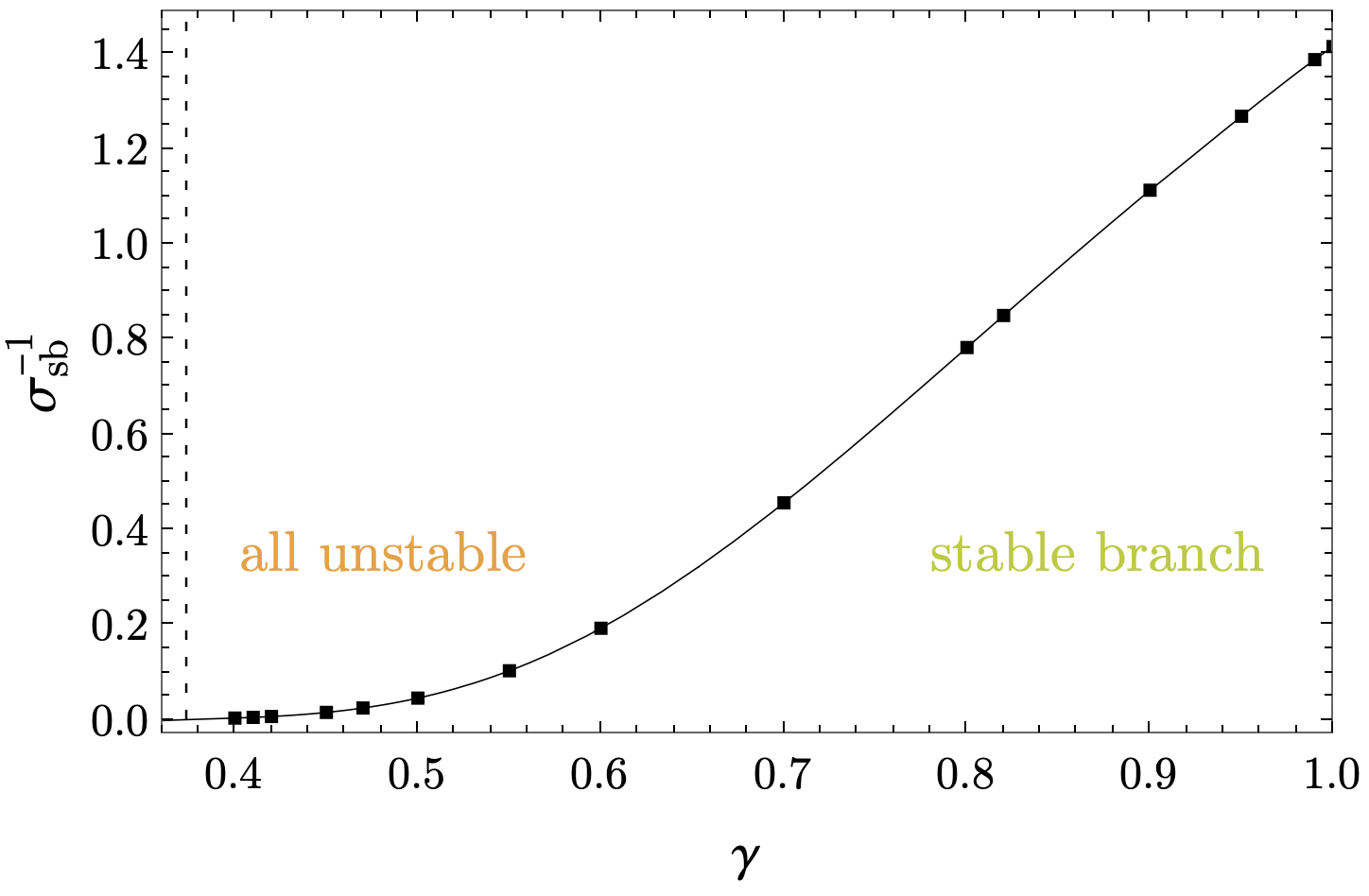} \includegraphics[width=.48\linewidth]{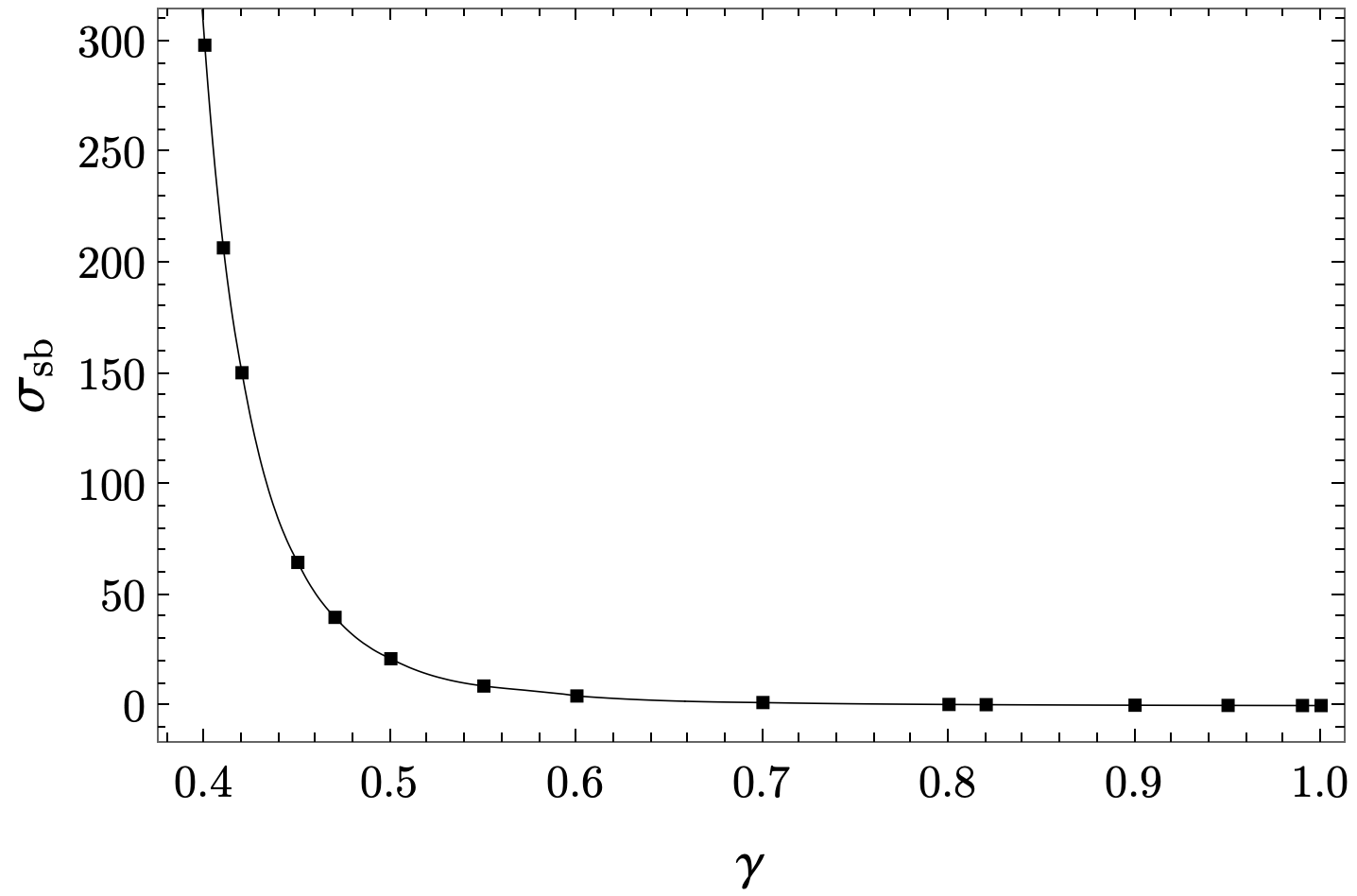} \caption{Dependence on $\gamma$ of the crossing point $\sigma_{\rm sb}^{-1}$ and its inverse. The curves vanish at $\gamma_c=0.373$ and $\gamma=1$, respectively. For $\gamma< \gamma_c$, for certain all uninvadable equilibria are linearly unstable.}\label{fig:AnnealedTrans}
\end{figure}

 Fig.~\ref{fig:GammaDependence} shows the comparison between $\phi_a^A$ and $\phi_{\rm May}$ as a function of $\sigma$, for two different values of the asymmetry parameter $\gamma$. One sees that for the smaller value of $\gamma$, all the equilibria are unstable in the plotted range of $\sigma$, while for the larger value of $\gamma$ there is a crossing value $\sigma_{\rm sb}$ such that for $\sigma< \sigma_{\rm sb}$ all equilibria are unstable, while for the larger values of $\sigma$ the annealed complexity is non-zero also in a window of diversities corresponding to linearly-stable equilibria. For the smaller values of $\gamma$, it is unclear from this plot whether such a crossing occurs at much larger values of variability $\sigma$; to determine this, we show in Fig \ref{fig:AnnealedTrans} the dependence on $\gamma$ of the inverse of the crossing point $\sigma_{\rm sb}^{-1}$ (respectively, $\sigma_{\rm sb}$), which is shown to vanish at a threshold value  $\gamma_c=0.373$ (respectively, at $\gamma=1$): we can therefore conclude that for $0 \leq \gamma \leq \gamma_c$, all the uninvadable equilibria are linearly unstable. For $\gamma> \gamma_c$, the annealed complexity suggests that some (exponentially many in $S$) linearly stable equilibria are present at large-enough $\sigma$; in the symmetric case $\gamma=1$, one has $\sigma_{\rm sb}=\sigma_c$, and therefore for all $\sigma$ in the multiple equilibria phase the annealed calculation predicts that some stable equilibria are present. For all $\gamma>\gamma_c$, however, the stable equilibria are much more rare with respect to the most numerous ones (corresponding to the maximum value of the complexity), which are always unstable. An illustration of this in given in Fig. \ref{fig:Annealed} ({\it left}). We remark that the fact that for $\gamma< \gamma_c$ all equilibria are unstable remains true even if a quenched calculation of the complexity is performed. On the other hand, the behaviour for $\gamma>\gamma_c$ obtained within the annealed framework is robust only in case the low-$\phi$ branch of the quenched complexity coincides with the annealed one, as it happens for $\gamma=0$.
It is also possible that the quenched complexity curve in this region has to be obtained beyond the replica symmetry assumptions considered in this work. This is suggested by the symmetric $\gamma=1$ case, where marginally stable equilibria are expected to dominate. We leave these checks to future work. 

\subsection{Topology trivialization transition for general $\gamma$}\label{sec:TopTrivGenGamma}
We have shown above that for $\gamma=0$ the total complexity $\Sigma_{\sigma}^{\rm tot}=\Sigma^{(Q)}_\sigma(\phi_{\rm max})$ vanishes quadratically when $\sigma \to \sigma_c^+$. We now show that this behaviour extends to $\gamma \neq 0$ within the annealed framework; on the other hand, if in the vicinity of $\sigma_c$ the maximum of the complexity curve (as a function of $\phi$) lies in a regime in which the quenched calculation has to be employed (as it happens for $\gamma=0$), then we can not exclude that the total complexity vanishes with a different power. In fact, our results suggest that this is the case for general $\gamma \neq 0$, as we argue below.

We consider the total variation \eqref{eq:TotVar}, and focus first on
the case $\gamma \neq 1$. The contribution to the total variation given by the determinants reads:
 \begin{equation}\label{eq:DerDete}
\partial_\sigma  \mathcal{d}(\phi)=   \begin{cases}
\frac{2 \gamma  \phi  \sigma ^2+\sqrt{1-4 \gamma  \phi  \sigma ^2}-1}{2 \gamma   
   \sigma ^3} &\text{  if  }\quad \phi< \frac{1}{\sigma^2 (1+ \gamma)^2}\\
 \frac{\phi}{\sigma} \tonde{1-\frac{1}{\sigma^2 \phi (1+\gamma)}}  &\text{  if  }\quad \phi> \frac{1}{\sigma^2 (1+ \gamma)^2}
     \end{cases}
 \end{equation}
  At the critical point $\sigma=\sigma_c= \sqrt{2}(1+ \gamma)^{-1}$, the diversity maximizing the complexity is $\phi_{\rm max}=\phi_{\rm May}= [\sigma(1+\gamma)]^{-2}$. The derivative above is continuous at this point, and equals to:
  \begin{equation}\label{eq:Const}
\partial_\sigma  \mathcal{d}(\phi) \Big|_{\sigma_c, \phi_{\rm max}}= - \frac{\gamma(1+\gamma)}{2 \sqrt{2}}.
  \end{equation}
To have a quadratic behavior of the total complexity, this term should be cancelled by the derivative of the term coming from the distribution of the forces. This is indeed what happens if for $\sigma \sim \sigma_c^+$ the  complexity at $\phi_{\rm max}$ is obtained within the annealed framework, and thus the contribution from the distribution of the forces is given by $\mathcal{p}_1({\bf x})$ in \eqref{eq.p1ann}. Then, for $\kappa=1$
\begin{equation}
\begin{split}
&\partial_\sigma \mathcal{p}_1({\bf x})=\frac{1}{\sigma^3 q_1^2 }\quadre{ (1 -\mu  m)^2  \tonde{q_1 -\frac{\gamma\, m^2 }{1+ \gamma} }-2  (1-\mu  m) q_1\tonde{p+\frac{m}{1+\gamma} }+ \xi_1  q_1} \\
 &-\frac{1}{\sigma}  +\frac{1}{ \sigma^3 (1+\gamma)}.
 \end{split}
\end{equation}
This expression has to be evaluated at the solution of the annealed saddle point equations; at $\sigma_c$, one finds $m=\mu^{-1}=-(1+\gamma)p$  and $q_1=(1+\gamma)^2 \xi_1$, which implies that
\begin{equation}
\partial_\sigma \mathcal{p}_1\Big|_{\sigma_c, \phi_{\rm max}}= \frac{\gamma (1+ \gamma)}{2 \sqrt{2}}.
\end{equation}
Therefore, the two contributions cancel exactly within the annealed approximation. On the other hand, if the total complexity at $\sigma \sim \sigma_c^+$ is quenched, one needs to make use of the expression \eqref{eq:Forc} and to determine:
\begin{equation}\label{eq:DerQuenched}
\begin{split}
&\partial_\sigma\bar{ \mathcal{p}}({\bf x})=-2\frac{\tonde{\kappa-\mu  m}}{\sigma^3 (1+\gamma)} \frac{m (q_1-q_0+z \gamma)}{(q_1-q_0)^2}-2\frac{\tonde{\kappa-\mu  m}}{\sigma^3} \frac{ p }{(q_1-q_0)}+ \frac{\gamma}{ \sigma^3 (1+\gamma)} \frac{ z^2 (q_1+ q_0) }{ (q_1-q_0)^3}\\
&+ \frac{\xi_1 }{  \sigma^3(q_1-q_0)}
+ \frac{q_0 (\xi_0- \xi_1)}{\sigma^3 (q_1-q_0)^2}
 +\frac{1}{ \sigma^3 (1+\gamma)} \quadre{1+ \frac{2 q_0 z}{(q_1-q_0)^2}}+\frac{1}{ \sigma^3} \frac{\left(\kappa-\mu 
   m\right)^2}{q_1-q_0} -\frac{1}{\sigma}.
   \end{split}
\end{equation}
To evaluate this expression, one should solve the quenched saddle point equations for general $\gamma$ at $\sigma= \sigma_c^+$. However, by assuming the continuity of the single-replica order parameters $m,p,q_1, \xi_1$ at $\sigma_c$, one can plug the corresponding values obtained from the annealed equations valid at $\sigma=\sigma_c^-$. 
By doing that, we see that 
the term \eqref{eq:DerQuenched} cancels exactly the contribution of the determinant provided that:
\begin{equation}
 \frac{ z}{(1+\gamma)(q_1-q_0)^2} \tonde{\frac{\gamma z (q_1+q_0)}{2 (q_1-q_0)} + q_0}=0.
\end{equation}
which has two possible solutions for $z$: $z=0$, or $z=2 q_0 (q_1-q_0)/[\gamma (q_1+q_0)]$.
Both these solutions however can be shown to be incompatible with the quenched self-consistent equations \footnote{The condition $z=0$ together with the other conditions on the single-replica order parameters would imply $\beta_1 \beta_2-\beta_3^2=0$. For $\gamma \neq 0$ one sees that this is not an admissible solution of the quenched saddle point equations obtained in the limit  $\beta_1 \beta_2-\beta_3^2 \to 0$: in particular, the limiting equation for $z$ is compatible with $z=0$ only for  $q_1=0=q_0$, which one knows from the annealed solution not to be the correct values at $\sigma_c$. On the other hand, the second choice for $z$ is also not compatible, as it gives rise to complex values of the conjugate parameters.}.Therefore, either for $\gamma \neq 0$ the total complexity at $\sigma \sim \sigma_c^+$ is annealed (and then it vanishes quadratically as $\sigma \to \sigma_c$), or it is quenched, in which case one should expect a different power law since the linear contribution is not vanishing. 

The case $\gamma=1$ is special since the derivative \eqref{eq:DerDete} for $\phi <\phi_{\rm May}$ converges to \eqref{eq:Const} for $\sigma \to \sigma_c^+$, with an additional term scaling as $(\sigma-\sigma_c)^{1/2}$ coming from the square root in \eqref{eq:DerDete}, whose argument vanishes when $\phi=\phi_{\rm May}, \sigma=\sigma_c$. Therefore, the total complexity is likely to have a non-analytic behaviour at the transition to the unique equilibrium phase, since the determinant has a contribution of the form $\mathcal{d}(\phi_{\rm max}) \sim (\sigma-\sigma_c)^{3/2}$.

\subsection{The symmetric case: comparison with the replica calculation}\label{sec:ComparisonReplicas}
In the symmetric case $\gamma=1$, the model is conservative and thus one can investigate the structure of the potential landscape associated to it by means of standard techniques developed within the theory of spin glasses. The potential landscape $L[\vec{N}]$  is defined by:
\begin{equation}\label{eq:PotentialLAnd}
F_i(\vec N)= - \frac{\partial L[\vec{N}] }{\partial N_i}, \quad \quad L[\vec{N}]= -\sum_{i=1}^S \tonde{\kappa_i N_i - \frac{N_i^2}{2}}+ \frac{1}{2} \sum_{i,j=1}^S \alpha_{ij}N_i  N_j.
\end{equation}
When the landscape has a simple structure (which in the language of replica theory corresponds to the so called 1-step Replica Symmetry-Breaking -- 1RSB ansatz) the complexity of certain of the landscape local minima can be obtained from the Lagrange transform of a generalized free-energy function, which is related to the partition function of several copies (or \emph{real replicas}) of the system weakly coupled to each others~\cite{monasson1995structural}. The outcome of the calculation is a curve $\Sigma_{\rm 1RSB}(l)$ giving the complexity of the typical (i.e., most numerous) local minima at fixed value $l=\lim_{S \to \infty} S^{-1} L[N^*]$ of the potential landscape \eqref{eq:PotentialLAnd}. We recall the essential steps of this procedure, which is known as the Monasson method, in Appendix~\ref{sec:Replicas}.  In Fig.~\ref{fig:Annealed} ({\it right}), we plot the resulting Monasson complexity as a function of the diversity $\phi$ of the minima contributing to it, in order to compare with Kac-Rice annealed complexity. One sees that the curve $\Sigma_{\rm 1RSB}(\phi)$ is contributed by two branches, one of which (the red dashed branch) has to be discarded, as we motivate in  Appendix~\ref{sec:Replicas}. This is confirmed by the fact that it gives a positive complexity in a range of diversity where the annealed complexity vanishes -- given that the annealed complexity is an upper bound to the quenched complexity, no local minima can exist in the region in which it is negative. The second branch (green) gives a positive complexity in the region of diversity corresponding to stable equilibria ($\phi<\phi_{\rm May}$); this is consistent with the fact that the replica method allows to find stable local minima and not unstable saddles in the energy landscape. One sees moreover that  $\Sigma_{\rm 1RSB}(\phi)$ not only is quite smaller than the annealed complexity (which might be motivated by the fact that the annealed calculation is not correct and overestimates the complexity), but has a quite different shape. This is due to the fact that the Kac-Rice complexity counts the dominating minima \emph{at fixed diversity}, while $\Sigma_{\rm 1RSB}$ counts the dominating minima \emph{at fixed value of the potential}: the two different constraints imposed in the complexity calculation are not interchangeable. The curve $\Sigma_{\rm 1RSB}(\phi)$ vanishes at $\phi=0.2494$, which corresponds to the 1RSB prediction of the diversity of the equilibrium local minima, i.e. of the ground state. This value is slightly smaller than $\phi_{\rm May}=0.25$, the value corresponding to marginally stable minima which are expected to be the equilibrium ones for $\gamma=1$ and $\sigma>\sigma_c$: this discrepancy is due to the fact that a different (full-RSB) equilibrium calculation is required to capture the correct diversity. However, one sees that the calculation performed within the 1RSB framework is quantitatively quite accurate.
Conversely, the maximum of $\Sigma_{\rm 1RSB}(\phi)$  intercepts the annealed curve at the point where the stable branch turns into the unstable one: this suggests that in that range of diversities, the annealed calculation is correct, in the sense that it matches with the quenched one.

\begin{figure}[ht]
 \includegraphics[width=.48\linewidth]{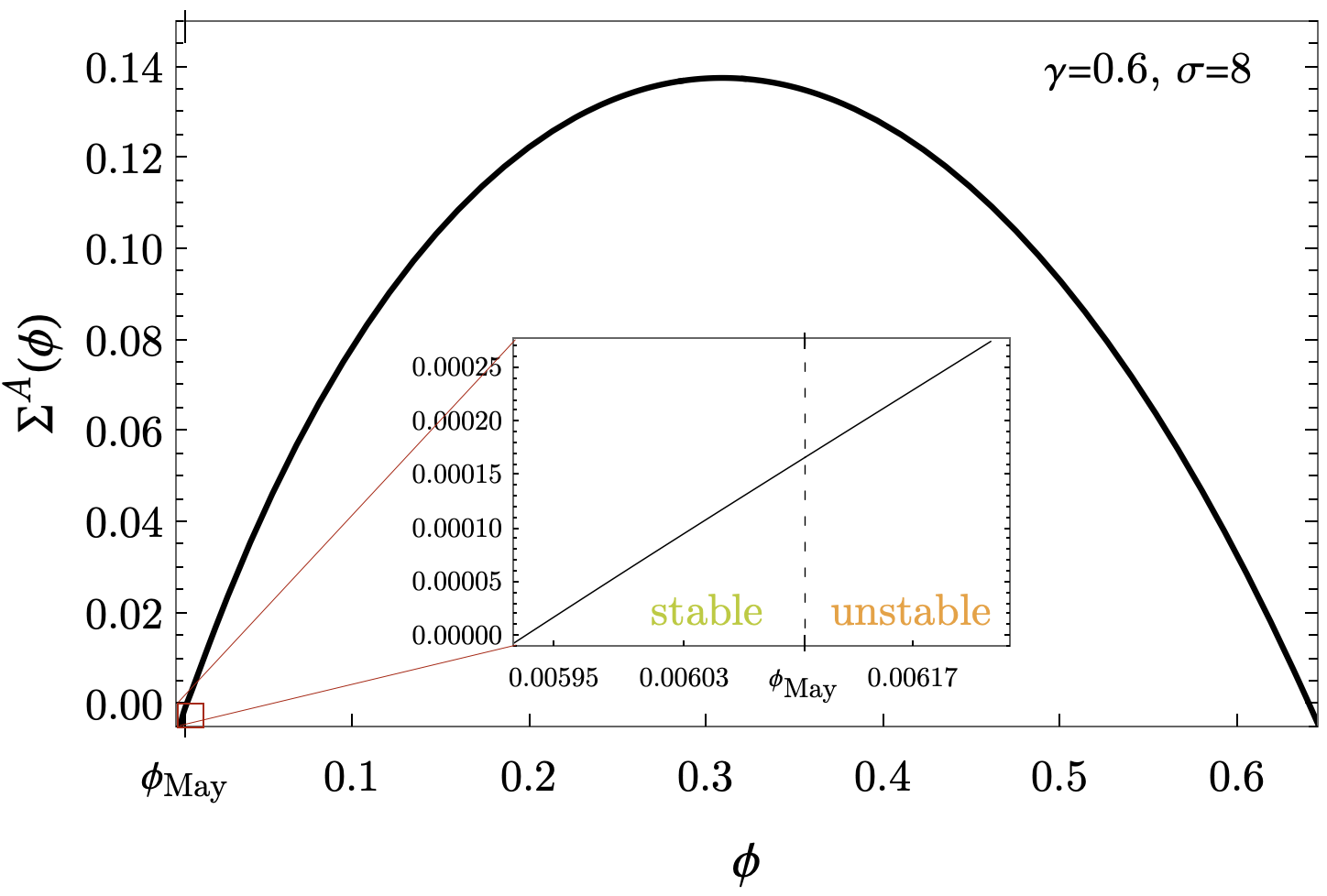}
 \includegraphics[width=.5\linewidth]{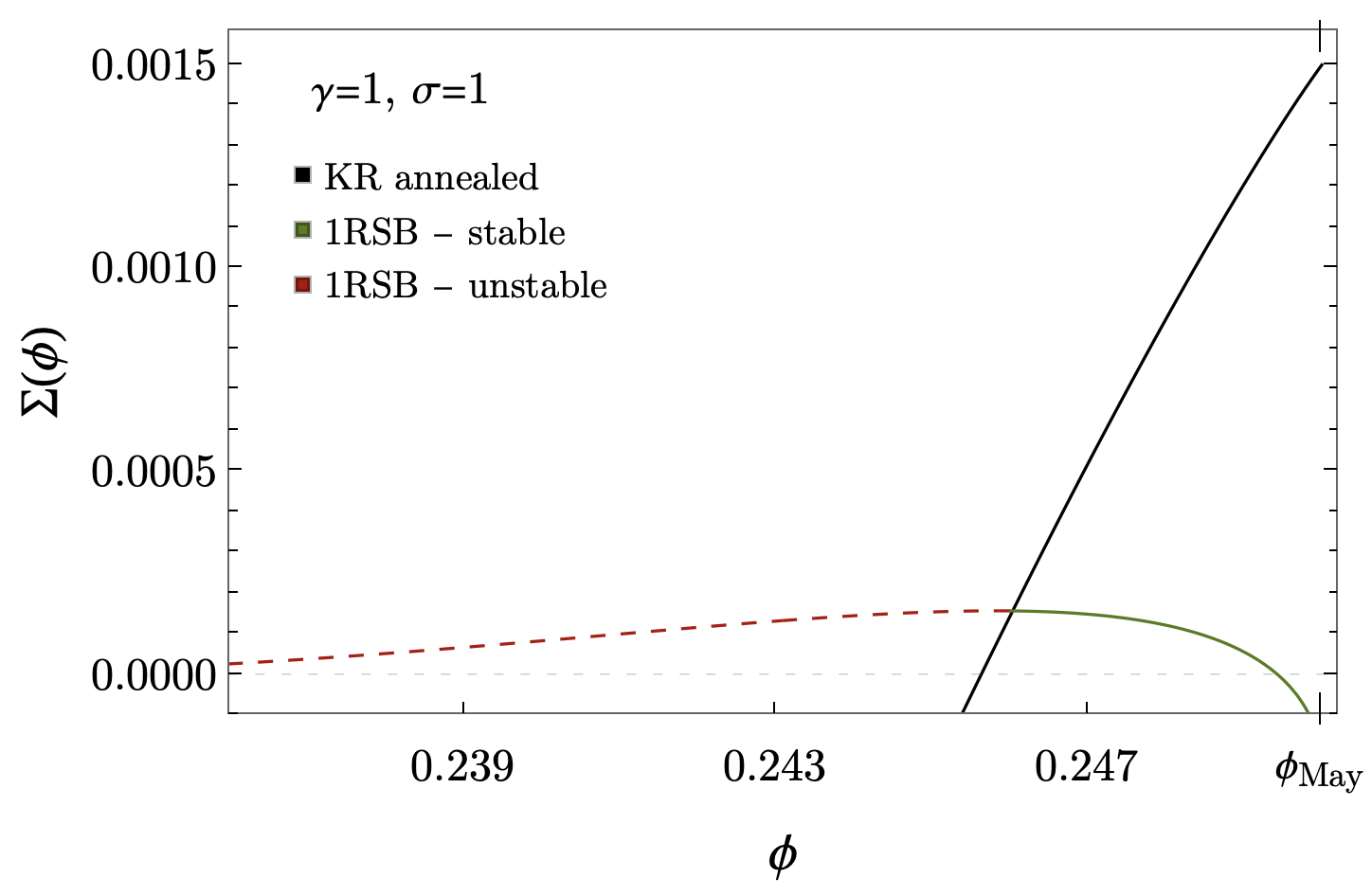} \caption{{\it Left. } Annealed complexity of equilibria for  $\gamma=0.6$ and $\sigma=8>\sigma_c$. A small part of the curve is contributed by stable equilibria with $\phi<\phi_{\rm May}$, see inset. {\it Right. } Comparison between the Kac-Rice annealed complexity and the complexity of local minima obtained with the replica calculation, for the symmetric case $\gamma=1$ and $\sigma=1>\sigma_c$.}\label{fig:Annealed}
\end{figure}

\section{Conclusions}\label{sec:Discussion}
In this work, we have determined the quenched complexity of equilibria of the Generalized Lotka-Volterra equations with random, asymmetric interactions between the species. The quenched complexity is defined in terms of order parameters  satisfying coupled self-consistent equations. We have derived such self-consistent equations for arbitrary values of the parameter $\gamma$, which controls the asymmetry in the statistics of the interactions. We have then discussed in details the strategy to solve these equations in the case of totally uncorrelated interactions, corresponding to $\gamma=0$, and we have presented the results for the associated complexity. 

Our results confirm the expectation that the typical number of equilibria is in general much smaller than the average number, that therefore a quenched calculation is necessary: indeed, the annealed complexity gives a non-tight upper bound to the quenched complexity, at least for most values of diversities of the equilibria (for general $\sigma$ in the multiple  equilibria phase, only the number of equilibria at small diversity is correctly captured by the annealed approximation). For values of variability $\sigma$ quite close to the transition to the  unique equilibrium phase, $\sigma \sim \sigma_c^+$, the annealed approximation fails dramatically, as it predicts dominating equilibria having  a diversity for which typically \emph{no} equilibrium exists, meaning that the corresponding diversity lies outside of the support of the quenched complexity. This interesting phenomenology has been found in other random models, for example in problems of portfolio optimization~\cite{garnier2021new}. The comparison between the quenched and the annealed results for $\gamma=0$ also shows that the annealed approximation overestimates the diversity of the most numerous equilibria and thus their linear instability, which is directly related to the diversity. On the other hand, it gives a smaller value of the average abundance and of the self-overlap of the equilibria at given diversity.

For uncorrelated couplings ($\gamma=0$), we have compared our results with those obtained previously by means of the cavity method. We have shown that within the multiple equilibria phase, the cavity calculation captures a symmetry point for the saddle-point equations as a function of diversity, where quenched and annealed complexities become equal. The equilibria at the corresponding diversity are however sub-dominant for all values of $\sigma$, since they are exponentially less numerous than the typical ones (those at the diversity that maximizes the quenched complexity). Through the complexity calculation, we also got a more resolved description of the transition to the unbounded phase.
In the case of symmetric interactions ($\gamma=1$), we have compared the  annealed complexity obtained with the Kac-Rice method with the calculation of the complexity obtained with the so called Monasson method. This method allows one to obtain the number of stable minima of the potential landscape as a function of the value of the potential itself. We have shown that the equilibria identified with the two different approaches are not the same.

We have analyzed how the total complexity of equilibria vanishes at the value of variability $\sigma_c$ corresponding to the topology trivialization transition, i.e., to the transition to the unique equilibrium phase. We have shown explicitly that for $\gamma=0$ both the quenched and annealed complexity vanish as $\Sigma_\sigma \sim (\sigma- \sigma_c)^2$, an exponent previously found in other models studied within the annealed approximation \cite{LCTF22, arous2021landscape, Toboul}. Within the annealed approximation, this remains true for $\gamma \neq 0$; however, we provide evidence of the fact that for $\gamma \neq 0$, the total complexity, if quenched, should vanish with a different power. Solving the quenched self-consistent equations for arbitrary $\gamma$ will allow us to address these points in a thorough way: the corresponding analysis is ongoing.

There are several extensions of this work that we are leaving to future work as well. For instance, the generalization to randomly distributed carrying capacities $\kappa \to \kappa_i$ is straightforward: in the context of our calculation, it would just require to introduce additional order parameters defined by $k^a = \lim_{S \to 1\infty} S^{-1} (\vec{N}^a \cdot \vec{\kappa})$
where $\vec{\kappa}=(\kappa_1, \cdots, \kappa_S)$. In the case of homogeneous $\kappa_i$, this order parameter reduces to the average abundance $m^a$. It would also be natural to generalize this calculation to different types of interaction matrices, for instance imposing a fixed sign to the couplings: for purely competitive interactions, the number of stable equilibria has been explored in \cite{lischke2017finding} through a sampling algorithm. Considering a block structure of the matrix \cite{ipsen2020consequences} or some sparsity in its entries \cite{marcus2021local,mambuca2022dynamical} are also interesting directions to explore. 
We also remark that the stability of the symmetric assumption on the order parameters that we have made to perform this calculation should also be checked. This amounts to check that the variational manifold chosen to determine the solutions of the saddle point equations is stable; this analysis is particularly interesting in the case of asymmetric couplings, where no thermodynamic analogy can be exploited.

Let us conclude with a few comments on the implications of our results for the dynamics of the system. In the case of uncorrelated interactions, it follows from our solution that all the equilibria in the multiple equilibria phase are linearly unstable.  One may thus expect a complicated dynamical evolution, with the system that continuously approaches an equilibrium and then is driven away along the directions of instability. For larger $\gamma$ and in a certain range of $\sigma$, the annealed calculation predicts a very small fraction of stable equilibria having positive complexity: determining whether their complexity is non-zero also in the quenched formalism and, in that case, assessing their role in the dynamics is another interesting question.  Recent results also suggest that a relevant role in the dynamics is played by invadable equilibria: the calculation of the corresponding complexity is ongoing.

\section{Acknowledgements}
The authors would like to thank Ari M. Turner for discussions on this topic and for his contribution at the early stage of this work. 
 V. Ros thanks P.~Urbani, J. Garnier-Brun and J.-P. Bouchaud for discussions. V. Ros  acknowledges funding by the ``Investissements d’Avenir” LabEx PALM (ANR-10-LABX-0039-PALM).  G. Biroli is partially supported from the Simons Foundation (Grant No. 454935). G. Bunin acknowledges support from the Israel Science Foundation (ISF) Grant No. 773/18. \\

\appendix

\section{The Kac-Rice calculation of the moments: details}\label{sec:CalculationMoments}
In this appendix, we derive the explicit expressions of the various terms appearing in Eq.~\eqref{eq:KRmoments}, under the replica symmetric assumptions  \eqref{eq:RSa}. \\

\subsection{Joint distribution of the forces} 
We begin by computing the joint distribution $\mathcal{P}^{(n)}_{{\bf N}} \tonde{ {\bf f}}$ of the $S$-dimensional vectors $\vec{F}^{a}$ evaluated at $\vec{f}^{a}$, and by showing explicitly that it depends on the order parameters in \eqref{eq:OrderParameters}. 
From \eqref{eq:GradDef} we see that the component $F_i^{a}$ are linear combinations of the Gaussian variables $a_{ij}$, with average:
\begin{equation}
\langle F_i^a \rangle= \kappa- \frac{\mu}{S}\sum_{j=1}^S N_j^{a}-N_i^{a}= (\kappa-\mu m^a)- N_i^{a}.
\end{equation}
and covariance matrix:
\begin{equation}\label{eq:CovMatrix}
\frac{\sigma^2}{S} \hat C^{ab}_{ij} =\langle F_i^{a} F_j^{b}  \rangle-\langle F_i^{a} \rangle \langle F_j^{b}  \rangle= \frac{\sigma^2}{S}\tonde{{\vec N}^a \cdot {\vec N}^b \delta_{ij}+ \gamma \,  N^b_i N^a_j}.
\end{equation}
Therefore, it holds
\begin{equation}\label{eq:GaussianGradient}
\mathcal{P}^{(n)}_{{\bf N}} \tonde{ {\bf f}}= \frac{e^{-\frac{S}{2 \sigma^2}  ({\bf f}- \langle {\bf  F} \rangle)^T \, \hat {C}^{-1} \, ({\bf  f}- \langle {\bf  F} \rangle)}}{\sqrt{\tonde{\frac{2 \pi \sigma^2}{S}}^{Sn} \text{det} \,{\hat C}}}.
\end{equation}
In addition to ${\bf N}=(\vec N^1, \cdots, \vec N^n)$ and ${\bf f}=(\vec f^1, \cdots, \vec f^n)$, it is convenient to introduce:
{\medmuskip=.5mu
\thinmuskip=.5mu
\thickmuskip=.5mu
\begin{equation}
\begin{split}
{\bf w}_N =\tonde{\sum_{a \neq 1}\vec{ N}^a, \cdots, \sum_{a \neq n}\vec{ N}^a}, \quad {\bf w}_f =\tonde{\sum_{a \neq 1}\vec{f}^a, \cdots, \sum_{a \neq n}\vec{ f}^a},\quad
  {\bf v}= \tonde{\vec{ 1}, \cdots, \vec{ 1}}, \quad
   {\bf m}=\tonde{ \vec{m}^1 ,  \cdots, \vec{m}^n },
      \end{split}
\end{equation}}
 with the $S$-dimensional vectors  $\vec{1}=(1,1,\cdots, 1)$ and $\vec{m}^a= m^a \vec{1}$. These vectors are relevant as they form a closed set under the action of the covariance matrix \eqref{eq:CovMatrix}. In particular, under the assumptions \eqref{eq:RSa} and using that the equilibrium condition imposes $z^{aa}=0$, we find:
\begin{equation}\label{eq:ActionVecs}
 \begin{split}
\hat{C} {\bf N} &=(1+\gamma) [q_0 {\bf w}_N +  q_1 {\bf N}]\\ 
\hat{C} {\bf f}&= q_0 {\bf w}_f+ q_1 {\bf f}+ \gamma \, z {\bf w}_N\\ 
\hat{C} {\bf w}_N&= (1+\gamma)[q_1+(n-2)q_0] {\bf w}_N+ (1+\gamma) (n-1)q_0 {\bf N}\\ 
\hat{C} {\bf w}_f&= [q_1+ (n-2)q_0]{\bf w}_f+ \gamma \, (n-2)z {\bf w}_N+ (n-1)q_0 {\bf f} + \gamma \,(n-1)z {\bf N}\\ 
\hat{C} {\bf v}&= [q_1+ (n-1)q_0]{\bf v}+  \gamma \, m {\bf w}_N + \gamma \, m{\bf N}.
\end{split}
\end{equation}
Notice that the matrix elements of $\hat C$ on these vectors are only a function of the order parameters \eqref{eq:OrderParameters}.
Moreover, the quadratic form at the exponent in \eqref{eq:GaussianGradient} can be rewritten as a linear combination of matrix elements of $\hat C^{-1}$ on this restricted set of vectors: therefore, the exponent in \eqref{eq:GaussianGradient} is also a function of the order parameters only, which is determined by inverting the action of the matrix $\hat C$ on the subset spanned by these vectors. Introducing an orthonormal basis for this subspace and performing the inversion of the matrix $\hat C$ projected into the subspace (see Refs. \cite{MioSegnale, MioSelle} for similar calculations), we obtain:
\begin{equation}
 ({\bf f}- \langle {\bf  F} \rangle)^T \, \hat {C}^{-1} \, ({\bf  f}- \langle {\bf  F} \rangle)=\frac{n}{1+\gamma} \frac{ U_n({\bf x})}{ (q_1-q_0)^2 \, [q_1+(n-1) q_0]^2},
\end{equation}
with $U_n$ given in \eqref{eq:Fulln}. 
The determinant in the denominator of \eqref{eq:GaussianGradient} is dominated by the diagonal part of the covariance matrix, and under the assumptions \eqref{eq:RSa}:
\begin{equation}
 \text{det} \hat{C}= e^{S \log \quadre{S^n (q_1-q_0)^{n-1}(q_1+ (n-1)q_0)} + o(S)}.
\end{equation}
Combining these terms, we recover \eqref{eq:GradAsy}. \\

\subsection{Joint expectation of the linear stability matrices }\label{app:Determinants}
We now focus on the joint, conditional expectation value \eqref{eq:ExpectationJointDet}. The product is over $n$ determinants  \eqref{eq:Hessian} of size $N \phi \times N \phi$ taking the form:
\begin{equation}\label{eq:HessianDecomposed}
H_{ij}^a=\tonde{\frac{\partial F_i(\vec{N}^a)}{\partial N^a_j}}=
-\tonde{\frac{\sigma \sqrt{\phi}}{\sqrt{S \, \phi}} a _{ij}+\frac{\mu }{S} + \delta_{ij} } \quad \quad i,j \in I_a.
\end{equation}
Therefore, the matrices $H_{ij}^a$ of different replicas $a$ have the same statistic: they only differ by the components that are selected by the index sets $I_a$. We first recall how to deal with the expectation value in case of a single replica $n=1$: this calculation is a slight variation of that presented in Refs. \cite{FyodorovNonlinearAnalogue, arous2021counting}. We then generalize to arbitrary values of $n$, as it is necessary for the quenched calculation.

When $n=1$, a single configuration vector $\vec{N}$ is present, with $S \phi$ components $N_{i \in I}$ that are different from zero. 
As recalled in \Sref{sec:Ginibre}, the corresponding matrix  \eqref{eq:HessianDecomposed} prior to conditioning is a random matrix of the real elliptic type, with variance $ \sigma^2 \phi$ and with asymmmetry parameter $\gamma$. The third term in \eqref{eq:HessianDecomposed} only provides a global shift. The asymptotic eigenvalue density of \eref{eq:HessianDecomposed} reads:
\begin{equation}\label{eq:AsyDensS}
\overline \rho(\lambda) =\frac{1}{\pi \sigma^2 \phi (1-\gamma^2)}\,  \mathbbm{1}_{{\lambda \in \tilde S_{\sigma, \phi, \gamma} }}, \quad  \quad \tilde{S}_{\sigma,\phi, \gamma}= \left\{ \frac{(\Re \lambda+1)^2}{\sigma^2 \phi (1+ \gamma)^2} +\frac{(\Im \lambda)^2}{\sigma^2 \phi (1- \gamma)^2}   \leq 1 \right\}.
 \end{equation}
A rather straightforward exercise in Gaussian conditioning shows that conditioning to the event $\vec{F}(\vec{N})= \vec{f}$ modifies very weakly the statistics of the matrix $\hat a$. Indeed, the event $\vec{F}(\vec{N})= \vec{f}$ is equivalent to 
\begin{equation}\label{eq:ToCondition}
\begin{cases}
\kappa - N_i - \frac{\mu}{S}\sum_{j \in I}N_j- \frac{\sigma}{\sqrt{S}}\sum_{j \in I} a_{ij} N_j=0 &\text{ if } i \in I\\
\kappa-  f_i-  \frac{\mu}{S}\sum_{j \in I}N_j- \frac{\sigma}{\sqrt{S}}\sum_{j \in I} a_{ij} N_j=0 &\text{ if } i \notin I, 
\end{cases}
\end{equation}
where we used that $N_{i \notin I}=0$ and $f_{i \in I}=0$. From \eqref{eq:ToCondition} it follows that conditioning to $\vec F(\vec N)= \vec f$ amounts to fixing the action of the $S \times S$ random matrix $\hat a$ on the $S$-dimensional vector $\vec{N}$. If one rotates the matrix $\hat a$ in such a way that the components are expressed in a new orthonormal basis $\vec{e}_i$ such that $\vec{e}_1=\vec{N}/\sqrt{\vec{N} \cdot \vec{N}}$ and $\vec{e}_{i \neq 1}$ are a completion of the space, then the event \eqref{eq:ToCondition} corresponds to fixing to a deterministic vector the first column of the rotated matrix $\hat a$, with components $a_{k1}$ for $k \geq 1$.  For $\gamma \neq 0$, because of the non-zero correlations, also the statistics of the matrix element $a_{1k}$ with $k >1$ will be affected; in particular, the average of these components is modified and their variance is reduced by a factor $1-\gamma^2$ (in the symmetric case $\gamma=1$, the variance of these entries vanishes as well, consistently with the fact that $a_{1k}= a_{k1}$).  We recall that the determinant \eqref{eq:ExpectationJointDet} is that of the projection of the matrix on the $S \phi \times S \phi$ dimensional subspace spanned by the species that are present. The vector $\vec{e}_1$ belongs to this subspace. Therefore, the relevant block of the matrix $\hat a$ contributing to  \eqref{eq:HessianDecomposed}  will still have a special line and column (those corresponding to the basis vector $\vec{e}_1$), with entries whose statistics is perturbed with respect to that of the original, unconditioned elliptic matrix. This perturbation is however of finite rank, and thus it does not affect the asymptotic density $\overline \rho(\lambda)$ to leading order in $S$. \\
We now argue that the density \eqref{eq:AsyDensS} is the only quantity needed to compute the conditional expectation value of the determinant to leading order in $S$. To this aim, one needs to recall that the convergence of the empirical measures $\mu_M$ of real elliptic matrices happens on a scale that is quadratic in the size $M$ of the matrix. More precisely, 
the empirical spectral measures $\mu_M$ satisfy a large deviation principle with rate $M^2$, meaning that for large $M$ the probability $P_M[\mu]$ of observing a spectral measure $\mu$ scales as $P_M[\mu] \sim e^{- M^2 \, I[\mu] +o(M^2)}$, where the rate function $I[\mu]$ is minimized precisely by the asymptotic measure $\overline{\mu}$ with density $\overline \rho (\lambda)$ in \eqref{eq:AsyDensS}. For the real elliptic ensemble, the rate functional $I[\mu]$ has been derived for generic $\gamma$ in \cite{arous2021counting} (see Sec. 3 of the Supplementary Information), generalizing the result for the special case  $\gamma=0$ given in \cite{arous1998large}. The $M^2$-scaling of the large deviation principle is quite generic in random matrix theory \cite{arous1997large, petz1998logarithmic}, and it is essentially determined by the scaling of the number of independent entries in the matrix: small-rank perturbations of the statistics of the matrix such as those described above are not sufficient to modify the speed of convergence of the large deviation principle nor, as pointed out above, the minimizer of the rate functional $ I[\mu]$. Keeping this in mind, it is then straightforward to write the conditional expectation value of the determinant of $H +\mathbb{I} $ as an expectation over the spectral measure $\mu(\lambda)$ of the matrix, as:
\begin{equation}
\left \langle  \left| \text{det} \tonde{H+ \mathbb{I}}\right|  \; \; \Big| \; {\vec  F}({\vec N})= {\vec f} \right \rangle = \int   \mathcal{D} \mu \; \tilde{P}_{S \phi} [\mu] \; e^{ S \phi \int  d \mu (\lambda) \, \log |\lambda| + o(S)},
\end{equation} 
where $\tilde{P}_{S \phi} [\mu]$ is the probability of observing an empirical measure $\mu$ for matrices with the same statistics as the conditional matrix $H+ \mathbb{I}$. Using that $\tilde{P}_{S \phi} [\mu] \sim  e^{- S^2 \phi^2 I[\mu]+ o(S^2)}$ and that $\vec \mu (\lambda)$ minimizes the rate functional $I[\mu]$, via a saddle point calculation in the space of measures we obtain:
\begin{equation}
\begin{split}
\left \langle  \left| \text{det} \tonde{H+ \mathbb{I}}\right|  \; \; \Big| \; {\vec  F}({\vec N})= {\vec f} \right \rangle &= \int   \mathcal{D} \mu \;  e^{- S^2 \phi^2 I[\mu]+ o(S^2)+ S \phi \int  d \mu (\lambda) \, \log |\lambda| + o(S)}\\
&= e^{ S \phi \int  d\lambda \, \vec \rho(\lambda) \log |\lambda| + o(S)},
\end{split}
\end{equation} 
where the last equality follows from the fact that $\tilde{P}_{S \phi} [\mu]$ is normalized to one, implying that the linear-order term in $S$ at the exponent of $\tilde{P}_{S \phi} [\mu]$ must vanish as well when computed at the saddle point $\vec \mu$. These identities imply that:
\begin{equation}
  \mathcal{D}^{(n=1)}_{{\vec N}} =\left \langle  \left| \text{det} \tonde{H}\right|  \; \; \Big| \; {\vec  F}({\vec N})= {\vec f} \right \rangle =e^{S \phi \int_{-1}^{1} \frac{dx}{\pi} \, 2 \int_0^{\sqrt{1- x^2}}  dy\,  \log \sqrt{\quadre{\sigma \sqrt{\phi} (1+ \gamma) x+1}^2+ \sigma^2 \phi (1-\gamma) y^2}}.
  \end{equation}

 We now discuss how to generalise this result to the case $n >1$: the arguments in this case follow closely those presented in Refs. \cite{MioSegnale,MioSelle}, which we summarize here very briefly. For $n>1$, one has to compute the joint expectation of the product of $n$ matrices that are correlated with each others, due to the fact that the $S \phi \times S \phi$ matrices associated to different replicas share a finite fraction of lines and columns.  Following the same line of reasoning as above, we can set:
\begin{equation}\label{eq:Colonne}
\left \langle  \prod_{a=1}^n\left| \text{det} \tonde{H^a+ \mathbb{I}}\right|  \; \; \Big| \; {\bf  F}({\bf N})= {\bf f} \right \rangle = \int   \prod_{a=1}^n \mathcal{D} \mu^a \; \tilde{P}_{S \phi} [\left \{\mu^a \right \}] \; e^{ S \phi  \sum_{a=1}^n \int  d \mu^a (\lambda) \, \log |\lambda| + o(S)},
\end{equation} 
where now $ \tilde{P}_{S \phi} [\left \{\mu^a \right \}]$ is the joint distribution of the spectral measures of the $n$ matrices, which must exhibit the same large-$S$ scaling as its reduced distribution, $ \tilde{P}_{S \phi} [\left \{\mu^a \right \}] \sim O(e^{S^2})$; therefore, the expression \eqref{eq:Colonne} can again be computed in a saddle point approximation. The saddle-point solutions are determined just by the minimization of the term scaling quadratically with $S$: as such, they must coincide with the marginals of the joint distribution in the space of measures (see the argument around Eq. (43) in \cite{MioSegnale}). Thus, correlations between the matrices are not relevant for computing \eqref{eq:Colonne} to leading exponential order in $S$: what remains to determine is just the asymptotic eigenvalue density of each conditional matrix $H^a$. We remark that each  $H^a$ has to be conditioned to $\vec F(\vec N^b)= \vec f^b$ for all $b=1, \cdots, n$. Following the same arguments as above, one sees that this conditioning amounts again to a finite-rank  perturbation to the original elliptic ensemble statistics: for $n>1$, the conditional matrices will contain $n$ lines and $n$ columns having a modified statistics with respect to the original one. These are the lines and columns corresponding to the subspace spanned by the unit vectors $\vec{e}_b= \vec N^b/\sqrt{\vec{N}^b \cdot \vec{N}^b}$. 
The presence of these special lines does not affect the bulk of the density of states (provided that the number of special lines and columns is not of $O(S)$). We can therefore conclude that
\begin{equation}\label{eq:ColonneFinal}
  \mathcal{D}^{(n)}_{{\vec N}} = e^{S n  \phi \int_{-1}^{1} \frac{dx}{\pi}  \int_0^{\sqrt{1- x^2}}  dy\,  \log \grafe{\quadre{\sigma \sqrt{\phi} (1+ \gamma) x+1}^2+ \sigma^2 \phi (1-\gamma)^2 y^2}}= e^{S n \mathcal{d}(\phi)}.
\end{equation}

\subsection{Explicit expression for the double integral }\label{app:DeterminantsExplicit}
 The double integral in \eqref{eq:ColonneFinal} can be evaluated explicitly. 
For any $\gamma \neq  \pm 1$, performing the inner integration we get:
\begin{equation}\label{eq:2}
\begin{split}
    \mathcal{d}(\phi)&= \frac{\phi}{\pi} \int_{-1}^1 dx \grafe{\frac{2}{\sigma \sqrt{\phi}(1-\gamma)}[1+ \sigma \sqrt{\phi}(1+\gamma) x] \arctan \quadre{\frac{\sigma \sqrt{\phi}(1-\gamma) \sqrt{1-x^2}}{{1+ \sigma \sqrt{\phi}(1+\gamma) x}}}}\\
        &+ \frac{\phi}{\pi} \int_{-1}^1 dx \grafe{ \sqrt{1-x^2} \log \quadre{\tonde{1+ \sigma \sqrt{\phi}(1+\gamma) x}^2+ \sigma^2 \phi(1-\gamma)^2(1-x^2)}}    \\
    &-2 \frac{\phi}{\pi} \int_{-1}^1 dx  \sqrt{1-x^2}.
   \end{split}
\end{equation}
  We discuss some special values of $\gamma$ first, and then the result for general $\gamma$. We set $A= \sigma \sqrt{\phi} (1+\gamma)$ and $B= \sigma \sqrt{\phi} (1-\gamma)$ to simplify the notation. \\

 \underline{Case $\gamma=0$. }
The argument of the logarithm in \eqref{eq:2} is a quadratic function of $x$ except for $\gamma=0$.
In this case we have $A= \sigma \sqrt{\phi}=B$ and
\begin{equation}\label{eq:Gamma0}
       \mathcal{d}(\phi)= \frac{\phi}{\pi} \int_{-1}^1 dx \grafe{\frac{2 (1+ A x)}{A} \arctan \tonde{\frac{A \sqrt{1-x^2}}{{1+ A x}}}-2 \sqrt{1-x^2}+ \sqrt{1-x^2} \log \quadre{1+ A^2 + 2 A x}}.
\end{equation}
Both the first and the last integrals have a different behavior depending on whether $A>1$ or $A<1$. The first integral gives:
\begin{equation}
I_1 \equiv   \frac{\phi}{\pi} \int_{-1}^1 dx \frac{2 (1+ A x)}{A} \arctan \tonde{\frac{A \sqrt{1-x^2}}{{1+  A x}}}=  \frac{\phi}{\pi} 
   \begin{cases}
   \pi \tonde{1- \frac{A^2}{4}} &\text{  if  }\quad 0<A <1\\
  \frac{\pi}{2} \tonde{1+ \frac{1}{2 A^2}} &\text{  if  }\quad A >1
   \end{cases}
    \end{equation}
For $x \in [-1,1]$ the argument of the logarithm in \eqref{eq:Gamma0} is always non-negative,
since the root $x^*=-(1+ A^2)/2A$ is always smaller than $-1$. The singularity is hit for $A=1$, when $x^*=-1$.  The corresponding integral has two different behaviors for $A<1$ and $A>1$, because it involves functions having a branch-cut at $A=1$. One finds explicitly:
\begin{equation}
I_3 \equiv   \frac{\phi}{\pi} \int_{-1}^1 dx \sqrt{1-x^2} \log \quadre{1+ A^2 + 2 A x}=  \frac{\phi}{\pi} 
   \begin{cases}
   \frac{\pi A^2}{4} &\text{  if  }\quad 0<A <1\\
  \frac{\pi}{4} \tonde{\frac{1}{A^2} + 2 \log A^2} &\text{  if  }\quad A >1
   \end{cases}
    \end{equation}
It is convenient to obtain this result integrating by parts, using that for $z<1$:
\begin{equation}
F(z)= \int_{-1}^z dx    \sqrt{1-x^2}= \frac{\pi}{4}+ \frac{1}{2} \tonde{z \sqrt{1-z^2} + \arcsin z}.
\end{equation}
Finally
\begin{equation}
I_2 \equiv  -2 \frac{\phi}{\pi} \int_{-1}^1 dx \sqrt{1-x^2} = -\phi.
    \end{equation}
 Combining these formulas and using that $A= \sigma \sqrt{\phi}$ we find for $\gamma=0$ :
    \begin{equation}
       \mathcal{d}(\phi)= \begin{cases}
  0 &\text{  if  }\quad 0<\sigma \sqrt{\phi} <1\\
 \frac{1}{2 \sigma^2}-\frac{\phi}{2}+\frac{\phi}{2}\log(\sigma^2 \phi)&\text{  if  }\quad \sigma \sqrt{\phi} >1.
   \end{cases}
\end{equation}

 \underline{Case $\gamma=1$. } 
In this case $B=0$ and the integrand in \eqref{eq:2} is singular. Plugging $\gamma=1$ directly in \eqref{eq:ColonneFinal} we find:
\begin{equation}
       \mathcal{d}(\phi)= \frac{2 \phi}{\pi} \int_{-1}^1 dx  \sqrt{1-x^2} { \log |1+ A x|}=  \frac{2 \phi}{\pi} \int_{-1}^1 dx  \sqrt{1-x^2}  \log |1+ A x|.
\end{equation}
Again, the integral of the logarithm has a different behaviour depending on whether $A>1$ or $A<1$. In particular, 
\begin{equation}
       \mathcal{d}(\phi)=\phi  \begin{cases}
    \frac{1}{A^2}-\log 2 - \frac{1}{2}-\frac{\sqrt{1-A^2}}{A^2} + \log \tonde{1+ \sqrt{1-A^2}} &\text{  if  }\quad 0<A <1\\
 \log A + \frac{1}{A^2}-\log 2 - \frac{1}{2} &\text{  if  }\quad A >1
   \end{cases}
\end{equation}
Using that $A= 2 \sigma \sqrt{\phi}$ we get
\begin{equation}
       \mathcal{d}(\phi)=  \phi  \begin{cases}
    \frac{1}{4 \sigma^2 \phi}-\log 2 - \frac{1}{2}-\frac{\sqrt{1-4 \sigma^2 \phi}}{4 \sigma^2 \phi} + \log \tonde{1+ \sqrt{1-4 \sigma^2 \phi}} &\text{  if  }\quad 0<\phi <\frac{1}{4 \sigma^2}\\
        \log (\sigma \sqrt{\phi})+ \frac{1}{4 \sigma^2 \phi} - \frac{1}{2} &\text{  if  }\quad \phi >\frac{1}{4 \sigma^2}.
   \end{cases}
\end{equation}

 \underline{Case $\gamma=-1$. } In this case $A=0$ and it is convenient to re-write \eqref{eq:ColonneFinal} as:
\begin{equation}
\begin{split}
       \mathcal{d}(\phi)&=\frac{\phi}{\pi}\int_{-1}^1 dy \sqrt{1-y^2} \log [1+ B^2 y^2]\\
       &= \phi  \log [1+\sqrt{B^2+1}]- \frac{\phi}{2 B^2} \tonde{2-2 \sqrt{B^2+1}+B^2 (1+\log
   (4))}
   \end{split}
\end{equation}
This integral has the same expression for all values of $B=-2 \sigma \sqrt{\phi}$.\\

 \underline{Case of general $\gamma$. }  In this case, it is convenient to compute the integral by expanding  the integrand in \eqref{eq:2} in powers of $a=\sigma \sqrt{\phi}$, integrate term by term the expansion  and then re-sum it. The final result is 
    \begin{equation}
   \mathcal{d}=   \begin{cases}
\frac{1}{4 \gamma \sigma^2} \tonde{1- \sqrt{1- 4 \gamma \sigma^2 \phi}}+ \phi \log \tonde{1+ \sqrt{1- 4 \gamma \sigma^2 \phi}}- \phi \tonde{\frac{1}{2}+\log 2} & \sigma \sqrt{\phi} (1+\gamma)<1\\
\frac{1}{2 \sigma^2} \frac{1}{1+\gamma} -\frac{\phi}{2}+\frac{\phi}{2} \log (\sigma^2 \phi) & \sigma \sqrt{\phi} (1+\gamma) >1
     \end{cases}
 \end{equation}
which coincided with \eqref{eq:DetSmall} in the main text. We see that this is consistent with the special cases discussed above; in particular, for $\gamma=-1$ only the first regime occurs for $\phi \in [0,1]$.

\subsection{The phase space volume term. } 
Let us now come to the computation of the phase space volume term \eqref{eq:startAppVol}. We begin by noticing that the introduction of the conjugate parameters allows us to decouple the various species and to set:  
\begin{equation}
\mathcal{V}_n({\bf x},\hat {\bf x}) = \tonde{\sum_{{\tau^a =0,1}}  e^{- \hat \phi_a \,  [\tau^a]^2}\int  \prod_{a=1}^n \, d N^a\, d f^a \,  \mathcal{j}(N^a, f^a)}^S,
\end{equation}
where
\begin{equation}
\begin{split}
\mathcal{j}(N^a, f^a)= &e^{-\sum_{a=1}^n ( \hat m_a \, N^a +\hat p_a \, f^a ) - \sum_{a,b=1}^n \tonde{\hat z_{ab} \,  N^a   f^b +\hat q_{ab} \,{ N}^a  { N}^b+\hat \xi_{ab} \,{ f}^a  f^b}  }  \\
&\times  \prod_{a:  \tau^a=1} \theta (N^a)  \delta( f^a) \prod_{a: \tau^a=0} \delta( N^a) \theta( -f^a).
\end{split}
\end{equation}
For $\hat \phi_a \equiv \hat \phi$,
this expression depends on $\tau^a$ only through the number $k \in \grafe{0, \cdots, n}$ of entries that are non-zero. Once $k$ is fixed, we can introduce ${\bf y}=(N^{1}, \cdots, N^{k}, f^{1}, \cdots, f^{n-k})$ and the matrix and vectors :
 \begin{equation}
 \mathbb{A}_k[\hat{\bf x}]= \begin{pmatrix}
 \stackrel{k \times k}{ \hat{\mathbb{Q}}} &   -\hat{\mathbb{Z}}\\
 - \hat{\mathbb{Z}} & \stackrel{(n-k) \times (n-k)}{\hat{\mathbb{X}}}
              \end{pmatrix}, \quad \quad {\bm \mu}_k[\hat{\bf x}]=\begin{pmatrix} \stackrel{k \times 1}{\hat{\bf m}} \\
             \stackrel{(n-k) \times 1}{- \hat{\bf p} }
              \end{pmatrix},
 \end{equation}
 with
 \begin{equation}
 \hat{\mathbb{Q}}_{ab} =\delta_{ab}\, 2 \hat{q}_{aa} + (1-\delta_{ab})\hat{q}_{ab}, \quad   \hat{\mathbb{Z}}_{ab}=\hat{z}_{ab}, \quad  \hat{\mathbb{X}}_{ab} =\delta_{ab}\, 2 \hat{\xi}_{aa} + (1-\delta_{ab})\hat{\xi}_{ab}.
  \end{equation}
Then:
\begin{equation}\label{eq:PFk}
 \begin{split}
\mathcal{V}_n({\bf x},\hat {\bf x}) = \tonde{ \sum_{k=0}^n \binom{n}{k} e^{- k \hat{\phi} } \int d{\bf y} \prod_{a=1}^n \theta(y_a) \text{exp} \grafe{-\frac{1}{2} \; {\bf y}^T \cdot  \mathbb{A}_k[\hat{\bf x}] \cdot {\bf y} -  {\bm \mu}_k[\hat{\bf x}] \cdot {\bf y} }}^S.\\
 \end{split}
\end{equation}

 To illustrate how to simplify this term, let us consider first the case $k=n$. In the RS assumptions, $\hat{\mathbb{Q}}_{ab}=\hat{q}_0+\delta_{ab}(2 \hat{q}_1-\hat{q}_0)$ and $\hat{m}_a=\hat{m}$, implying:
{\medmuskip=1mu
\thinmuskip=1mu
\thickmuskip=1mu
\begin{equation}
\int_0^\infty \prod_{a=1}^n dN^a e^{-\frac{1}{2}\sum_{a,b} N^a \hat{\mathbb{Q}}_{ab} N^b-  \sum_a \hat{m}_a N^a }=
\int_0^\infty \prod_{a=1}^n dN^a \, e^{-\frac{\hat{q}_0}{2} \tonde{\sum_{a=1}^n N^a}^2} \;   e^{- \sum_{a=1}^n\frac{(2 \hat{q}_1-\hat{q}_0)}{2} [N^a]^2-\sum_{a=1}^n \hat{m} N^a }
\end{equation}}
Assuming that $\hat q_0<0$ and $(2 \hat{q}_1-\hat{q}_0)>0$, and using the Gaussian identity:
\begin{equation}
e^{-\frac{\hat{q}_0}{2} \tonde{\sum_{a=1}^n N^a}^2}=\frac{1}{\sqrt{2 \pi [- \hat q_0]}} \int dz e^{-\frac{z^2}{2 [- \hat q_0]} + z \, \sum_{a=1}^n N^a},
\end{equation}
we see that
\begin{equation}
\begin{split}
\int_0^\infty \prod_{a=1}^n dN^a e^{-\frac{1}{2}\sum_{a,b} N^a \hat{\mathbb{Q}}_{ab} N^b-  \sum_a \hat{m}_a N^a } = \int_{-\infty}^{\infty} \frac{dz}{\sqrt{2 \pi [- \hat{q}_0]}} e^{\frac{-(z-\hat{m})^2}{2 [- \hat{q}_0]}} [g(z; 2 \hat q_1-\hat q_0)]^n,
 \end{split}
\end{equation}
where we introduced the function
\begin{equation}
\begin{split}
g(u; \hat a )=e^{\frac{u^2}{2  \hat a}} \sqrt{\frac{\pi}{2 \hat a}} \text{Erfc}\tonde{\frac{u}{\sqrt{2 \hat a}}}.
\end{split}
\end{equation}

 This expression can be easily expanded in powers of $n$. For $k$ generic and within the RS ansatz, we can proceed analogously. The relevant integral now reads:
\begin{equation}\label{eq:Intermediate}
\begin{split}
 \int_0^\infty \prod_{a=1}^k dN^a \prod_{b=k+1}^n dg^b \, e^{-\frac{1}{2} \hat{O}_k(N^a, g^b)} \; \prod_{a=1}^k  e^{- \frac{(2 \hat{q}_1- \hat{q}_0)}{2} [N^a]^2 -\hat{m}^a N^a}  \prod_{b=k+1}^n e^{-\frac{(2 \hat{\xi}_1-\hat{\xi}_0)}{2} [g^b]^2 + \hat{p}^b g^b} 
  \end{split}
\end{equation}
with the shorthand notation:
\begin{equation}
   \begin{split}
\hat{O}_k(N^a, g^b)= \hat{q}_0 \tonde{\sum_{a=1}^k  N^a}^2 + \hat{\xi}_0 \tonde{\sum_{a=k+1}^n g^a}^2-2\hat{z} \tonde{\sum_{a=1}^k N^a} \tonde{\sum_{b=k+1}^n g^b}.
 \end{split}
 \end{equation}
Assuming $\hat q_0, \hat \xi_0<0$, we can write:
\begin{equation}
e^{-\frac{1}{2} \hat{O}_k(N^a, g^b)}= \int \frac{du_1 du_2}{2 \pi \; \sqrt{\text{det} (\hat{A})}}  e^{-\frac{1}{2} \tonde{u_1,u_2 } \hat{A}^{-1}\tonde{u_1,u_2 }^T} \prod_{a=1}^ke^{u_1 N^a} \, \prod_{b=k+1}^n e^{-u_2 g^b}, 
\end{equation}
where we have introduced the $2 \times 2$ matrix  
\begin{equation}
\hat{A}= \begin{pmatrix}
-\hat{q}_0& -\hat{z}\\
-\hat{z} & -\hat{\xi}_0
\end{pmatrix}, \quad \quad \quad \hat{A}^{-1}= \frac{1}{\hat{q}_0 \hat{\xi}_0- \hat{z}^2}\begin{pmatrix}
-\hat{\xi}_0 & \hat{z}\\
\hat{z} & -\hat{q}_0
\end{pmatrix}
\end{equation}
and assumed that it is positive-definite. Performing the Gaussian integrations under the assumptions $ 2 \hat{q}_1- \hat{q}_0, 2 \hat{\xi}_1-\hat{\xi}_0>0$ , we find that \eqref{eq:Intermediate} is equivalent to:
\begin{equation}
\begin{split}
\int \frac{du_1 du_2}{2 \pi \; \sqrt{\text{det} (\hat{A})}}  e^{-\frac{1}{2} \tonde{u_1,u_2 } \hat{A}^{-1}\tonde{u_1,u_2 }^T}  \;   [g(\hat{m}-u_1; 2\hat{q}_1-\hat{q}_0)]^k [g(-\hat{p}+u_2; 2\hat{\xi}_1-\hat{\xi}_0)]^{n-k}.
  \end{split}
\end{equation}
Finally, performing the sum over $k$ we see that we can write the quantity inside the square brackets in \eqref{eq:PFk} as:
{\medmuskip=1mu
\thinmuskip=1mu
\thickmuskip=1mu
\begin{equation}\label{eq:PFkR00u}
 \begin{split}
\mathcal{V}_n=\quadre{ \int \frac{du_1 du_2}{2 \pi \; \sqrt{\text{det} (\hat{A})}}  e^{-\frac{1}{2} \tonde{u_1,u_2 } \hat{A}^{-1}\tonde{u_1,u_2 }^T}\tonde{e^{- \hat{\phi}}g(\hat{m}-u_1; 2\hat{q}_1- \hat{q}_0)+ g(-\hat{p}+u_2; 2\hat{\xi}_1- \hat{\xi}_0)}^n}^S.
  \end{split}
\end{equation}}
Expanding to linear order in $n$ we get 
\begin{equation}
 \begin{split}
\mathcal{V}_n({\bf x},\hat {\bf x})=e^{S n \bar{ \mathcal{J}} (\hat{\bf x}) + O(S n^2)},  \end{split}
\end{equation}
with $ \bar{ \mathcal{J}} (\hat{\bf x})$ given in \eqref{eq:VolumeQuenched}. For $n=1$, the above expression reduces to:
\begin{equation}
 \begin{split}
\mathcal{V}_1&=\quadre{ \int \frac{du_1 du_2}{2 \pi \; \sqrt{\text{det} (\hat{A})}}  e^{-\frac{1}{2} \tonde{u_1,u_2 } \hat{A}^{-1}\tonde{u_1,u_2 }^T}\,{e^{- \hat{\phi}}g(\hat{m}-u_1; 2\hat{q}_1- \hat{q}_0)+ g(-\hat{p}+u_2; 2\hat{\xi}_1- \hat{\xi}_0)}}^S\\
&=e^{S   \mathcal{J}_1 (\hat{\bf x})},
  \end{split}
  \end{equation}
where the expression for $ \mathcal{J}_1 (\hat{\bf x})$ in \eqref{eq:VolumeAnnealed} is obtained replacing the functions $g$ with their integral representation, and exchanging the order of integration. We stress that all these expressions are obtained under the hypothesis that: 
\begin{equation}
\hat{q}_0<0, \quad \hat \xi_0<0, \quad \hat{q}_0 \hat{\xi}_0- \hat z^2 >0, \quad 2 \hat{q}_1-\hat{q}_0>0, \quad 2 \hat{\xi}_1-\hat{\xi}_0>0.
\end{equation}

\section{ The saddle-point equations for general $\gamma$}\label{sec:VariationalEquations}
In this Appendix, we derive the saddle-point equations for the order parameters ${\bf x}, \hat{\bf x}$ for generic values of $\gamma$.

\subsection{The quenched saddle-point equations for general $\gamma$}\label{sec:VariationalEquationsQ}
The first set of equations $\hat {\bf x}=F_2[{\bf x}]$ is obtained differentiating $\bar{\mathcal{A}}({\bf x}, \hat{\bf x}, \phi)$ with respect to the order parameters ${\bf x}$. The corresponding equations read:
\begin{equation}\label{eq:HatPar1}
\begin{split}
&\hat p =-\frac{\kappa-\mu m}{\sigma^2 (q_1-q_0)},\\
   &\hat \xi_1= \frac{q_1-2 q_0}{2 \sigma ^2 (q_1-q_0)^2},\\
&   \hat \xi_0=- \frac{q_0}{ \sigma ^2 (q_1-q_0)^2},\\
&\hat m =- \frac{\mu  (\kappa-\mu  m)}{\sigma ^2 (q_1-q_0)}-\frac{(\kappa-\mu  m)
   }{(\gamma +1) \sigma ^2
   (q_1-q_0)}+\frac{\mu  m}{(\gamma +1) \sigma ^2 (q_1-q_0)}+\frac{\mu  p
   }{ \sigma ^2 (q_1-q_0)}-\frac{\gamma}{1+\gamma}\frac{(\kappa-2\mu  m)
     z}{ \sigma ^2
   (q_1-q_0)^2},\\
  &\hat z= \frac{\gamma  m (\kappa-\mu  m)}{(\gamma +1) \sigma ^2 (q_1-q_0)^2}-\frac{ \gamma  z (q_1+q_0)}{ (\gamma +1) \sigma ^2 (q_1-q_0)^3}- \frac{
   q_0 }{ (\gamma +1) \sigma ^2 (q_1-q_0)^2},
   \end{split}
\end{equation}
which gives immediately:
\begin{equation}
\sqrt{2 \hat \xi_1 -\hat \xi_0}= \frac{1}{\sqrt{\sigma^2 (q_1-q_0)}}, \quad \quad \frac{\hat p}{\sqrt{2 \hat \xi_1 -\hat \xi_0}}=- \frac{(\kappa-\mu m)}{\sqrt{\sigma^2 (q_1-q_0)}}.
\end{equation}
The equations for $\hat{q}_1$ and $\hat q_0$ are given by:
\begin{equation}\label{eq:HatPar2}
\begin{split}
\hat q_1&=-\frac{(\kappa-\mu  m) [m+(\gamma +1) p]}{(\gamma +1) \sigma ^2
   (q_1-q_0)^2}+\frac{2 (\kappa-\mu  m) [m (q_1-q_0+\gamma 
   z)+p(\gamma +1)  (q_1-q_0)]}{(\gamma +1) \sigma ^2
   (q_1-q_0)^3}\\
   &+\frac{q_0 \left(2 q_0 z-2 \gamma  z^2\right)}{ (\gamma +1) \sigma ^2
   (q_1-q_0)^4}-\frac{q_0^2  (3 \xi_1-2 \xi_0)}{2  \sigma ^2
   (q_1-q_0)^4}   -\frac{ q_1 q_0 (\xi_0-2
   \xi_1) }{  \sigma ^2
   (q_1-q_0)^4}-\frac{ q_1 [2 q_0 z+\gamma  z^2 ]}{ (\gamma +1) \sigma ^2
   (q_1-q_0)^4}\\
   &  -\frac{ \xi_1 q_1^2}{2 \sigma ^2
   (q_1-q_0)^4}-\frac{q_0}{2 (q_1-q_0)^2}+\frac{1}{2
   (q_1-q_0)}-\frac{(\kappa-\mu  m)^2}{2 \sigma ^2
   (q_1-q_0)^2},
   \end{split}
\end{equation}
and
\begin{equation}
\begin{split}\label{eq:HatPar3}
\hat q_0&= \frac{2(\kappa-\mu  m) (-m-(\gamma +1) p)}{(\gamma +1) \sigma ^2
   (q_1-q_0)^2}+\frac{4 (\kappa-\mu  m) (m (q_1-q_0+\gamma 
   z)+(\gamma +1) p (q_1-q_0))}{(\gamma +1) \sigma ^2
   (q_1-q_0)^3}\\
   &+\frac{-2 z \left(q_1^2+\gamma  z (2 q_1+q_0)-q_0^2\right)+2 (\gamma +1) \xi_1 q_0 (q_1-q_0)-(\gamma +1) \xi_0 (q_1-q_0)
   (q_1+q_0)}{(\gamma +1) \sigma ^2 (q_1-q_0)^4}    \\
   &-\frac{q_0}{(q_1-q_0)^2}-\frac{(\kappa-\mu  m)^2}{ \sigma ^2
   (q_1-q_0)^2},
   \end{split}
\end{equation}
from which one also gets:
\begin{equation}
2 \hat{q}_1- \hat q_0= \frac{1}{q_1-q_0}- \frac{\xi_1-\xi_0}{\sigma^2 (q_1-q_0)^2} +\frac{2 z}{(\gamma +1) \sigma ^2 (q_1-q_0)^2}+   \frac{2
   \gamma  z^2}{(\gamma +1) \sigma ^2 (q_1-q_0)^3}.
\end{equation}
The second set of equations ${\bf x}=F_1[\hat {\bf x}]$ is obtained differentiating  $\bar{\mathcal{A}}({\bf x}, \hat{\bf x}, \phi)$  with respect to the conjugate parameters $\hat {\bf x}$. We set:
\begin{equation}
R_{\hat {\bf x}}(u_1, u_2)=e^{\frac{(u_1-\hat m)^2}{2 (2 \hat q_1-\hat q_0)}} \text{Erfc} \tonde{\frac{\hat m -u_1}{\sqrt{2 (2 \hat q_1 -\hat q_0)}}} +e^{\hat{\phi}} \sqrt{\frac{2 \hat q_1-\hat q_0}{2 \hat \xi_1- \hat \xi_0}} e^{\frac{(u_2-\hat p)^2}{2 (2 \hat \xi_1-\hat \xi_0)}} \text{Erfc} \tonde{\frac{-[\hat p -u_2]}{\sqrt{2 (2 \hat \xi_1 -\hat \xi_0)}}},
\end{equation}
and:
\begin{equation}
G_{\hat {\bf x}}(u_1, u_2)=\frac{1}{2 \pi \; \sqrt{\hat q_0 \hat \xi_0- \hat z^2}}  e^{-\frac{1}{2} \tonde{u_1,u_2 } \hat{A}^{-1}\tonde{u_1,u_2 }^T},
\end{equation}

and obtain:
\begin{equation}
\begin{split}
&m=\int d{\bf u}\,  G_{\hat {\bf x}}(u_1, u_2)  \frac{1}{\sqrt{2 \hat q_1-\hat q_0}} \frac{\sqrt{\frac{2}{\pi}} - \frac{\hat m -u_1}{\sqrt{2 \hat q_1-\hat q_0}} e^{\frac{(u_1-\hat m)^2}{2 (2 \hat q_1-\hat q_0)}} \text{Erfc} \tonde{\frac{\hat m -u_1}{\sqrt{2 (2 \hat q_1 -\hat q_0)}}}}{ R_{\hat x}(u_1, u_2)}\\
&p=\int  d{\bf u}\,  G_{\hat {\bf x}}(u_1, u_2) \frac{ e^{\hat \phi}\sqrt{\frac{2 \hat q_1-\hat q_0}{2 \hat \xi_1-\hat \xi_0}}}{\sqrt{2 \hat \xi_1-\hat \xi_0}}\frac{-\sqrt{\frac{2}{\pi}} - \frac{\hat p -u_2}{\sqrt{2 \hat \xi_1-\hat \xi_0}} e^{\frac{(u_2-\hat p)^2}{2 (2 \hat \xi_1-\hat \xi_0)}} \text{Erfc} \tonde{-\frac{\hat p -u_2}{\sqrt{2 (2 \hat \xi_1 -\hat \xi_0)}}}}{ R_{\hat x}(u_1, u_2)}\\
&q_1=\int  d{\bf u}\,  G_{\hat {\bf x}}(u_1, u_2)  \frac{1}{2 \hat q_1-\hat q_0} \frac{-\sqrt{\frac{2}{\pi}}\frac{\hat m - u_1}{\sqrt{2 \hat q_1- \hat q_0}} + \tonde{1+\frac{(\hat m -u_1)^2}{{2 \hat q_1-\hat q_0}}} e^{\frac{(u_1-\hat m)^2}{2 (2 \hat q_1-\hat q_0)}} \text{Erfc} \tonde{\frac{\hat m -u_1}{\sqrt{2 (2 \hat q_1 -\hat q_0)}}}}{ R_{\hat x}(u_1, u_2)}\\
&\xi_1=\int  d{\bf u}\,  G_{\hat {\bf x}}(u_1, u_2) \frac{e^{\hat \phi}\sqrt{\frac{2 \hat q_1-\hat q_0}{2 \hat \xi_1-\hat \xi_0}}}{2 \hat \xi_1-\hat \xi_0} \frac{\sqrt{\frac{2}{\pi}}\frac{\hat p - u_2}{\sqrt{2 \hat \xi_1- \hat \xi_0}} + \tonde{1+\frac{(\hat p -u_2)^2}{{2 \hat \xi_1-\hat \xi_0}}} e^{\frac{(u_2-\hat p)^2}{2 (2 \hat \xi_1-\hat \xi_0)}} \text{Erfc} \tonde{-\frac{\hat p -u_2}{\sqrt{2 (2 \hat q_1 -\hat q_0)}}}}{ R_{\hat x}(u_1, u_2)}.
 \end{split}
\end{equation}

 The derivatives with respect to $\hat q_0, \hat \xi_0, \hat z$ involve also the Gaussian measure. We obtain:
\begin{equation}
\begin{split}
&q_0=\int  d{\bf u}\,  G_{\hat {\bf x}}(u_1, u_2)\tonde{\frac{\sqrt{\frac{2}{\pi}}\frac{1}{\sqrt{2 \hat q_1-\hat q_0}} - \frac{\hat m -u_1}{{2 \hat q_1-\hat q_0}} e^{\frac{(u_1-\hat m)^2}{2 (2 \hat q_1-\hat q_0)}} \text{Erfc} \tonde{\frac{\hat m -u_1}{\sqrt{2 (2 \hat q_1 -\hat q_0)}}}}{R_{\hat {\bf x}}(u_1, u_2)}}^2 \\
&\xi_0=\int d{\bf u}\,  G_{\hat {\bf x}}(u_1, u_2) \frac{2 \hat q_1-\hat q_0}{2 \hat \xi_1-\hat \xi_0}e^{2 \hat \phi}  \tonde{\frac{\sqrt{\frac{2}{\pi}}\frac{1}{\sqrt{2 \hat \xi_1- \hat \xi_0}} + \frac{\hat p -u_2}{{2 \hat \xi_1-\hat \xi_0}} e^{\frac{(u_2-\hat p)^2}{2 (2 \hat \xi_1-\hat \xi_0)}} \text{Erfc} \tonde{-\frac{\hat p -u_2}{\sqrt{2 (2 \hat \xi_1 -\hat \xi_0)}}}}{R_{\hat {\bf x}}(u_1, u_2)}}^2
\end{split}
\end{equation}
and 
\begin{equation}
\begin{split}
&z=\int  d{\bf u}\,  G_{\hat {\bf x}}(u_1, u_2) \frac{\sqrt{\frac{2}{\pi}}\frac{1}{\sqrt{2 \hat q_1-\hat q_0}} - \frac{\hat m -u_1}{{2 \hat q_1-\hat q_0}} e^{\frac{(u_1-\hat m)^2}{2 (2 \hat q_1-\hat q_0)}} \text{Erfc} \tonde{\frac{\hat m -u_1}{\sqrt{2 (2 \hat q_1 -\hat q_0)}}}}{R_{\hat {\bf x}}(u_1, u_2)}  \times\\
& \times \frac{-\sqrt{\frac{2 \hat q_1-\hat q_0}{2 \hat \xi_1-\hat \xi_0}} e^{\hat \phi} \sqrt{\frac{2}{\pi}}\frac{1}{\sqrt{2 \hat \xi_1- \hat \xi_0}} - \sqrt{\frac{2 \hat q_1-\hat q_0}{2 \hat \xi_1-\hat \xi_0}} e^{\hat \phi}\frac{\hat p -u_2}{{2 \hat \xi_1-\hat \xi_0}} e^{\frac{(u_2-\hat p)^2}{2 (2 \hat \xi_1-\hat \xi_0)}} \text{Erfc} \tonde{-\frac{\hat p -u_2}{\sqrt{2 (2 \hat \xi_1 -\hat \xi_0)}}}}{R_{\hat {\bf x}}(u_1, u_2)} 
\end{split}
\end{equation}
Finally, the equation obtained deriving with respect to $\hat \phi$ is given by:
\begin{equation}
\phi=\int  d{\bf u}\,  G_{\hat {\bf x}}(u_1, u_2) \frac{ e^{\frac{(u_1-\hat m)^2}{2 (2 \hat q_1-\hat q_0)}} \text{Erfc} \tonde{\frac{\hat m -u_1}{\sqrt{2 (2 \hat q_1 -\hat q_0)}}}}{ R_{\hat {\bf x}}(u_1, u_2)}.
\end{equation}

\subsection{The annealed saddle-point equations for general $\gamma$}\label{sec:VariationalEquationsA}

In this case, the set of equations $\hat {\bf x}=F_2[{\bf x}]$  obtained differentiating $\mathcal{A}_1({\bf x}, \hat{\bf x}, \phi)$ with respect to the order parameters ${\bf x}$ takes the simpler form:
\begin{equation}
\begin{split}
&\hat{p}=- \frac{(\kappa-\mu m)}{\sigma^2 \; q_1}\\
&\hat{\xi_1}= \frac{1}{2\sigma^2 \; q_1}\\
&\hat{m} =\frac{\mu p}{\sigma^2 q_1} + \frac{\mu m}{\sigma^2 q_1(1+\gamma)} - \frac{\mu(\kappa-\mu m)}{\sigma^2 q_1} + \frac{ \gamma m (\kappa-\mu m) \quadre{\mu m - (\kappa-\mu m)} }{\sigma^2 (1+\gamma) q_1^2}-\frac{(\kappa-\mu m)}{ \sigma^2 q_1 (1+ \gamma)} \\
&\hat{q}_1 =-\frac{\xi_1}{ 2 \sigma^2 q_1^2} + \frac{2(\kappa-\mu m) [(\gamma+1) p+m]}{2 \sigma^2(1+\gamma)q_1^2}-\frac{(\kappa-\mu m)^2}{2 \sigma^2 q_1^2}+ 2 \frac{\gamma}{1+\gamma} \frac{m^2 (\kappa-\mu m)^2}{2 \sigma^2 q_1^3} + \frac{1}{2 q_1}.
\end{split}
\end{equation}
The derivative with respect to $\hat \phi$ reads
{\medmuskip=0mu
\thinmuskip=0mu
\thickmuskip=0mu
\begin{equation}
\phi= \frac{e^{- \hat{\phi}}}{2} \sqrt{\frac{\pi}{\hat{q}_1}} e^{\frac{\hat{m}^2}{4 \hat{q}_1}} \text{Erfc} \tonde{\frac{\hat{m}}{2 \sqrt{\hat{q}_1}}} \tonde{\frac{1}{2} \sqrt{\frac{\pi}{\hat{\xi}_1}} e^{\frac{\hat{p}^2}{4 \hat{\xi}_1}} \quadre{1+\text{Erf} \tonde{\frac{\hat{p}}{2 \sqrt{\hat{\xi}_1}}}}+\frac{e^{- \hat{\phi}}}{2} \sqrt{\frac{\pi}{\hat{q}_1}} e^{\frac{\hat{m}^2}{4 \hat{q}_1}} \text{Erfc} \tonde{\frac{\hat{m}}{2 \sqrt{\hat{q}_1}}}}^{-1}.
\end{equation}}

 Exploiting these identities, the equations $ {\bf x}=F_1[\hat {\bf x}]$ can be written as two pairs of decoupled equations, given by: 
\begin{equation}\label{eq:Dec1}
\begin{split}
&m= -\phi \frac{\hat{m}}{2 \hat{q}_1}+ \frac{\phi}{\sqrt{\pi \; \hat{q}_1 }}\frac{e^{-\frac{\hat{m}^2}{4 \hat{q}_1}}}{ \text{Erfc} \tonde{\frac{\hat{m}}{2 \sqrt{\hat{q}_1}}}},\\
&q_1=\frac{ \phi}{2 \hat{q}_1}+\frac{\phi \hat{m}^2}{ 4\; \hat{q}_1^2} -\frac{\phi \hat{m}}{2 \sqrt{\pi} \hat{q}_1^{\frac{3}{2}}}\frac{ e^{-\frac{\hat{m}^2}{4 \hat{q}_1}}}{ \text{Erfc} \tonde{\frac{\hat{m}}{2 \sqrt{\hat{q}_1}}}},
\end{split}
\end{equation}
and by 
\begin{equation}\label{eq:Dec2}
\begin{split}
p&= -(1- \phi)\frac{\hat{p}}{2 \hat{\xi}_1} - \frac{(1- \phi)}{\sqrt{\pi \hat{\xi}_1}} \frac{e^{-\frac{\hat{p}^2}{4 \hat{\xi}_1}}}{\quadre{1+\text{Erf} \tonde{\frac{\hat{p}}{2 \sqrt{\hat{\xi}_1}}}}}\\
\xi_1&= \frac{(1- \phi)}{2 \hat{\xi}_1}+\frac{(1- \phi) \hat{p}^2}{ 4\; \hat{\xi}_1^2} +\frac{(1- \phi) \hat{p}}{2 \sqrt{\pi} \hat{\xi}_1^{\frac{3}{2}}}\frac{ e^{-\frac{\hat{p}^2}{4 \hat{\xi}_1}}}{ \quadre{1+\text{Erf} \tonde{\frac{\hat{p}}{2 \sqrt{\hat{\xi}_1}}}}}.
\end{split}
\end{equation}

\section{Rescaled conjugate parameters and useful identities}\label{sec:IdentityByParts}

The quenched saddle point equations ${\bf x}= F_1[\hat {\bf x}]$ for generic $\gamma$  presented in \ref{sec:VariationalEquationsQ} are conveniently expressed in terms of the following rescaled variables:
\begin{equation}\label{eq:NewParametersApp}
\begin{split}
x_1=\frac{\hat m}{\sqrt{2 \hat q_1- \hat q_0}}, \quad 
&x_2=\frac{\hat p}{\sqrt{2 \hat \xi_1- \hat \xi_0}}, \quad 
y={\sqrt{2 \hat \xi_1- \hat \xi_0}}, \quad 
r =\sqrt{\frac{2 \hat q_1- \hat q_0}{2 \hat \xi_1- \hat \xi_0}}, \\
& \beta_1=\frac{\hat q_0}{y^2}, \quad \beta_2=\frac{\hat \xi_0}{y^2}, \quad \beta_3=\frac{\hat z}{y^2},
\end{split}
\end{equation}
see also \eqref{eq:NewParametersQ}. They are equivalent to:
\begin{equation}\label{eq:AppConv1}
{\begin{split}
&m y=\int \frac{du_1 du_2}{2 \pi} \frac{r \; e^{-\frac{1}{2}\frac{\tonde{- u_1^2 \beta_2 r^2 + 2 \beta_3 r u_1 u_2 - u_2^2 \beta_1}}{\beta_1 \beta_2- \beta_3^2} } }{ \sqrt{\beta_1 \beta_2- \beta_3^2}} \quadre{  \frac{1}{r} \frac{\sqrt{\frac{2}{\pi}} - (x_1 -u_1) e^{\frac{(x_1-u_1)^2}{2}} \text{Erfc} \tonde{\frac{x_1 -u_1}{\sqrt{2 }}}}{ \mathcal{R}_{\hat x}(u_1, u_2)}}\\
&p y =\int \frac{du_1 du_2}{2 \pi} \frac{r \; e^{-\frac{1}{2}\frac{\tonde{- u_1^2 \beta_2 r^2 + 2 \beta_3 r u_1 u_2 - u_2^2 \beta_1}}{\beta_1 \beta_2- \beta_3^2} } }{ \sqrt{\beta_1 \beta_2- \beta_3^2}} \quadre{- r e^{\hat \phi} \frac{\sqrt{\frac{2}{\pi}} + (x_2 -u_2) e^{\frac{(x_2-u_2)^2}{2}} \text{Erfc} \tonde{-\frac{x_2 -u_2}{\sqrt{2 }}}}{ \mathcal{R}_{\hat x}(u_1, u_2)}},
 \end{split}}
\end{equation}
and  
\begin{equation}\label{eq:AppConv2}
{\begin{split}
q_1 y^2=&\int \frac{du_1 du_2}{2 \pi} \frac{r \; e^{-\frac{1}{2}\frac{\tonde{- u_1^2 \beta_2 r^2 + 2 \beta_3 r u_1 u_2 - u_2^2 \beta_1}}{\beta_1 \beta_2- \beta_3^2} } }{ \sqrt{\beta_1 \beta_2- \beta_3^2}} \\
\times &\quadre{  \frac{1}{r^2} \frac{-\sqrt{\frac{2}{\pi}}(x_1 -u_1) +[1+ (x_1 -u_1)^2] e^{\frac{(x_1-u_1)^2}{2}} \text{Erfc} \tonde{\frac{x_1 -u_1}{\sqrt{2 }}}}{ \mathcal{R}_{\hat x}(u_1, u_2)}}\\
\xi_1 y^2=&\int \frac{du_1 du_2}{2 \pi} \frac{r \; e^{-\frac{1}{2}\frac{\tonde{- u_1^2 \beta_2 r^2 + 2 \beta_3 r u_1 u_2 - u_2^2 \beta_1}}{\beta_1 \beta_2- \beta_3^2} } }{ \sqrt{\beta_1 \beta_2- \beta_3^2}}\\
&\times e^{\hat \phi} r \frac{\sqrt{\frac{2}{\pi}}(x_2 - u_2) +[1+(x_2 -u_2)^2]e^{\frac{(u_2- x_2)^2}{2}} \text{Erfc} \tonde{-\frac{x_2 -u_2}{\sqrt{2 }}}}{ \mathcal R_{\hat x}(u_1, u_2)}.\\
 \end{split}}
\end{equation}

The last three equations give:
\begin{equation}\label{eq:AppConv3}
{\begin{split}
&q_0 y^2=\int \frac{du_1 du_2}{2 \pi} \frac{r \; e^{-\frac{1}{2}\frac{\tonde{- u_1^2 \beta_2 r^2 + 2 \beta_3 r u_1 u_2 - u_2^2 \beta_1}}{\beta_1 \beta_2- \beta_3^2} } }{ \sqrt{\beta_1 \beta_2- \beta_3^2}} \quadre{  \frac{1}{r} \frac{\sqrt{\frac{2}{\pi}} - (x_1 -u_1) e^{\frac{(x_1-u_1)^2}{2}} \text{Erfc} \tonde{\frac{x_1 -u_1}{\sqrt{2 }}}}{ \mathcal{R}_{\hat x}(u_1, u_2)}}^2\\
&\xi_0 y^2 =\int \frac{du_1 du_2}{2 \pi} \frac{r \; e^{-\frac{1}{2}\frac{\tonde{- u_1^2 \beta_2 r^2 + 2 \beta_3 r u_1 u_2 - u_2^2 \beta_1}}{\beta_1 \beta_2- \beta_3^2} } }{ \sqrt{\beta_1 \beta_2- \beta_3^2}} \quadre{- r e^{\hat \phi} \frac{\sqrt{\frac{2}{\pi}} + (x_2 -u_2) e^{\frac{(x_2-u_2)^2}{2}} \text{Erfc} \tonde{-\frac{x_2 -u_2}{\sqrt{2 }}}}{ \mathcal{R}_{\hat x}(u_1, u_2)}}^2
 \end{split}}
\end{equation}
and 
\begin{equation}\label{eq:AppConv4}
{\begin{split}
&z y^2=\int \frac{du_1 du_2}{2 \pi} \frac{r \; e^{-\frac{1}{2}\frac{\tonde{- u_1^2 \beta_2 r^2 +2 \beta_3 r u_1 u_2 - u_2^2 \beta_1}}{\beta_1 \beta_2- \beta_3^2} } }{ \sqrt{\beta_1 \beta_2- \beta_3^2}} \quadre{  \frac{1}{r} \frac{\sqrt{\frac{2}{\pi}} - (x_1 -u_1) e^{\frac{(x_1-u_1)^2}{2}} \text{Erfc} \tonde{\frac{x_1 -u_1}{\sqrt{2 }}}}{ \mathcal{R}_{\hat x}(u_1, u_2)}}\times \\
& \times \quadre{- r e^{\hat \phi} \frac{\sqrt{\frac{2}{\pi}} + (x_2 -u_2) e^{\frac{(x_2-u_2)^2}{2}} \text{Erfc} \tonde{-\frac{x_2 -u_2}{\sqrt{2 }}}}{ \mathcal{R}_{\hat x}(u_1, u_2)}}
 \end{split}}
\end{equation}

Finally:
\begin{equation}\label{eq:AppConv5}
{\phi=\int \frac{du_1 du_2}{2 \pi} \frac{r \; e^{-\frac{1}{2}\frac{\tonde{- u_1^2 \beta_2 r^2 + 2 \beta_3 r u_1 u_2 - u_2^2 \beta_1}}{\beta_1 \beta_2- \beta_3^2} } }{ \sqrt{\beta_1 \beta_2- \beta_3^2}} \frac{ e^{\frac{(u_1-x_1)^2}{2 }} \text{Erfc} \tonde{\frac{x_1 -u_1}{\sqrt{2 }}}}{ \mathcal{R}_{\hat {\bf x}}(u_1, u_2)}}.
\end{equation}

 These convolutions are not independent, but can be related by integration by parts. Indeed, using the above expression it is straightforward to show that the following identities hold for all values of the conjugate parameters:
\begin{equation}
\begin{split}
r^2 (q_1 y^2)&= -r x_1 (m y) + \phi - \beta_1 \quadre{(q_1 y^2)-(q_0 y^2)} + \beta_3 (z y^2)\\
 (\xi_1 y^2)&= - x_2 (p y) +(1- \phi) - \beta_2 \quadre{(\xi_1 y^2)- (\xi_0 y^2)} + \beta_3 (z y^2),
\end{split}
\end{equation}
where the brackets denote the integral representations for the corresponding parameters. 
Summing the equations, one derives the identity:
{\medmuskip=1mu
\thinmuskip=.5mu
\thickmuskip=1mu
\begin{equation}\label{EqSum}
r^2 (q_1 y^2)+  (\xi_1 y^2)=1 -r x_1 (m y)- x_2 (p y)- \beta_1 \quadre{(q_1 y^2)- (q_0 y^2)} - \beta_2 \quadre{(\xi_1 y^2)- (\xi_0 y^2)}  + 2 \beta_3 (z y^2),
\end{equation}}
while the difference gives:
\begin{equation}\label{EqDiff}
r^2 (q_1 y^2)- (\xi_1 y^2)= -r x_1 (m y)+ x_2 (p y) + 2 \phi- 1 - \beta_1 \quadre{(q_1 y^2)- (q_0 y^2)} + \beta_2 \quadre{(\xi_1 y^2)- (\xi_0 y^2)}.
\end{equation}
In the case $\gamma=0$, the identity \eqref{EqSum}  entails $x_2= r x_1$, as we discuss in the main text.\\

 In the annealed case, it is straightforward to check that the equations \eqref{eq:Dec1}, \eqref{eq:Dec2} are equivalent to \eqref{eq:Aann}, \eqref{eq:Bann}, once the parameters  \eqref{eq:NewParametersA} are introduced. Moreover, by inspecting  \eqref{eq:Aann} and \eqref{eq:Bann} one sees that the following two relations hold:
\begin{equation}
\begin{split}
\mathcal{r}^2 (q_1 \mathcal{y}^2)&= -\mathcal{r} \mathcal{x}_1 (m \mathcal{y}) + \phi \\
 (\xi_1 \mathcal{y}^2)&= - \mathcal{x}_2 (p \mathcal{y}) +(1- \phi),
\end{split}
\end{equation}
which are equivalent to 
\begin{equation}
\begin{split}
\mathcal{r}^2 (q_1 \mathcal{y}^2)+ (\xi_1 \mathcal{y}^2)&= -\mathcal{r} \mathcal{x}_1 (m \mathcal{y}) - \mathcal{x}_2 (p \mathcal{y}) + 1 \\
\mathcal{r}^2 (q_1 \mathcal{y}^2)-  (\xi_1 \mathcal{y}^2)&= -\mathcal{r} \mathcal{x}_1 (m \mathcal{y})+ \mathcal{x}_2 (p \mathcal{y}) + 2\phi-1.
\end{split}
\end{equation}
For $\gamma=0$, the first identity entails again $\mathcal{x}_2= r \mathcal{x}_1$.

\section{The self-consistent equations in the uncorrelated $\gamma=0$ case}\label{sec:EqsUncorrelated}

Setting $\gamma=0$ in the equations given in \ref{sec:VariationalEquationsQ} we obtain:
\begin{equation}\label{eq:DirectGamma0}
\begin{split}
\hat p=&-\frac{(\kappa-\mu m)}{\sigma^2(q_1-q_0)}\\
\hat \xi_1=& \frac{q_1-2 q_0}{2 \sigma^2(q_1-q_0)^2}\\
\hat \xi_0=&\hat z=-\frac{q_0}{ \sigma^2(q_1-q_0)^2}\\
\hat m=&\frac{1}{\sigma^2(q_1-q_0)}\quadre{\mu(m+p) -(\kappa-\mu m)(1+\mu)}\\
\hat q_1=&\frac{(\kappa-\mu m)(m+p)}{\sigma^2(q_1-q_0)^2}-\frac{\xi_1}{2 \sigma^2(q_1-q_0)^2}- \frac{q_0[\xi_0-\xi_1+ 2 z]}{\sigma^2(q_1-q_0)^3}-\frac{(\kappa-\mu m)^2}{2\sigma^2(q_1-q_0)^2}-\frac{q_0}{2(q_1-q_0)^2}\\
&+\frac{1}{2(q_1-q_0)}
\end{split}
\end{equation}
and finally 
\begin{equation}
\begin{split}
\hat q_0&=\frac{2(\kappa-\mu m)(m+p)}{\sigma^2(q_1-q_0)^2}-\frac{(\kappa-\mu m)^2}{\sigma^2(q_1-q_0)^2}-\frac{q_0}{(q_1-q_0)^2}-\frac{ \xi_1}{ \sigma^2(q_1-q_0)^2}+ \frac{(q_1+q_0)(\xi_1-\xi_0- 2 z)}{\sigma^2(q_1-q_0)^3}
\end{split}
\end{equation}
from which it follows that
\begin{equation}
\begin{split}
2 \hat q_1- \hat q_0&= \frac{1}{\sigma^2 (q_1-q_0)} \quadre{\sigma^2+ \frac{\xi_0-\xi_1+ 2 z}{q_1-q_0}}= (2 \hat \xi_1-\hat \xi_0) \quadre{\sigma^2+ \frac{\xi_0-\xi_1+ 2 z}{q_1-q_0}}\\
2 \hat \xi_1- \hat \xi_0&= \frac{1}{\sigma^2 (q_1-q_0)}.
\end{split}
\end{equation}
In terms of the parameters \eqref{eq:NewParametersApp}, these equations read:
\begin{equation}
\begin{split}
&x_2=-\kappa y+ \mu y \, m \\
&r \, x_1= \kappa y+ (2+\mu)x_2+\mu \,y\, p\\
&1=\sigma^2 y^2 (q_1-q_0)\\
&r^2=\sigma^2(\xi_0-\xi_1+ 2z) y^2+ \sigma^2\\
&\beta_3=\beta_2\\
&\beta_2=1- \sigma^2 y^2 q_1= - \sigma^2 y^2 q_0\\ 
&\beta_1=  r^2 \beta_2 + \quadre{ \sigma^4 q_1 -  \sigma^2 r^2 q_1- \sigma^2 \xi_1 }y^2+\frac{\mu + 2}{\mu} \sigma^2 x_2^2 -\frac{2}{\mu}\sigma^2 x_1 x_2 r.
\end{split}
\end{equation}
The factor $y$ multiplies the order parameter in such a way that the resulting expressions do not depend on $y$, as one can see from Eqs. \eqref{eq:AppConv1},\eqref{eq:AppConv2},\eqref{eq:AppConv3},\eqref{eq:AppConv4},\eqref{eq:AppConv5}. This implies that
the variable $y$ can be fixed at the end of the calculation, via the identity $\kappa y=-x_2 + \mu \; m y$.
The remaining equations are those given in \eqref{eq:SetA}, with the expressions multiplying factors of $y$ given by the integral convolutions in the above section (with $\beta_3 = \beta_2$). 
The derivation of the annealed equations \eqref{eq:SetAann} is completely analogous.

\section{The cavity solution: a reminder}\label{sec:Cavity}
As we remarked in the main text, the one equilibrium phase of the Lotka-Volterra model can be characterized via the so called cavity method \cite{mezard1987sk}. In essence, the method consists in introducing a new species in the interacting system, and in relating the properties of the system with $S+1$ species to that with $S$ species, under the hypothesis that a unique equilibrium exists. The cavity analysis of the Lotka-Volterra equations has been performed in \cite{GuyCavity} (see also \cite{barbier2017cavity} for a discussion of this method in the context of ecology), and analogous results have been obtained in \cite{RoyDMFT, galla2018dynamically} via a dynamical formalism. In particular, the cavity treatment allows to derive the value of the three parameters characterizing the unique, stable equilibrium attracting the dynamics: the diversity $\phi$, and the first two moments $m, q_1$ of the configuration vector; the result is incorporated into a self-consistent equation for the variable $(\kappa- \mu m)/ [\sigma \sqrt{q_1}]$, which we recognise to coincide (up to a sign) with the parameter $\mathcal{x}_2$ in our annealed formalism, see Eqs. (a) and (c) in \eqref{eq:NewParametersA}. We set here $\kappa=1$, and follow the notation of \cite{RoyDMFT}. The self-consistent equation for $\mathcal{x}_2$ obtained within the cavity approximation reads:
\begin{equation}\label{eq:SelfConsCav}
\sigma^2 [w_2(-\mathcal{x}_2) + \gamma w_0(-\mathcal{x}_2)]^2=w_2(-\mathcal{x}_2), \quad \quad w_n(x)= \int_{-\infty}^{x} ds  \frac{e^{-\frac{s^2}{2}}}{\sqrt{2 \pi}}\, (x-s)^n.
\end{equation}
From the solution $\mathcal{x}_2^{\rm cav}$ to this equation, the second moment $q_1^{\rm cav}$ and the diversity $\phi^{\rm cav}$ are obtained from:
\begin{equation}\label{eq:cavParBun}
\begin{split}
q_1^{\rm cav}&= \frac{\sigma^2 [w_2(-\mathcal{x}_2^{\rm cav}) + \gamma w_0(-\mathcal{x}_2^{\rm cav})]^2}{\grafe{ \mu w_1(-\mathcal{x}_2^{\rm cav})+ \mathcal{x}_2^{\rm cav} \sigma^2[w_2(-\mathcal{x}_2^{\rm cav}) + \gamma w_0(-\mathcal{x}_2^{\rm cav})]}^2}\\
\phi^{\rm cav}&=w_0(-\mathcal{x}_2^{\rm cav}).
\end{split}
\end{equation}
For $\gamma=0$, the equation \eqref{eq:SelfConsCav} becomes:
\begin{equation}\label{eq:SelfConsCav2}
\sigma^2 \tonde{-\frac{e^{-\frac{\mathcal{x}_2^2}{2}} \mathcal{x}_2  }{\sqrt{2 \pi}} + \frac{1+  \mathcal{x}_2 ^2}{2} \text{Erfc}\tonde{\frac{\mathcal{x}_2}{\sqrt{2}}}}=1,
\end{equation}
which is exactly the equation \eqref{eq:CavAn} that one gets in the annealed scheme, for $\mathcal{r}=1$ (and thus $\mathcal{x}_1 = \mathcal{x}_2$). For $\mathcal{r}=1$ and $\hat \phi=0$, moreover, \eqref{eq:CavAnInutile} is also satisfied. Finally, using that $\mathcal{y}=[\sigma \sqrt{q_1}]^{-1}$, one sees that the equations \eqref{eq:AnnealedPhi} and \eqref{eq:AnnealedY} are equivalent to \eqref{eq:cavParBun}.\\
To summarize, when $\hat \phi=0$ and $\mathcal{r}=1$ the annealed self-consistent equations reproduce the solution obtained with the cavity method. Moreover, the quenched equation map into the annealed equations at this ‘‘cavity matching point", as we showed explicitly in Sec. \ref{sec:CavPoint}. The cavity matching point describes different things depending on whether one is in the unique equilibrium phase $\sigma \leq \sigma_c$, or in the multiple equilibria phase $\sigma >\sigma_c$. For $\sigma \leq \sigma_c$, the cavity solution describes the properties of the unique equilibrium attracting the dynamics of the system. At the corresponding value of diversity $\phi^{\rm cav}$, the complexity $\Sigma^{(A)}_\sigma(\phi)$ reaches its maximum, and it is equal to zero. For $\sigma > \sigma_c$,  $\phi^{\rm cav}$ only marks the diversity value below which the complexity can be computed within the annealed approximation; in particular, $\phi^{\rm cav}$ does not give the diversity of the most numerous equilibria, for which  $\Sigma^{(Q)}_\sigma(\phi)$ is maximal.
At the critical point $\sigma=\sigma_c$, all the equation match and the unique equilibrium has parameters (for $\kappa=1$): 
\begin{equation}
m=\frac{1}{\mu}, \quad q_1=\frac{\pi}{\mu^2}, \quad p=-\frac{m}{1+\gamma}=-\frac{1}{(1+\gamma) \mu}, \quad \xi_1 = \frac{q_1}{(1+ \gamma)^2}= \frac{\pi}{(1+\gamma)^2 \mu^2}.
\end{equation}

\section{The replica calculation of the complexity: a reminder}\label{sec:Replicas}

\subsection{The Monasson recipe for the complexity}
The complexity curve $\Sigma_{\rm 1RSB}$ discussed in the main text is obtained within the so called Monasson method \cite{monasson1995structural}. This method requires the system to be conservative, and thus to be associated to a potential landscape. In the Lotka-Volterra symmetric case the potential landscape reads:
\begin{equation}\label{eq:Potential}
L({\vec N})= - \sum_{i=1}^S \tonde{\kappa_i N_i- \frac{N_i^2}{2}}+\frac{1}{2}\sum_{i,j=1}^S \alpha_{ij} N_i N_j,
\end{equation}
and the method allows to obtain the complexity $\Sigma_{\rm 1RSB}(l)$ of the typical (i.e., most numerous) local minima ${\vec N}^*$ of \eqref{eq:Potential} such that $l= \lim_{S \to \infty} S^{-1} L({\vec N}^*)$.
The main idea of \cite{monasson1995structural} is that $\Sigma_{\rm 1RSB}(l)$ can be obtained as a Legendre transform of the free-energy of $m$ copies of the system evolving in the same random landscape, weakly-coupled in such a way that they explore the same state (basin of attraction of a local minimum of the free-energy). The object to compute is then the modified free energy function: 
\begin{equation}
\beta \Phi(m, \beta)= \lim_{S \to \infty} \lim_{n \to 0}  -\frac{1}{n \, S} \log \langle {Z}_m^n \rangle= \beta m f_{\rm 1RSB}(m, \beta),
\end{equation}
where ${Z}_m^n$ is the partition function of the $m$ copies and $f_{\rm 1RSB}(m, \beta)$ is the free energy of one single copy of the system computed within the 1RSB ansatz, with $m$ being the variational parameter in the 1RSB ansatz for the overlap matrix -- the parameter measuring the size of the inner blocks of the overlap matrix \cite{mezard1987spin}. In the zero-temperature limit $\beta \to  \infty$, the free energy becomes a function of the scaled parameter $\tilde{m}= \beta m$, i.e. $f_{\rm 1RSB}(m, \beta) \to \tilde{f}_{\rm 1RSB}(\tilde m)$. In terms of these quantities, the complexity curve $\Sigma_{\rm 1RSB}(l)$ is obtained parametrically though the coupled system of equations:
\begin{equation}\label{eq:EqsMonasson}
    \begin{split}
     l =\partial_{\tilde m} \tonde{ \tilde m \, \tilde{f}_{\rm 1RSB}(\tilde m)}, \quad \quad \Sigma= \tilde m^2 \partial_{\tilde m} \tilde{f}_{\rm 1RSB}(\tilde m),
    \end{split}
\end{equation}
by tuning the parameter $\tilde{m}$ which parametrizes for the value $l$ of the potential \eqref{eq:Potential}.

\subsection{The structure of the replica calculation}
Performing the 1RSB free-energy calculation \cite{LVMarginality}, one gets:
\begin{equation}
  \tilde{f}_{\rm 1RSB}(\tilde m)= \min_{m, q_0, q_1, \chi} F\tonde{\tilde{m};m, q_0,q_1, \chi},  \end{equation}
where $q_0, q_1, m$ are variational parameters having the same meaning as in the replicated Kac-Rice calculation, while $ \chi$ is a parameter related to the properties of the Hessian matrix (the matrix of second derivatives of the potential \eqref{eq:Potential}) at a local minimum.  
For $\kappa_i=\kappa$ one finds:
\begin{equation}\label{eq:ReplicaAction}
\begin{split}
&F\tonde{\tilde{m};m, q_0,q_1, \chi}= \frac{\sigma^2}{4}\quadre{ \tilde{m} (q_1^2-q_0^2)+ \frac{2 q_1}{\sigma^2 \chi}} - \frac{\mu m^2}{2}\\
&-\frac{1}{\tilde m} \int \frac{dz}{\sqrt{2 \pi}}e^{-\frac{z^2}{2}}\log \quadre{\frac{e^{\frac{\alpha \Delta^2(z)}{2(1-\alpha)}} \;\text{Erfc} \tonde{ \frac{-\Delta(z)}{\sqrt{2(1-\alpha)}}}+ \sqrt{1-\alpha}\; \text{Erfc} \tonde{ \frac{\Delta(z)}{\sqrt{2}}}}{2 \sqrt{1-\alpha}}}
\end{split}
\end{equation}
with
\begin{equation}\label{eq:DefDeltaAlfa}
    \Delta(z)=\frac{\kappa-\mu m}{\sqrt{\sigma^2(q_1-q_0)}}+ \sqrt{\frac{q_0}{q_1-q_0}}z, \quad \quad \alpha=\frac{\tilde{m} \sigma^2(q_1-q_0)}{1-\chi^{-1}}.
\end{equation}
The equations \eqref{eq:EqsMonasson} give:
\begin{equation}\label{eq:FreeEnergyMonasson}
\begin{split}
&l= \frac{\sigma^2}{2}\quadre{ \tilde{m} (q_1^2-q_0^2)+  \frac{q_1}{\sigma^2 \chi} } - \frac{\mu m^2}{2}-\frac{\sigma^2 (q_1-q_0)}{2 (1-\alpha) (1-\chi^{-1})} \times\\
&\times\int \frac{dz}{\sqrt{2 \pi}}e^{-\frac{z^2}{2}} \quadre{\frac{\sqrt{\frac{2}{\pi}} \frac{\Delta(z)}{\sqrt{1-\alpha}} + e^{\frac{ \Delta^2(z)}{2(1-\alpha)}}\tonde{1+ \frac{\Delta^2(z)}{1-\alpha}} \text{Erfc} \tonde{ \frac{-\Delta(z)}{\sqrt{2(1-\alpha)}}}}{e^{\frac{ \Delta^2(z)}{2(1-\alpha)}} \;\text{Erfc} \tonde{ \frac{-\Delta(z)}{\sqrt{2(1-\alpha)}}}+ \sqrt{1-\alpha}e^{\frac{ \Delta^2(z)}{2}}\; \text{Erfc} \tonde{ \frac{\Delta(z)}{\sqrt{2}}}}}
\end{split}
\end{equation}
and 
\begin{equation}\label{eq:ComplexityMonasson}
\begin{split}
&\Sigma_{\rm 1RSB}= \frac{\sigma^2}{4} \tilde{m} (q_1^2-q_0^2)+ \int \frac{dz}{\sqrt{2 \pi}}e^{-\frac{z^2}{2}}\log \quadre{\frac{e^{\frac{\alpha \Delta^2(z)}{2(1-\alpha)}} \;\text{Erfc} \tonde{ \frac{-\Delta(z)}{\sqrt{2(1-\alpha)}}}+ \sqrt{1-\alpha}\; \text{Erfc} \tonde{ \frac{\Delta(z)}{\sqrt{2}}}}{2 \sqrt{1-\alpha}}}\\
&-\frac{\sigma^2 (q_1-q_0) \tilde{m}}{2 (1-\alpha) (1-\chi^{-1})} 
\int \frac{dz}{\sqrt{2 \pi}}e^{-\frac{z^2}{2}} \quadre{\frac{\sqrt{\frac{2}{\pi}} \frac{\Delta(z)}{\sqrt{1-\alpha}} + e^{\frac{ \Delta^2(z)}{2(1-\alpha)}}\tonde{1+ \frac{\Delta^2(z)}{1-\alpha}} \text{Erfc} \tonde{ \frac{-\Delta(z)}{\sqrt{2(1-\alpha)}}}}{e^{\frac{ \Delta^2(z)}{2(1-\alpha)}} \;\text{Erfc} \tonde{ \frac{-\Delta(z)}{\sqrt{2(1-\alpha)}}}+ \sqrt{1-\alpha}e^{\frac{ \Delta^2(z)}{2}}\; \text{Erfc} \tonde{ \frac{\Delta(z)}{\sqrt{2}}}}},
\end{split}
\end{equation}
where the order parameters $q_1, q_0, m$ and $ \chi$ have to be determined taking the variation of \eqref{eq:ReplicaAction}. The resulting saddle-point equations have a structure that is rather simple to interpret \cite{LVMarginality}. Indeed, the replica calculation is formulated in terms of an effective single-species potential:
\begin{equation}\label{eq:SinglePartH}
L_{\rm eff}(N, \xi, z)= - \frac{N^2}{2 \, \chi}
+\tonde{-\kappa N + \frac{N^2}{2}}+ \tonde{\mu m- \xi \sigma \sqrt{q_1-q_0}- z \sigma \sqrt{q_0}}N
\end{equation}
depending on two fields $\xi, z$ introduced performing the Hubbard-Stratonovich transformations to decouple the quartic terms in the overlaps arising after averaging over the random couplings $\alpha_{ij}$. 
The self-consistent equations for the order parameters $m, q_0, q_1$ can be expressed as a triple average:
\begin{equation}\label{eq:TripleAv}
\begin{split}
m&=\int \frac{dz}{\sqrt{2 \pi}} e^{-\frac{z^2}{2}} \int d\nu_{\tilde m}(\xi; z) \;  \mathbb{E}_{0} \quadre{N}\\
q_0&=\int \frac{dz}{\sqrt{2 \pi}} e^{-\frac{z^2}{2}} \tonde{\int d\nu_{\tilde m}(\xi; z) \;\mathbb{E}_{0} \quadre{N} }^2\\
q_1&=\int \frac{dz}{\sqrt{2 \pi}} e^{-\frac{z^2}{2}} \int d\nu_{\tilde m}(\xi; z) \; \tonde{\mathbb{E}_{0} \quadre{N}}^2
\end{split}
\end{equation}
where the internal average reads
\begin{equation}
\mathbb{E}_{0} \quadre{N}=\lim_{\beta \to \infty} \frac{\int_{0}^\infty dN e^{- \beta L_{\rm eff}(N, \xi, z)} N}{\int_0^\infty dN e^{- \beta L_{\rm eff}(N, \xi, z)} } = N_{\rm sp}(\xi,z)= \max \left\{0, \frac{ \xi  \sigma \sqrt{q_1-q_0}  + \kappa-\mu m + \sigma \sqrt{q_0} z}{1- \chi^{-1}}  \right\}
\end{equation}
as it follows from a saddle-point calculation, while the outer average is taken with respect to the measure:
\begin{equation}
d \nu_{\tilde m}(\xi; z) = \frac{1}{Z[z]}  \frac{d \xi}{\sqrt{2 \pi}}e^{-\frac{\xi^2}{2}} e^{-\tilde  m \;  L_{\rm eff}(N_{\rm sp}(\xi,z), \xi, z)} 
 \end{equation}
 with $Z[z]$ a normalization. Similarly, one finds the following equation for $\chi$:
 \begin{equation}
 \chi^2 \, \sigma^2 \int \frac{dz}{\sqrt{2 \pi}} e^{-\frac{z^2}{2}} \int d\nu_{\tilde m}(\xi; z) \;  \mathbb{E}_{0} \quadre{\theta(N)}- \chi+1=0,
 \end{equation}
 and thus by analogy with the above equation one can make the identification:
 \begin{equation}\label{eq:EqChi}
 \chi^2 \,  \sigma^2 \phi - \chi+1=0,
 \end{equation}
 where $\phi$ is the diversity of the counted local minima, which is not a free-parameter in the replica calculation but it is fixed as a function of $q_1, q_0, m, \chi$. 
The order parameters are therefore obtained as double averages of the (moments of the) truncated Gaussian variable $N_{\rm sp}(z, \xi)$, 
  \begin{equation}
\begin{split}
m&=\int \frac{dz}{\sqrt{2 \pi}} e^{-\frac{z^2}{2}} \frac{1}{Z[z]} \int \frac{d \xi}{\sqrt{2 \pi}}e^{-\frac{\xi^2}{2}} e^{- L_{\rm eff}(N_{\rm sp,} \xi, z)} N_{\rm sp}\\
q_0&=\int \frac{dz}{\sqrt{2 \pi}} e^{-\frac{z^2}{2}}\tonde{\frac{1}{Z[z]} \int \frac{d \xi}{\sqrt{2 \pi}}e^{-\frac{\xi^2}{2}} e^{- L_{\rm eff}(N_{\rm sp,} \xi, z)} N_{\rm sp}}^2\\
q_1&=\int \frac{dz}{\sqrt{2 \pi}} e^{-\frac{z^2}{2}} \frac{1}{Z[z]} \int \frac{d \xi}{\sqrt{2 \pi}}e^{-\frac{\xi^2}{2}} e^{- L_{\rm eff}(N_{\rm sp,} \xi, z)} N_{\rm sp}^2,
\end{split}
\end{equation}
  which in turn is obtained as the global minimum of an effective single particle potential \eqref{eq:SinglePartH}, where the interactions between species are encoded self-consistently in the Gaussian fields $\xi$ and $z$.  We remark that given the effective potential \eqref{eq:SinglePartH}, one can introduce an effective force:
\begin{equation}\label{eq:EffectiveForce}
f_{\rm eff}(N, \xi,z)=- \frac{\partial L_{\rm eff}}{\partial N}= - (1-  \chi^{-1}) \quadre{N- \frac{ \xi  \sigma \sqrt{q_1-q_0}  + \kappa-\mu m + \sigma \sqrt{q_0} z}{1- \chi^{-1}} }.
\end{equation} 
One sees that when $N_{\rm sp}>0$, then $f_{\rm eff}(N_{\rm sp}, \xi,z)=0$; similarly, when $N_{\rm sp}=0$ because $ \xi + \Delta (z)<0$, then $f_{\rm eff}(N_{\rm sp}, \xi,z)<0$. Therefore, the uninvadability constraint is encoded naturally structure of the replica calculation.  Notice also that while \eqref{eq:ComplexityMonasson} does not depend on $\mu$, \eqref{eq:FreeEnergyMonasson} does.

Finally, we remark that using the equation \eqref{eq:EqChi}, the quantity $\chi$ can be related to the resolvent of the matrix \eqref{eq:Hessian} evaluated at the counted local minima. Indeed, in the symmetric case $\gamma=1$ the resolvent of the matrix \eqref{eq:Hessian} evaluated in a local minimum of diversity $\phi$, defined as
\begin{equation}
    G^H(z)= \lim_{S \to \infty} \frac{1}{S \phi} \text{Tr} \grafe{\frac{1}{z-H}},
\end{equation}
in the large-$S$ limit equals to:
\begin{equation}
    G^H(z)=G_{\sigma \sqrt{\phi}}(z+1),
\end{equation}
where $G_{\sigma \sqrt{\phi}}(z)= (z- \text{sign}(z) \sqrt{z^2- 4 \sigma^2 \phi})/(2 \sigma^2 \phi)$ is the resolvent of a matrix with GOE statistics, with variance $\sigma^2 \phi$. Comparing with \eqref{eq:EqChi} we see that it holds
\begin{equation}
    G^H(0)= \frac{1}{2 \sigma^2 \phi} \tonde{1- \sqrt{1-4 \sigma^2 \phi}}= \chi.
\end{equation}

\subsection{The resulting complexity and the relation with the Kac-Rice quenched calculation}
\begin{figure}[ht]
  \centering
 \includegraphics[width=.48\linewidth]{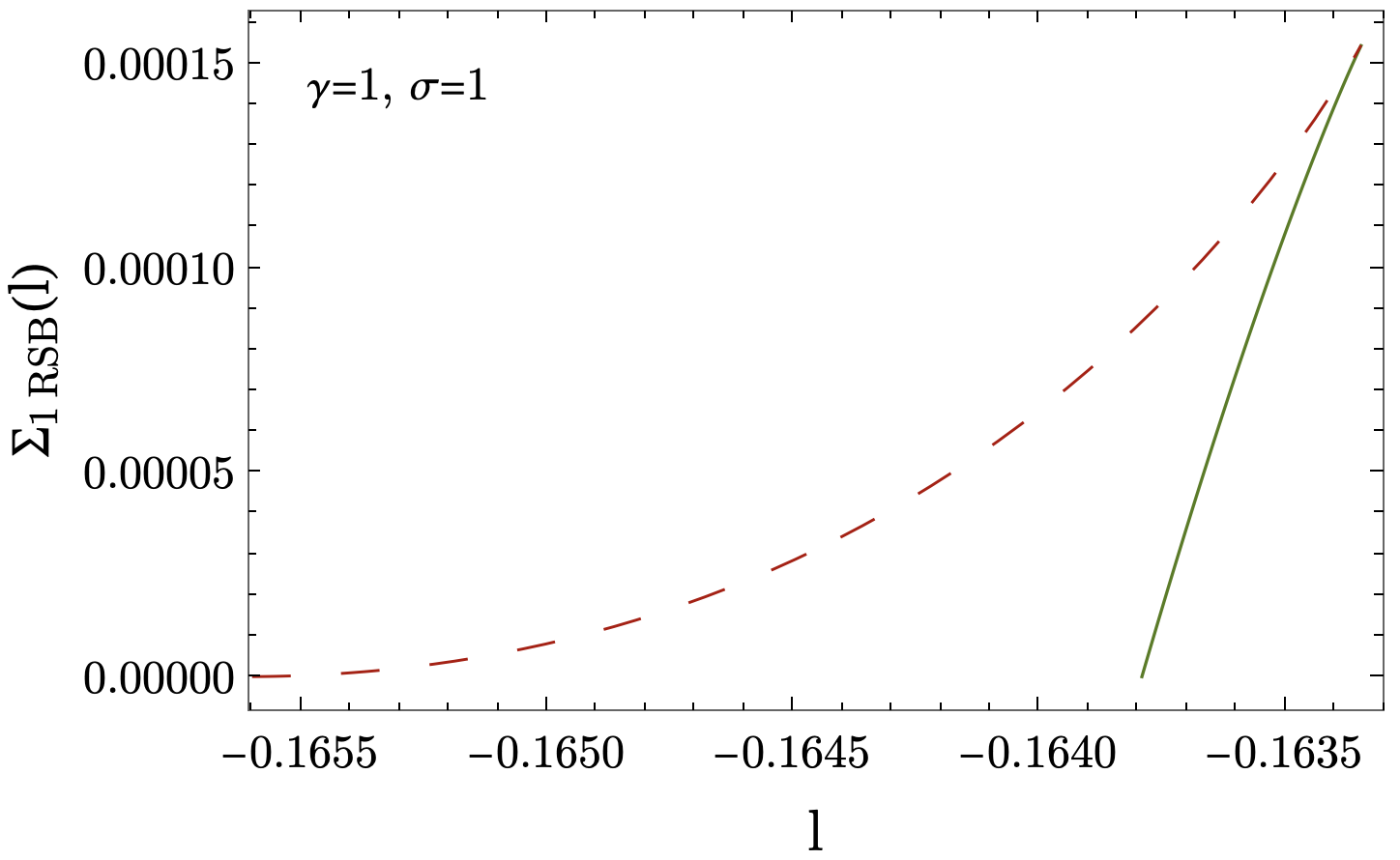}
 \includegraphics[width=.5\linewidth]{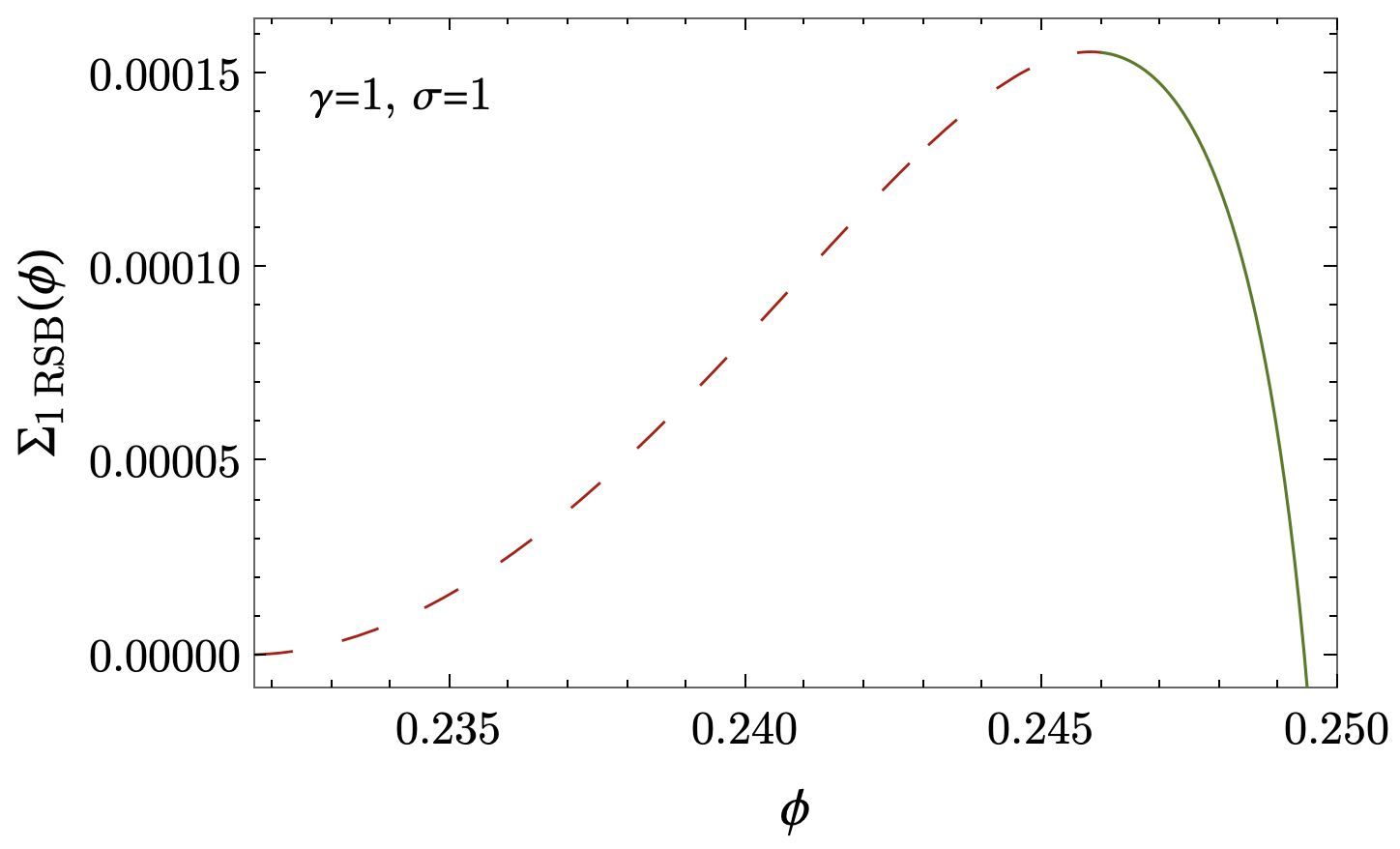}
\caption{1RSB complexity for $\sigma=1=\gamma$ and $\mu=5$ as a function of the intensive value of the potential $l$ and of the diversity $\phi$, respectively.}\label{fig:Monasson}
\end{figure}

In Fig. \ref{fig:Monasson} ({\it left}) we show the complexity \eqref{eq:ComplexityMonasson} as a function of the intensive value of the potential \eqref{eq:FreeEnergyMonasson} for one value of diversity in the multiple equilibria phase. By varying the parameter $\tilde m$, one obtains two branches in the curve $\Sigma_{\rm 1RSB}(l)$: the red branch is clearly unstable, as the resulting complexity does not have the right convexity properties; the green branch is instead stable. One sees that the complexity increases with the value of the potential $l$ of the counted local minima, as it usually happens in disordered landscapes; it vanishes at $l=-0.1638$, which gives an estimate of the ‘ground state energy' of the model within the 1RSB approximation. 
Given the solution of the replica self-consistent equations, to each value of $l$ one can associate a unique value of diversity 
\begin{equation}
    \phi= \int \frac{dz}{\sqrt{2 \pi}} e^{-\frac{z^2}{2}} \int d\nu_{\tilde m}(\xi; z) \;  \mathbb{E}_{0} \quadre{\theta(N)}.
\end{equation}
 In Fig. \ref{fig:Monasson} ({\it right}) we show the 1RSB complexity as a function on diversity. The comparison with the annealed Kac-Rice complexity is given in Fig. \ref{fig:Annealed} ({\it right}) and discussed in the main text. 
 
One sees from the expressions in this Appendix that there is an apparent similarity between the self-consistent equations obtained within the replicated Kac-Rice formalism, and those obtained within the replica framework. Therefore, it is natural to wonder in which limit the replica solution can be recovered within the quenched Kac-Rice framework. First, we remark that an analogous version of the Kac-Rice order parameters $p, \xi_0$ and $\xi_1$
can be obtained as moments of the effective force \eqref{eq:EffectiveForce}, in analogy with \eqref{eq:TripleAv} with $N$ replaced by $f_{\rm eff}(N)$ (similarly for $z$). 
Comparing the expressions in this Appendix with those in Eq. \eqref{eq:NewParametersApp} and in the following ones, we see that the integral expressions for the order parameters coincide formally provided that the following conditions hold in the Kac-Rice framework: $ x_2/x1= r e^{\hat \phi}= r \beta_2/\beta_1=(1-\alpha)^{-1/2}$
with $\alpha$ defined in \eqref{eq:DefDeltaAlfa}, and $ \beta_1 \beta_2-\beta_3^2=0$, which is the singular limit of the Gaussian measure in Eq. \eqref{eq:NewParametersApp} and in the following ones. However, we find that the replica solutions for the order parameters do not solve \emph{all} of the quenched Kac-Rice self-consistent equations under the assumptions above at the value of diversity determined by the replica solution. This is compatible with the fact that imposing a fixed value of potential $l$ of the counting minima or imposing a fixed diversity $\phi$ is not equivalent: the typical local minima at a given level-set of the potential have a certain diversity, but they are not the typical (most numerous) minima at that diversity (which are those picked up by the Kac-Rice calculation).

\newpage

\bibliographystyle{unsrt}
\bibliography{BibliographyLV.bib}

\end{document}